Center for Long-term Artificial Intelligence

International Research Center for AI Ethics and Governance
Institute of Automation, Chinese Academy of Sciences

# AI Governance InternationaL Evaluation Index

## AGILE Index 2024

(Released on Feb 4th 2024)

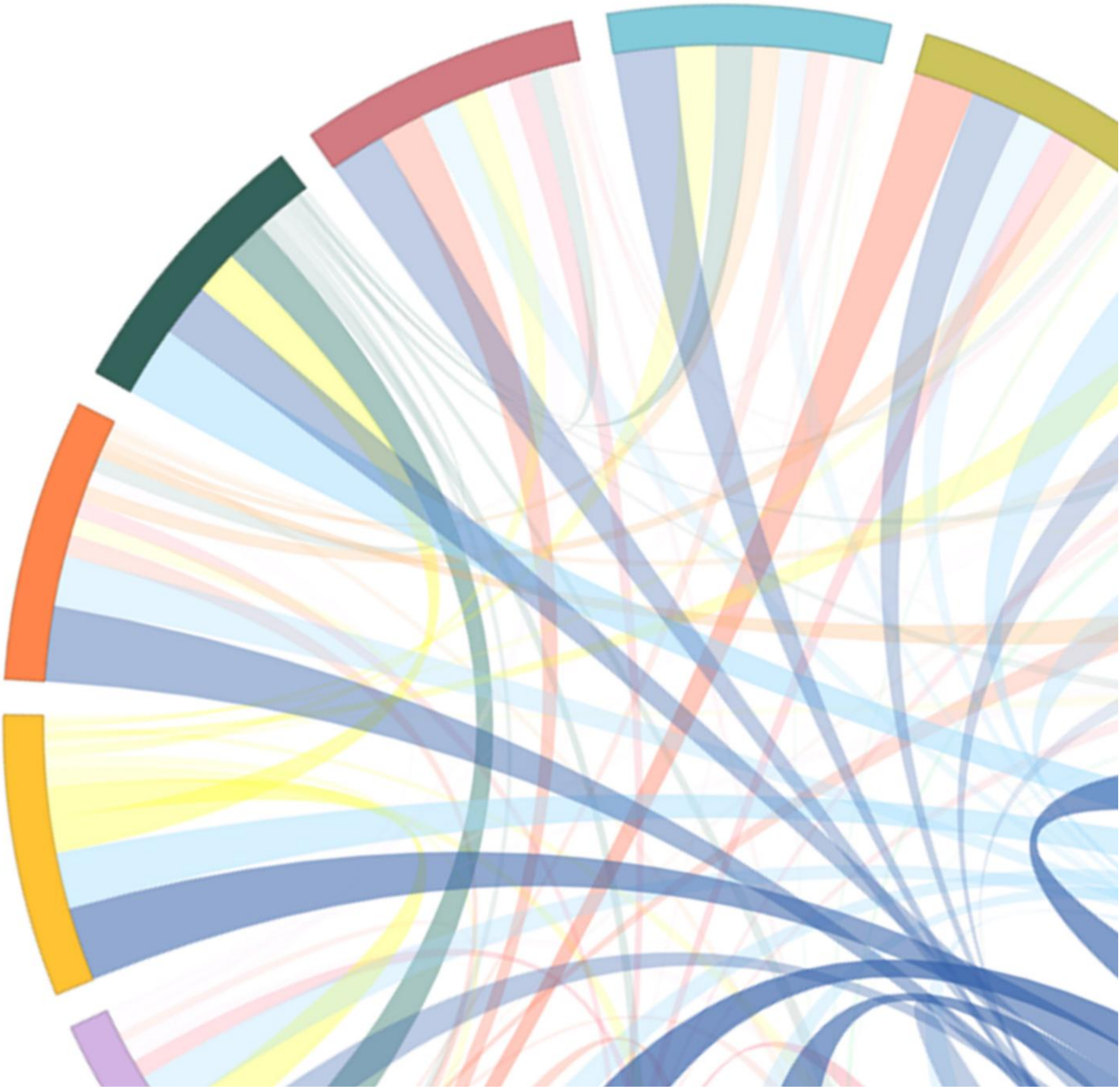

**Citation Information:**

**Website information:**

https://agile-index.ai/

**Institutes information:**

Center for Long-term Artificial Intelligence (CLAI)
https://long-term-ai.center/
International Research Center for AI Ethics and Governance,
Institute of Automation, Chinese Academy of Sciences
https://ai-ethics-and-governance.institute/

**Funding information:**

This research was supported by the National Science and Technology Major Project of China (Grant No. 2022ZD0116202).

**Contact information:**

Please contact contact@long-term-ai.cn for more information or with any comments.

**Report Version:**

**Copyright information:**

# Executive Summary

The rapid advancement of Artificial Intelligence (AI) technology is profoundly transforming human society and concurrently presenting a series of ethical, legal, and social issues. The effective governance of AI has become a crucial global concern. During the past year, the extensive deployment of generative AI, particularly large language models, marked a new phase in AI governance. Continuous efforts are being made by the international community in actively addressing the novel challenges posed by these AI developments. As consensus on international governance continues to be established and put into action, the practical importance of conducting a global assessment of the state of AI governance is progressively coming to light.

In this context, **the Center for Long-term Artificial Intelligence (CLAI), in collaboration with the International Research Center for AI Ethics and Governance hosted at the Institute of Automation, Chinese Academy of Sciences, jointly initiated the development of the AI Governance InternationaL Evaluation Index (AGILE Index).** The index is utilized to delve into the status of AI governance to date in 14 countries for the first batch of evaluation. The aim is to depict the current state of AI governance in these countries through data scoring, assist them in identifying their governance stage and uncovering governance issues, and ultimately offer insights for the enhancement of their AI governance systems.

Adhering to the design principle, "the level of governance should match the level of development," the inaugural evaluation of the AGILE Index commences with an exploration of four foundational pillars: the development level of AI, the AI governance environment, the AI governance instruments, and the AI governance effectiveness. **It covers 39 indicators across 18 dimensions to comprehensively assess the AI governance level of 14 representative countries globally**. The countries evaluated include the Group of Seven (G7, namely the United States, United Kingdom, France, Germany, Japan, Canada, Italy), the BRICS countries (Brazil, Russia, India, China, South Africa), and representative countries from specific regions (Singapore, the United Arab Emirates), totalling 14 countries.

## The first AGILE Index evaluation reveals many noteworthy findings:

### Overall,

1. The United States, which scoring slightly above 70, heads the first tier, followed by China, Singapore, Canada, Germany, and the United Kingdom, all scoring above 60. (Page 17)
2. There is a strong positive correlation between AGILE Index score and the per capita GDP. (Page 17)
3. BRICS countries show slightly better performance in effective governance. (Page 18)
4. Differences in the governance environment categorize 14 countries' score distribution into three types. Singapore, Canada, Germany, Japan, and France scored more evenly in all pillars. (Page 19)

### In terms of AI development,

1. The United States exhibits notable performance in the field of AI development compared to other countries. (Page 22)
2. China exhibits significant progress in AI development alongside opportunities for further AI infrastructure enhancement. (Page 23)
3. Beyond the United States and China, a global mosaic of strengths emerges. (Page 23)

### In terms of AI governance environment,

1. There was a sharp 12-fold increase in documented AI risk incidents in 2023, underscoring the pressing need for AI governance to keep pace with rapid technological advancements. (Page 25)
2. Among the 14 evaluated countries, especially the USA, face a significant proportion of documented AI risk incidents globally, highlighting the collective pressure on AI governance worldwide. (Page 26)
3. Although high-income countries often demonstrate a higher level of preparedness for AI governance, it's important to recognize the opportunity for all countries to excel in AI governance, regardless of their overall governance readiness. (Page 27)

## In terms of AI governance instruments,

1. The 14 evaluated countries showed relatively strong performance in AI strategy, AI governance bodies, and participation in international AI governance engagement. (Page 29)
2. Between 2020 and 2023, the governance instruments of AI have evolved from setting broad principles in the preceding five years to the development of tangible measures, including AI legislation, AI standards, and AI impact assessment tools. (Page 31)
3. In the context of international participation in AI governance among the 14 evaluated countries, the United Kingdom, France, and Japan have demonstrated significant involvement. (Page 33)
4. AI legislation varies globally, with some countries adopting comprehensive state-level laws, while others integrate AI-specific amendments into existing frameworks or follow a more fragmented approach with state-specific initiatives and federal guidelines. (Page 34)

## In terms of AI governance effectiveness,

1. The public in BRICS countries generally express higher levels of trust in AI compared to their counterparts in high-income countries. (Page 36)
2. A stark gender imbalance permeates AI researchers across all 14 evaluated countries, with only about one in five researchers being female. (Page 37)
3. Further analysis on gender gap underscores a critical challenge: almost no countries excel at both AI gender inclusivity and broader societal gender equality, which calls for further research attention. (Page 38)
4. While all 14 countries actively engage in global developer communities, the United States, China, and India stand out for their substantial contributions and impact. (Page 39)
5. Approximately 3% to 4% of all AI-related publications focus on AI governance. (Page 41)
6. The total volume of literature on AI governance has shown an exponential acceleration in recent years, with a growth rate reaching 45% in 2022. (Page 42)
7. Security, safety, and collaboration were consistently the most researched topics related to AI governance, with long-term AI and accountability receiving significantly less research focus. (Page 43)
8. There are visible collaborations on AI governance among all countries, indicating that AI governance is globally connected and indivisible. (Page 44)
9. China and the United States together contribute more than half of the papers in nearly all AI for SDGs research directions. SDG3 (Good health and well-being), SDG9 (Industry, innovation, and infrastructure) and SDG4 (Quality Education) are the three most popular research topics for all 14 countries. (Page 46)

10. In AI for SDGs application, while SDG3, SDG9, and SDG4 remain popular, there are also notable number of projects for SDG11 (Sustainable Cities and Communities), SDG12 (responsible consumption and production) and SDG13 (Climate actions). (Page 48)

# Table of Contents

# Research Team

## Chief Scientist

**Yi Zeng**
Professor and Director
Brain-inspired Cognitive AI Lab
Chinese Academy of Sciences

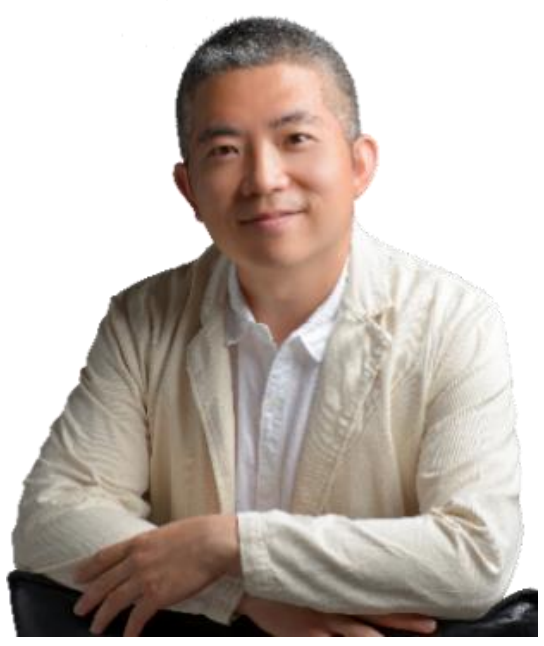

Professor Yi Zeng is a Professor and Director at the Brain-inspired Cognitive AI Lab, as well as the founding Director of the International Research Center for AI Ethics and Governance, both situated at the Institute of Automation, Chinese Academy of Sciences. He is also the founding Director of the Center for Long-term AI and leads the AI for SDGs Cooperation Network and the Defense AI and Arms Control Network. In addition, Yi serves in several key roles, including as Director of the Professional Committee of Information Technology and Artificial Intelligence at the Committee for Ethics in Science and Technology, Chinese Academy of Sciences. Furthermore, Yi chairs the Professional Committee on Mind Computation at the Chinese Association for AI, is a board member of the National Governance Committee for the New Generation AI in China and serves on the Committee on AI at the National Committee on Science and Technology Ethics in China. He holds positions as a member of the UN High-level Advisory Body on AI, the UNESCO Ad Hoc Expert Group on AI Ethics, and the WHO Expert Group on AI Ethics/Governance for Health. His areas of interest include Brain and Mind-inspired AI, AI Safety, Ethics, Governance, and AI for Sustainable Development.

## Core Team

- Enmeng Lu — Institute of Automation, Chinese Academy of Sciences
- Xin Guan — Center for Long-term AI
- Cunqing Huangfu — Institute of Automation, Chinese Academy of Sciences
- Zizhe Ruan — Institute of Automation, Chinese Academy of Sciences
- Ammar Younas — Institute of Philosophy, Chinese Academy of Sciences

## Data & Content Support

- Kang Sun — Center for Long-term AI
- Xuan Tang — Institute of Automation, Chinese Academy of Sciences
- Yuwei Wang — Institute of Automation, Chinese Academy of Sciences
- Hongjie Suo — Center for Long-term AI
- Dongqi Liang — Institute of Automation, Chinese Academy of Sciences
- Zhengqiang Han — Institute of Philosophy, Chinese Academy of Sciences
- Aorigele Bao — Institute of Philosophy, Chinese Academy of Sciences
- Xiaoyang Guo — Institute of Philosophy, Chinese Academy of Sciences
- Jin Wang — Institute of Philosophy, Chinese Academy of Sciences
- Jiawei Xie — Institute of Philosophy, Chinese Academy of Sciences
- Yao Liang — Institute of Automation, Chinese Academy of Sciences

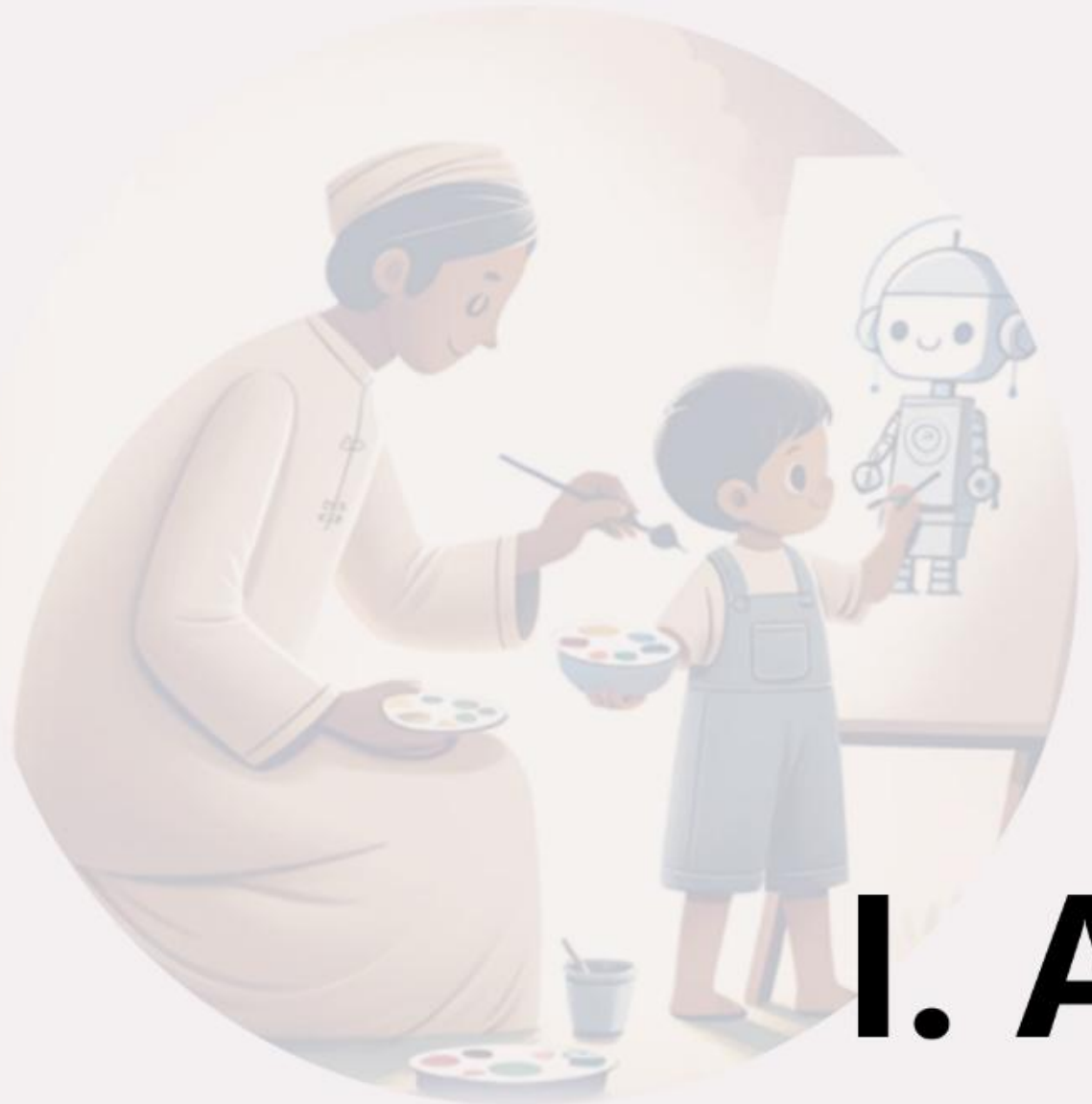

# I. AGILE Index

## 1.1. Evaluation Framework

The design philosophy of the AI Governance InternationaL Evaluation Index (AGILE Index) aligns with the principle that **"the level of governance should match the level of development."** Specifically, a country should align its AI governance level with its overall AI development level to ensure the healthy and sustainable growth of Artificial Intelligence. The principle's essence is to avoid both uncontrolled development due to insufficient governance, which can harm society, and the stifling of technological innovation from excessive governance. **The goal of governance is to achieve a beneficial healthy interaction between technological innovation and social welfare, so as to maximize the benefits of AI and minimize its potential risks.** Based on this, the AGILE Index has four pillars for AI governance evaluation: development level, governance environment, governance instruments, and governance effectiveness. Through a thorough assessment of these aspects, the AGILE Index aims to provide each country with a comprehensive and clear analysis of the status of AI governance, thereby offering robust support for future development and governance.

***Figure 1** Four pillars for AI governance evaluation*

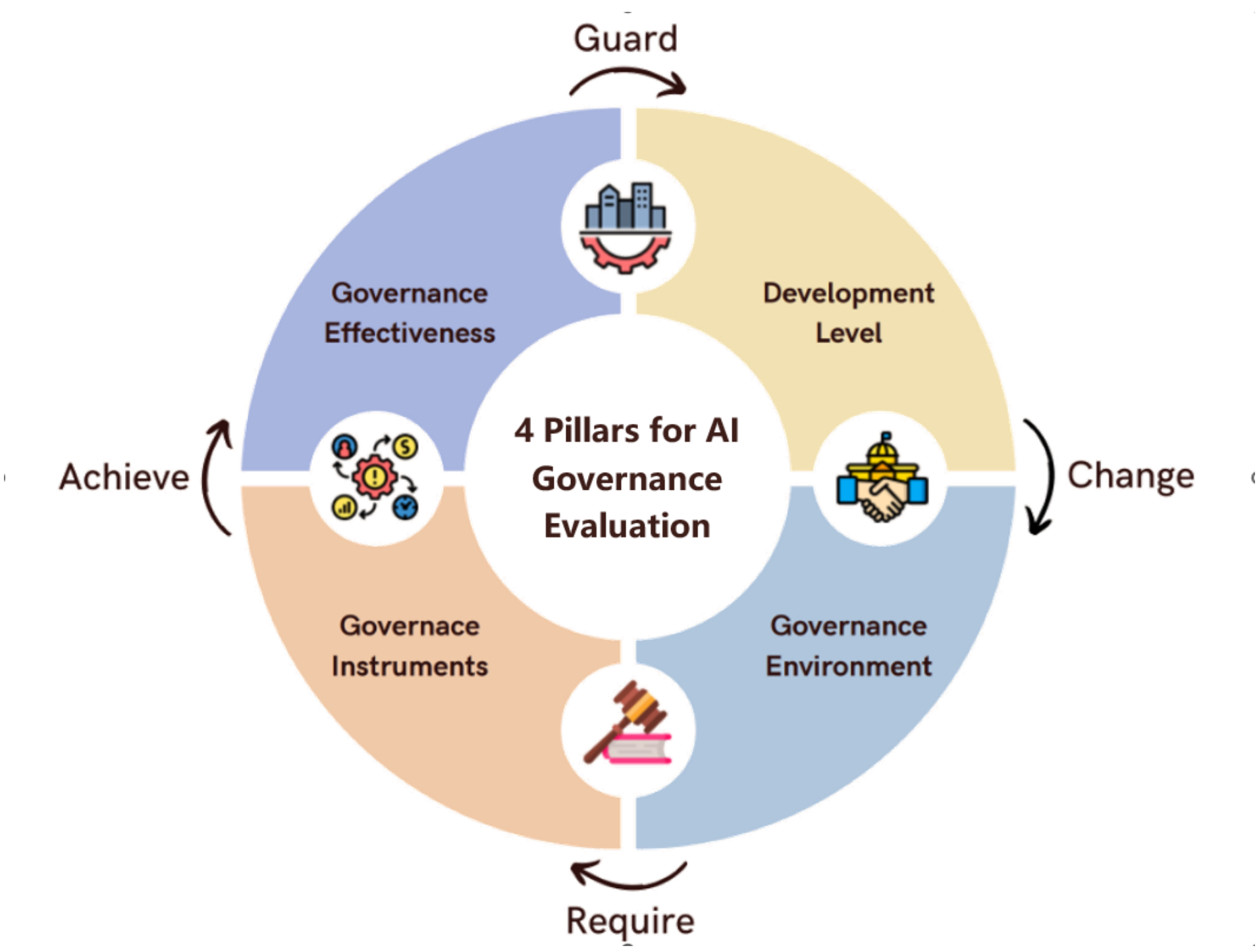

## Development Level

The pillar score regarding development level reflects a country's scale of AI research and development, infrastructure, and industry. A higher score signifies AI R&D at a larger scale, enhanced infrastructure, and a more mature industry. As the development level increases, new issues and risks will emerge, necessitating a reassessment of the country's current status.

## Governance Environment

The pillar score for the governance environment reflects the technological and political landscape in AI Governance. A higher score suggests fewer governance challenges, more advanced support, and a reduced governance burden. As the governance environment evolves, the country must respond swiftly and implement new governance instruments.

## Governance Instruments

The pillar score for governance instruments reflects how comprehensive the AI governance tools are in a country. A high score signifies that the country possesses a robust and varied array of instruments to guarantee safe and ethical AI applications. The quality of implementation of these governance instruments will be reflected in governance effectiveness.

## Governance Effectiveness

The pillar score for governance effectiveness reflects the effectiveness of AI governance in practice. A high score indicates great public trust in AI, considerable transparency in data algorithms, and vigorous research and application in AI governance and AI for SDGs. These elements lay the groundwork for the continued healthy development of AI.

## 1.2. Indicator system

The AGILE Index, built on **four pillars**, comprises **18 dimensions and 39 indicators**, offering a comprehensive framework for evaluating the AI governance status in various countries. The table below outlines these dimensions and indicators. For an in-depth understanding of each indicator, including their data sources and the methodology used for the index score calculation, please see **Appendices 1 and 2**.

*Table 1 AGILE Index Dimensions and Indicators*

| Pillars | Dimensions | Indicators |
|---|---|---|
| **P1. Development Level** | **D1. AI Research and Development Activity** | D1.1. Number of **publications in AI**-related journals/conferences & the per capita ratio |
| | | D1.2. Number of **professionals** in the field of AI & the per capita ratio |
| | | D1.3. Number of granted **AI patents** & the per capita ratio |
| | | D1.4. Number of **AI systems** developed & the GDP ratio |
| | **D2. AI Infrastructure** | D2.1. Number of colocation **data centers** & the per capita ratio |
| | | D2.2. Non-distributed **supercomputers** floating point operations per second & the per capita ratio |
| | **D3. AI Industry Scale** | D3.1. AI companies' total **funding** & the GDP ratio |
| | | D3.2. Number of **AI startups** & the GDP ratio |
| | | D3.3. Number of **AI companies listed** on stock exchanges & the GDP ratio |
| | | |
| **P2. Governance Environment** | **D4. AI Risk Exposure** | D4.1. Number of **AI-related risk cases/incidents** & the GDP ratio |
| | **D5. AI Governance Readiness** | D5.1. Overall assessment of the **level of governance** in the country |
| | | D5.2. The overall process of achieving **sustainable development goals** in the *country* |
| | | |
| **P3. Governance Instruments** | **D6. AI Strategy & Planning** | D6.1. Whether an **AI strategy** has been released in the country |
| | | D6.2. Whether the AI strategy has **measurable goals** |
| | | D6.3. Whether the AI strategy mentions **training or skills upgrading** |
| | | D6.4. **Budget scale** & the GDP ratio for AI-specific expenditure |
| | **D7. AI Governance Bodies** | D7.1. Whether **AI governance bodies** have been established in the country |
| | **D8. AI Principles & Norms** | D8.1. Whether governments have issued **AI principles or norms** |
| | **D9. AI Impact Assessment** | D9.1. Whether governments have introduced **AI impact assessment mechanisms** |
| | | D9.2. Number of **regulatory sandboxes** for safety test of (financial) AI |
| | **D10. AI Standards & Certification** | D10.1. Whether governments have developed **standards and certification mechanisms** for AI |
| | **D11. AI Legislation Status** | D11.1. Whether countries have enacted a **national-level law** regarding AI |
| | | D11.2. Whether countries have implemented **data protection legislation** specifically addressing AI |
| | | D11.3. Whether countries have enacted **consumer protection legislation** specifically tailored to AI |
| | | D11.4. Whether countries are working on **AI legal instruments** which are at a later stage of enactment |
| | **D12. Global AI Governance Engagement** | D12.1. The participation level in **international AI governance mechanisms** |
| | | D12.2. The participation level in **ISO AI standardization** |
| **P4. Governance Effectiveness** | **D13. Public Understanding of AI** | D13.1. The **AI-related skill proficiencies** of the public |
| | | D13.2. The level of the public's **awareness of AI's impact** |
| | **D14. Public Trust in AI** | D14.1. The level of the **public's positive attitude** towards AI's development |
| | | D14.2. The level of **enterprises' positive attitudes** towards AI's adoption |
| | **D15. AI Development Inclusivity** | D15.1. Gender ratio of **AI literature authors** |
| | | D15.2. Gender ratio of **graduates in AI-related majors** |
| | | D15.3. The level of AI accessibility by **disadvantaged groups** |
| | **D16. Data & Algorithm Openness** | D16.1. Number of impactful **open AI models and datasets** released |
| | | D16.2. The level of contributions in the **AI developer community** |
| | **D17. AI Governance Research Activity** | D17.1. Total number & the proportion of **literature on AI governance** topics |
| | **D18. AI for SDGs (AI4SDGs) Activity** | D18.1. Total number & the proportion of **literature on AI and SDGs** topics |
| | | D18.2. Number of reported cases of **AI applications for SDGs** & the GDP ratio |

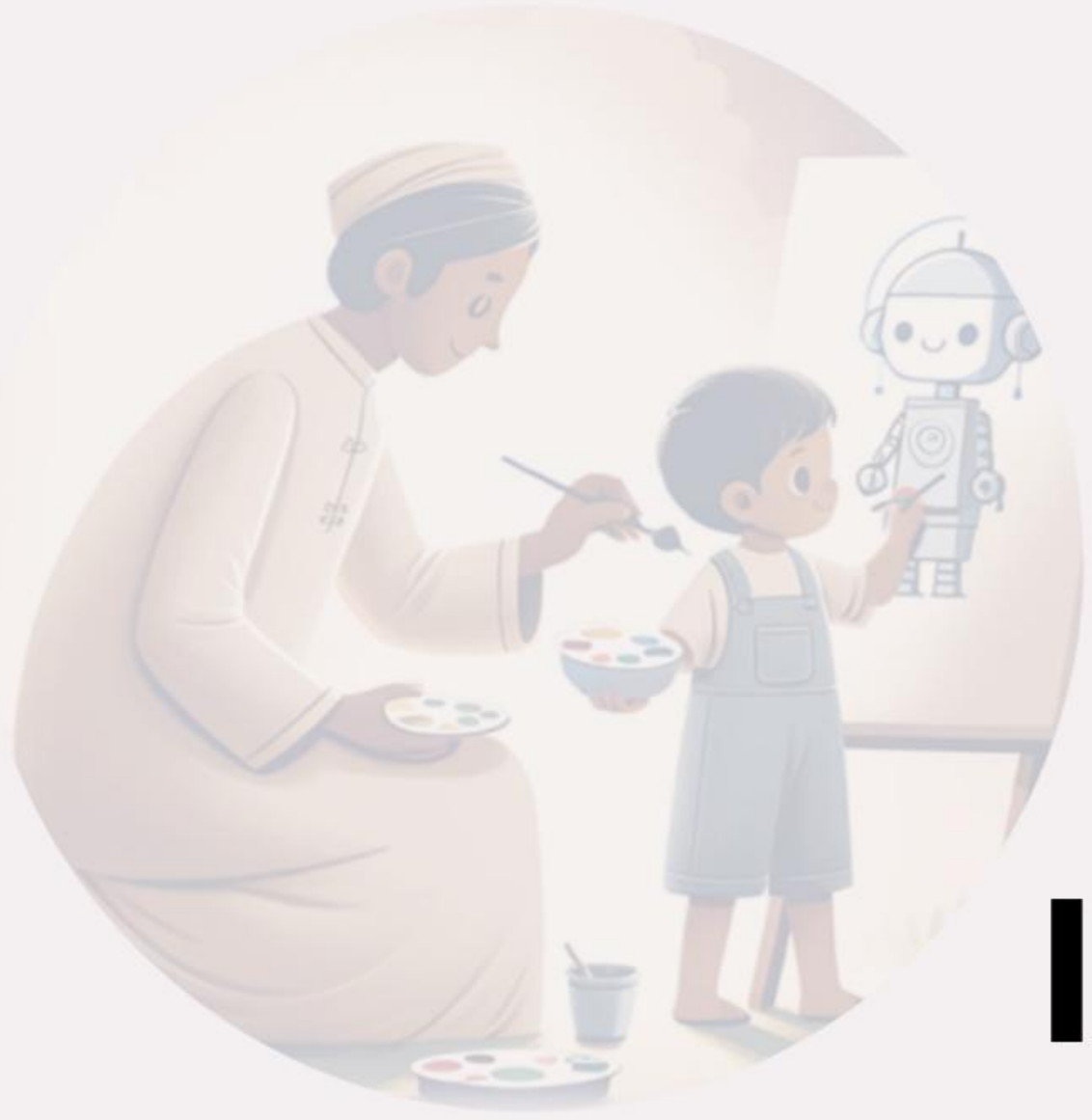

# II. Overview

# 2.1. Score Composition

*Table 2* ***AGILE Index Total Score, Pillar Score, and Dimension Score***

| | | US | China | Singapore | Canada | Germany | UK | Japan | France | UAE | Italy | India | Russia | Brazil | South Africa |
|---|---|---|---|---|---|---|---|---|---|---|---|---|---|---|---|
| T | Total Score | 72.3 | 68.5 | 66.4 | 64.9 | 62.6 | 61.4 | 56.5 | 55.3 | 49.4 | 49.4 | 48.1 | 42.7 | 37.9 | 34.4 |
| R | Ranking | 1 | 2 | 3 | 4 | 5 | 6 | 7 | 8 | 9 | 10 | 11 | 12 | 13 | 14 |
| D1 | AI R&D Activity | 99 | 72 | 68 | 66 | 65 | 71 | 53 | 50 | 31 | 29 | 34 | 33 | 16 | 11 |
| D2 | AI Infrastructure | 100 | 40 | 52 | 56 | 69 | 63 | 60 | 60 | 32 | 62 | 25 | 26 | 28 | 15 |
| D3 | AI Industry Scale | 100 | 80 | 59 | 77 | 41 | 74 | 39 | 51 | 25 | 38 | 13 | 4 | 14 | 20 |
| P1 | Development Level | 100 | 64 | 60 | 66 | 58 | 69 | 51 | 54 | 29 | 36 | 31 | 21 | 19 | 15 |
| D4 | AI Risk Exposure | 100 | 69 | 28 | 57 | 52 | 88 | 36 | 31 | 40 | 28 | 43 | 61 | 40 | 19 |
| D5 | AI Gov. Readiness | 77 | 59 | 80 | 84 | 85 | 83 | 84 | 82 | 72 | 74 | 59 | 51 | 61 | 56 |
| P2 | Governance Environment | 38 | 45 | 76 | 63 | 67 | 48 | 74 | 76 | 66 | 73 | 58 | 45 | 60 | 69 |
| D6 | AI Strategy & Planning | 91 | 98 | 87 | 88 | 90 | 93 | 85 | 91 | 79 | 89 | 59 | 86 | 81 | 0 |
| D7 | AI Gov. Bodies | 100 | 100 | 100 | 100 | 100 | 100 | 100 | 100 | 100 | 100 | 0 | 100 | 0 | 0 |
| D8 | AI Principles & Norms | 100 | 100 | 100 | 100 | 100 | 100 | 100 | 0 | 100 | 0 | 0 | 100 | 0 | 0 |
| D9 | AI Impact Assess. | 100 | 17 | 82 | 67 | 0 | 67 | 17 | 50 | 82 | 0 | 40 | 17 | 32 | 17 |
| D10 | AI Standards | 100 | 100 | 0 | 100 | 100 | 0 | 0 | 0 | 0 | 0 | 0 | 100 | 0 | 0 |
| D11 | AI Legislation | 50 | 75 | 50 | 75 | 75 | 75 | 75 | 75 | 50 | 75 | 75 | 50 | 75 | 25 |
| D12 | Global AI Gov. Engage | 93 | 93 | 93 | 86 | 93 | 100 | 100 | 100 | 54 | 93 | 93 | 79 | 68 | 54 |
| P3 | Governance Instruments | 90 | 83 | 73 | 88 | 80 | 76 | 68 | 59 | 66 | 51 | 38 | 76 | 37 | 14 |
| D13 | Public AI Underst. | 31 | 100 | 99 | 44 | 48 | 36 | 46 | 43 | 5 | 45 | 65 | 43 | 39 | 57 |
| D14 | Public AI Trust | 14 | 100 | 71 | 21 | 43 | 23 | 54 | 20 | 51 | 58 | 80 | 48 | 56 | 58 |
| D15 | AI Dev. Inclusivity | 33 | 100 | 68 | 50 | 28 | 51 | 21 | 30 | 79 | 43 | 74 | 31 | 21 | 71 |
| D16 | Data & Alg. Openness | 87 | 59 | 59 | 55 | 33 | 56 | 16 | 41 | 19 | 9 | 90 | 27 | 40 | 6 |
| D17 | AI Gov. Research | 100 | 81 | 34 | 53 | 63 | 67 | 38 | 34 | 7 | 39 | 32 | 10 | 21 | 13 |
| D18 | AI4SDGs Activity | 100 | 51 | 8 | 29 | 61 | 80 | 27 | 28 | 56 | 32 | 54 | 17 | 35 | 39 |
| P4 | Governance Effectiveness | 61 | 82 | 56 | 42 | 46 | 52 | 34 | 32 | 36 | 38 | 66 | 29 | 35 | 41 |

## 2.2. Overall Observations

### Key observation 1: Four Tiers of AGILE Index Scores

The index categorizes 14 countries into **four tiers** based on scores above 60, 50, and 40 points. The United States, which scoring slightly above 70, heads the first tier, followed by China, Singapore, Canada, Germany, and the United Kingdom, all scoring above 60. Japan and France, with 56.5 and 55.3 points, are in the second tier. The Italy, United Arab Emirates, India, and Russia, scoring between 40 to 50, fall in the third tier. Brazil and South Africa, scoring below 40, are in the fourth tier.

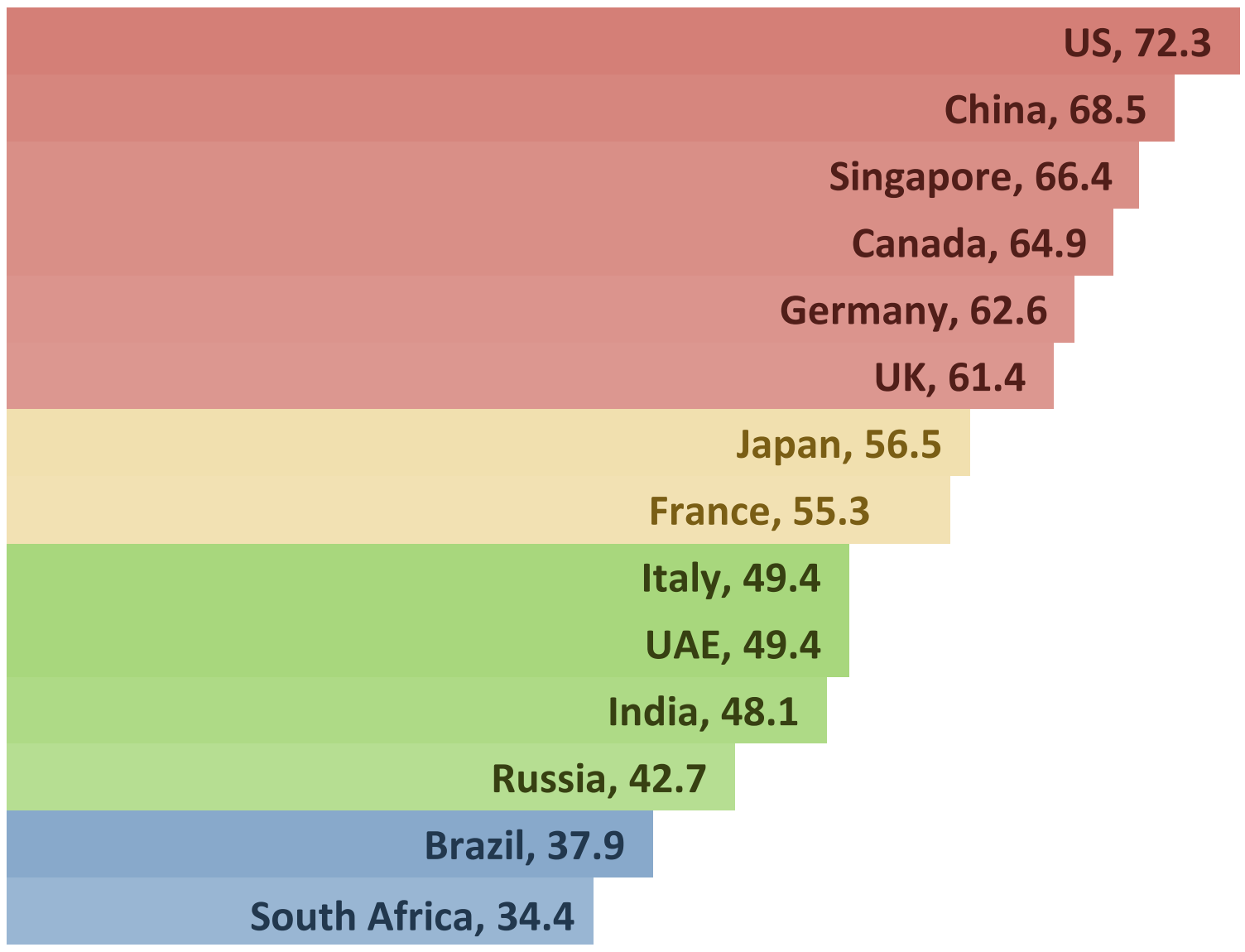

*Figure 2 14 Countries' AGILE Index Scores and Tiers*

---

### Key observation 2: There is a strong positive correlation between AGILE Index score and the per capita GDP.

The AGILE Index score demonstrates a strong positive correlation with the per capita GDP of various countries. For analytical purposes, we categorize the 14 nations into **two groups** based on per capita GDP: a high-income group, including nine countries such as the G7 nations, Singapore, and the UAE, and a BRICS group of five middle-income countries pre-2023 expansion. Notably, the UAE is treated as high-income in this report, despite its BRICS membership. The data clearly shows that higher-income countries typically score above BRICS nations in the AGILE Index. However, China and India's AGILE Index scores (68.5;48.1) are significantly higher than their per capita GDP levels, while the United Arab Emirates' AGILE Index score (49) is observably lower than its per capita GDP level.

*Figure 3* ***Positive relationship between AGILE Index Score and the per capita GDP***

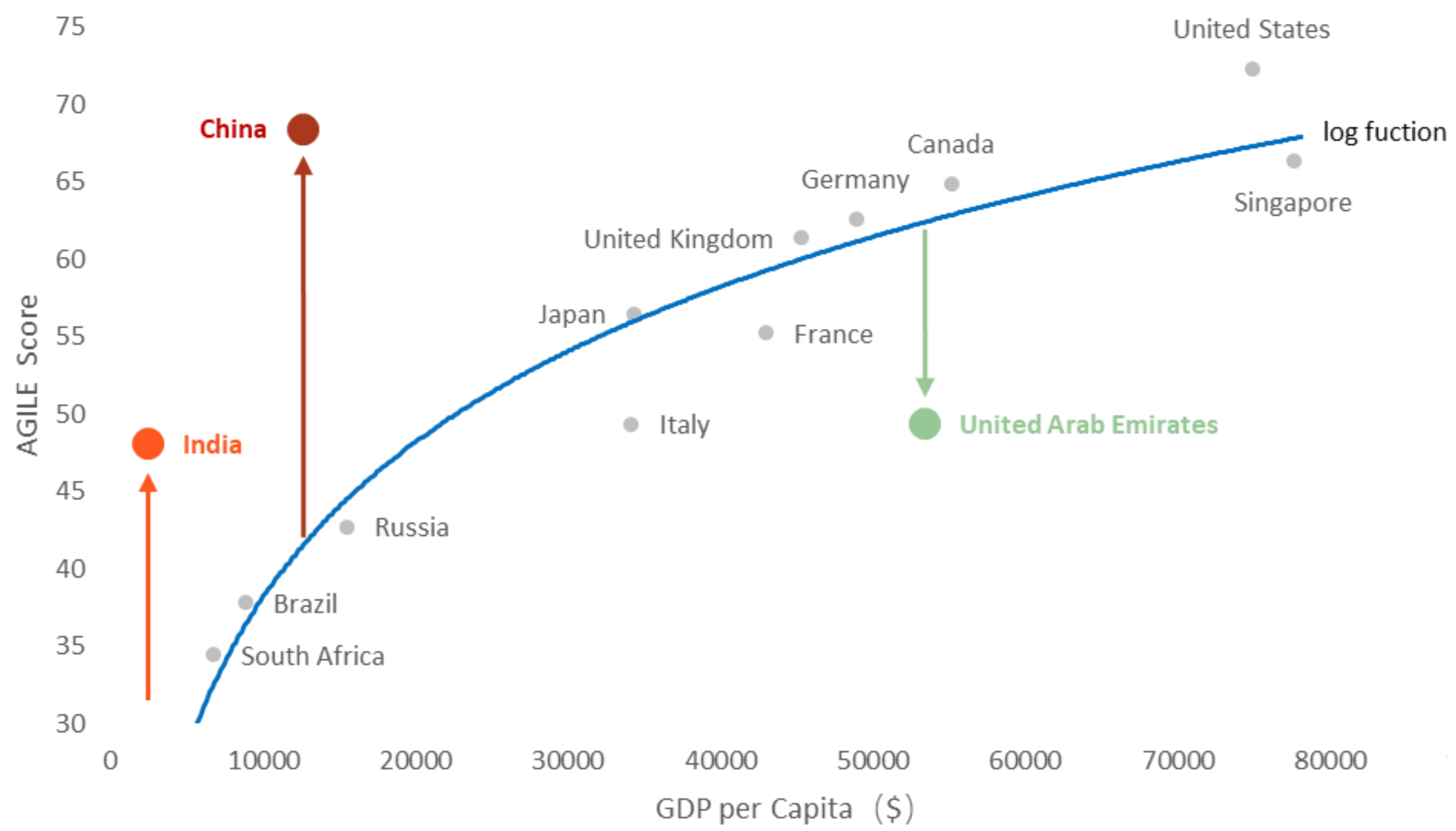

The clear positive correlation between the AGILE Index score and per capita GDP is a further indication of **the indispensability of development as a foundation for governance**. The higher overall scores of China and India compared to their per capita GDP levels can partly be attributed to their higher levels of AI development relative to their per capita GDP. Additionally, both countries have good performance on AI governance effectiveness pillar, with China and India scoring 82 and 66 respectively, ranking first and second among the 14 countries.

---

## Key observation 3：BRICS countries show slightly better performance in effective governance, as they exhibit better performance in dimensions of public understanding, trust, and development inclusivity of AI.

The high-income country group's average scores in the four pillars of development level, governance environment, governance instruments, and governance effectiveness are 58, 65, 72, and 44 respectively. In contrast, BRICS countries score 30, 55, 50, and 51. The high-income country group leads the BRICS countries by 28.1, 9.2, and 22.7 points in the first three pillars, while the BRICS countries lead by 6.5 points in governance effectiveness. This suggests high-income countries excel in development and governance tools, while BRICS countries show slightly better performance in governance effectiveness. In specific, compared to high-income countries, BRICS countries exhibit observably better performance in dimensions of effective AI governance, particularly in public understanding, trust, and development inclusivity of AI.

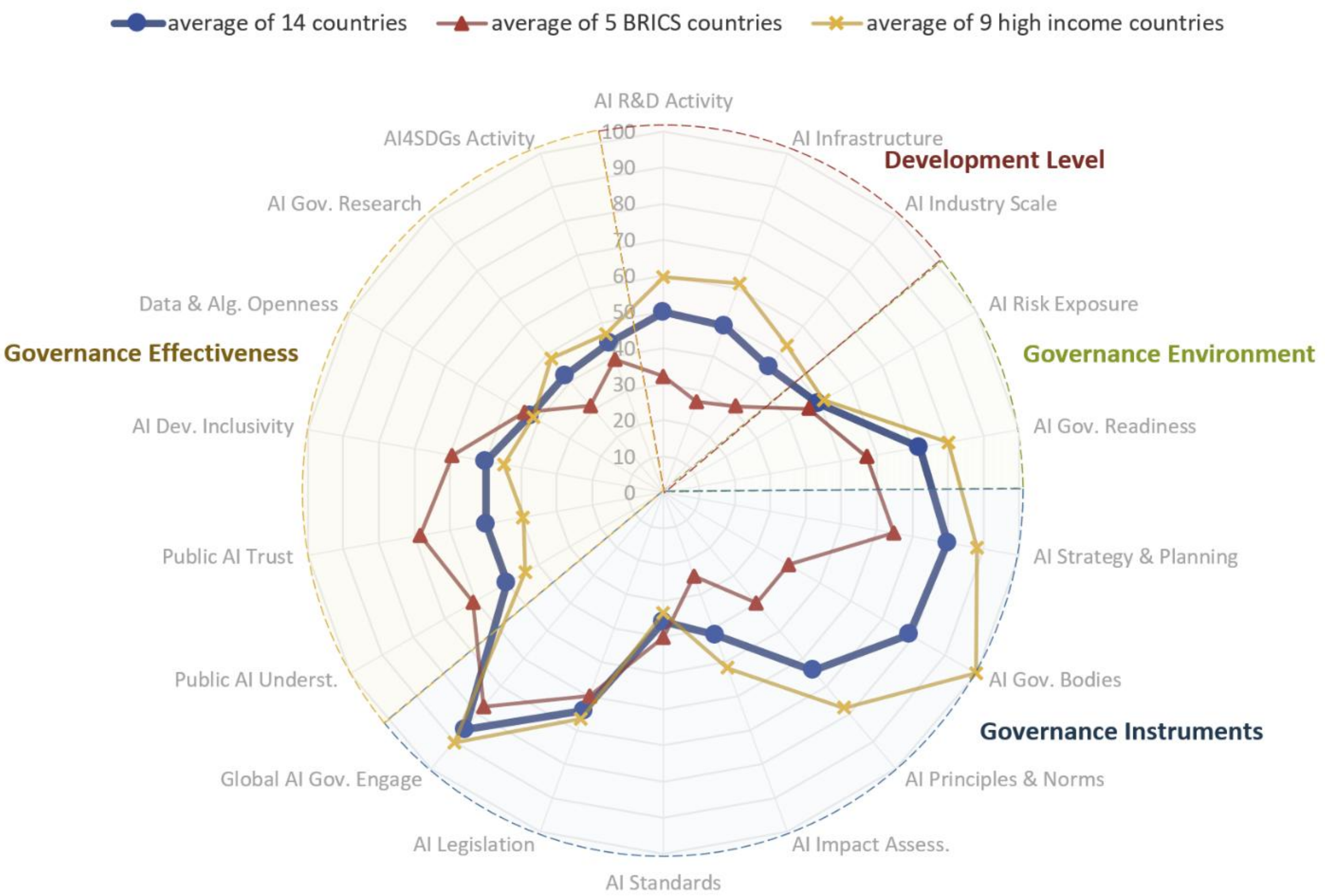

*Figure 4* ***Dimension average for BRICS, high-income countries, and both groups.***

---

## Key observation 4： Differences in the governance environment categorized the 14 countries into three governance types. Singapore, Canada, Germany, Japan, and France scored more evenly in all pillars.

Further analysis of AGILE Index scores across different pillars enables us to categorize the 14 assessed countries into **three types**. China, the United States, and the United Kingdom scored higher overall in the first AGILE Index evaluation, but lower in the governance environment pillar. This reflects that although these three countries are more advanced in AI development, have invested significantly in AI governance, and achieved relatively good effectiveness, they also face significantly **higher AI governance pressures** and challenges than other countries. Singapore, Canada, Germany, Japan and France scored more evenly in all pillars, indicating that these countries have implemented comprehensive governance instruments under lesser governance pressure. In contrast, India, Brazil, Italy, and the United Arab Emirates, despite scoring higher in the governance environment, have relatively lower overall AGILE Index scores.

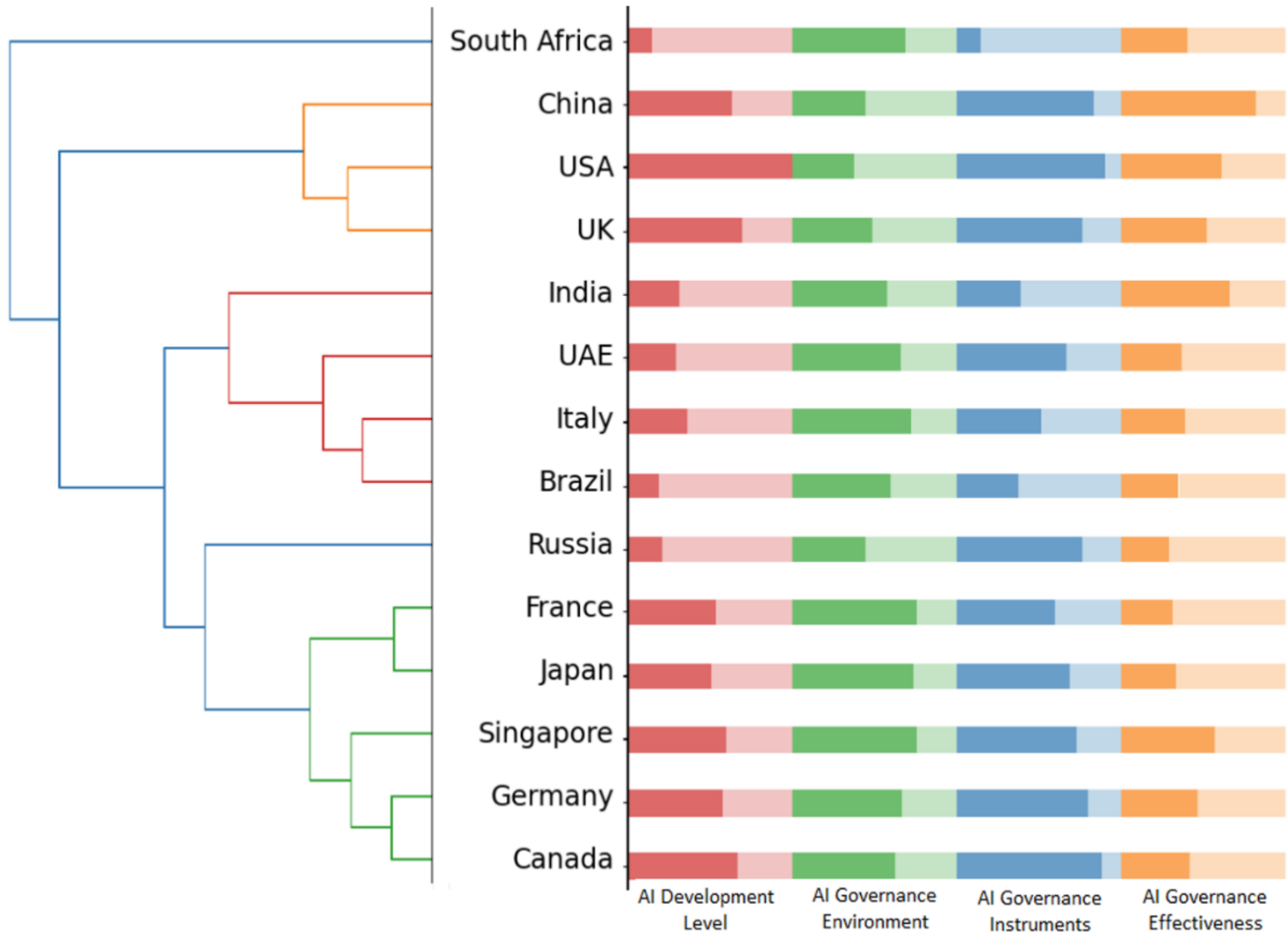

***Figure 5 Three types of the 14 countries' AGILE Index Pillar score distribution***

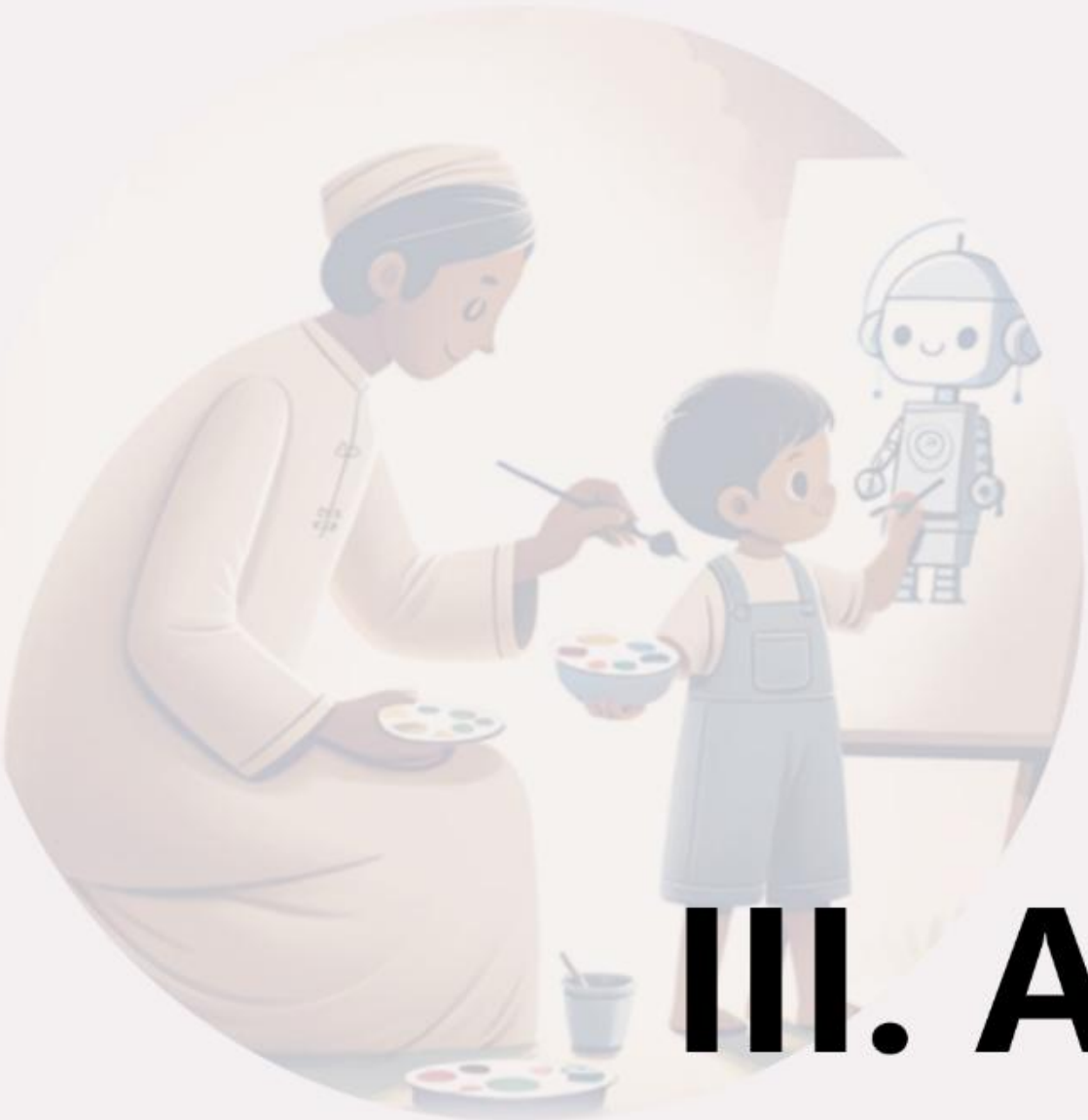

# III. Analysis and Observations

## 3.1 Pillar 1: AI Development Level

### Pillar 1 overview: The AGILE Index assesses each country's level of AI development across three dimensions: AI research and development (R&D), AI infrastructure construction, and AI industry scale.

In total, the evaluation of 14 countries reveals more than two million AI professionals, over one million research papers, nearly one hundred thousand patents, and the publication of nearly 500 significant machine learning systems. Collectively, these countries possess a combined supercomputer computing power exceeding 9,000 pFLOP/s, and houses over 3,000 colocation data centers, facilitating diverse AI R&D activities.

***Table 3*** ***Total figures in the development pillar***

(Data Source: *Tortoise Media, Epoch.AI, DBLP, Top500 List and Data Center Map*)

| AI-related **researchers** | AI-related **articles** | AI-related **patents** | Significant AI **systems** | (pFLOP/s [Rpeak]) **Supercomputer operations** | Colocation **data centers** |
|---|---|---|---|---|---|
| 2 million | 1 million | 100k | 500 | 9k | 3k |

---

### Observation 1.1: The United States holds a clear lead over other countries in AI development level.

Figure 6 visually summarizes key indicators of AI activity across 14 countries. This analysis excludes per capita ratios to provide a high-level comparison of total numbers in various metrics, including research papers, researchers, patents granted, machine learning systems, supercomputing power, data centers, and AI company funding and startups. Notably, the United States holds a clear lead, contributing over one-third of the combined research papers and professionals and exceeding half the total in many other areas.

## Observation 1.2: China demonstrates notable strengths in AI development, while facing challenges in AI infrastructure building.

China demonstrates notable strengths in AI, ranking second overall in a pool of 14 countries. Notably, it leads the BRICS countries by a significant margin. Across five out of eight key indicators, China secures the second spot, with contributions ranging from 10% to 33%. When it comes to developed AI systems, China stands neck-and-neck with the United Kingdom, occupying the third position with nearly 10% of the total. Supercomputing power sees China close behind second-placed Japan, contributing 9% of operations. However, a concerning gap emerges in data center infrastructure. With only 2.5% of the total across the 14 countries, China falls to ninth place, highlighting a potential area for future development.

## Observation 1.3: Beyond the United States and China, a global mosaic of strengths emerges.

Beyond the US, several countries contribute significantly to the global AI landscape. The UK, Germany, and Canada consistently rank within the top five for six out of eight indicators. Germany leads in research papers (around 10%), researchers (around 9%), and supercomputing power (around 4.4%). The UK shines in machine learning systems (around 10%), data centers (around 8%), AI funding (around 4%), and startups (around 9%). Meanwhile, Japan secures third place in patents (around 5%) and supercomputer operations (around 10%), showcasing its unique strengths.

Among the BRICS countries, India stands out with impressive ranks in seven out of eight areas, only trailing Russia in AI patents. Notably, it also ranks fourth in the overall startup landscape, with roughly 5% of the total number. It's worth noting that the overall development levels tend to be higher in the high-income group compared to the BRICS group.

***Figure 6 Share of 8 key indicators in Development pillars among 14 countries.***

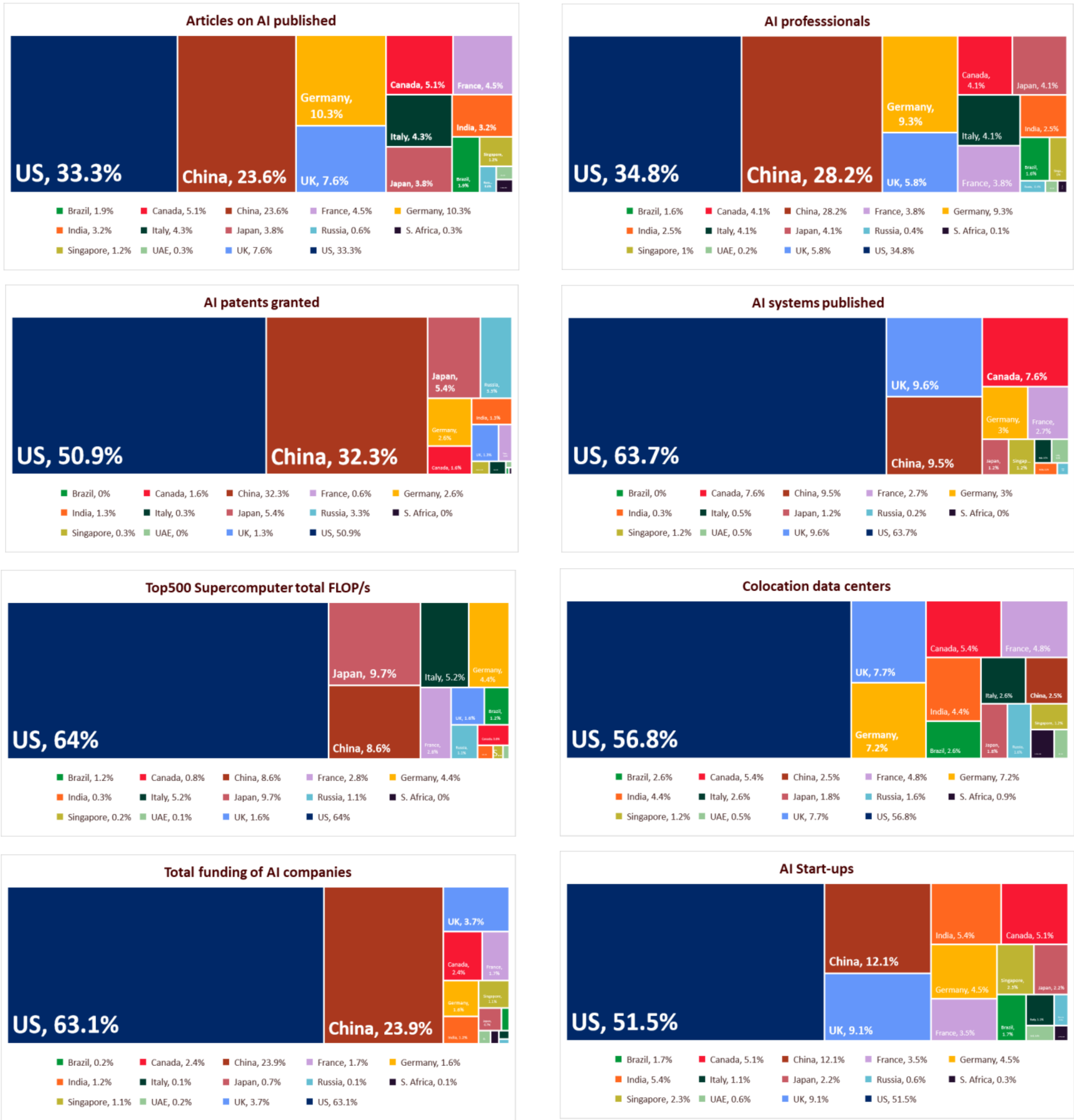

Data source: *Tortoise Media, Epoch.AI, DBLP, Top500 List and Data Center Map*

## 3.2. Pillar 2: AI Governance Environment

### Observation 2.1: There was a sharp 12-fold increase in documented AI risk incidents in 2023, underscoring the pressing need for AI governance to keep pace with rapid technological advancements.

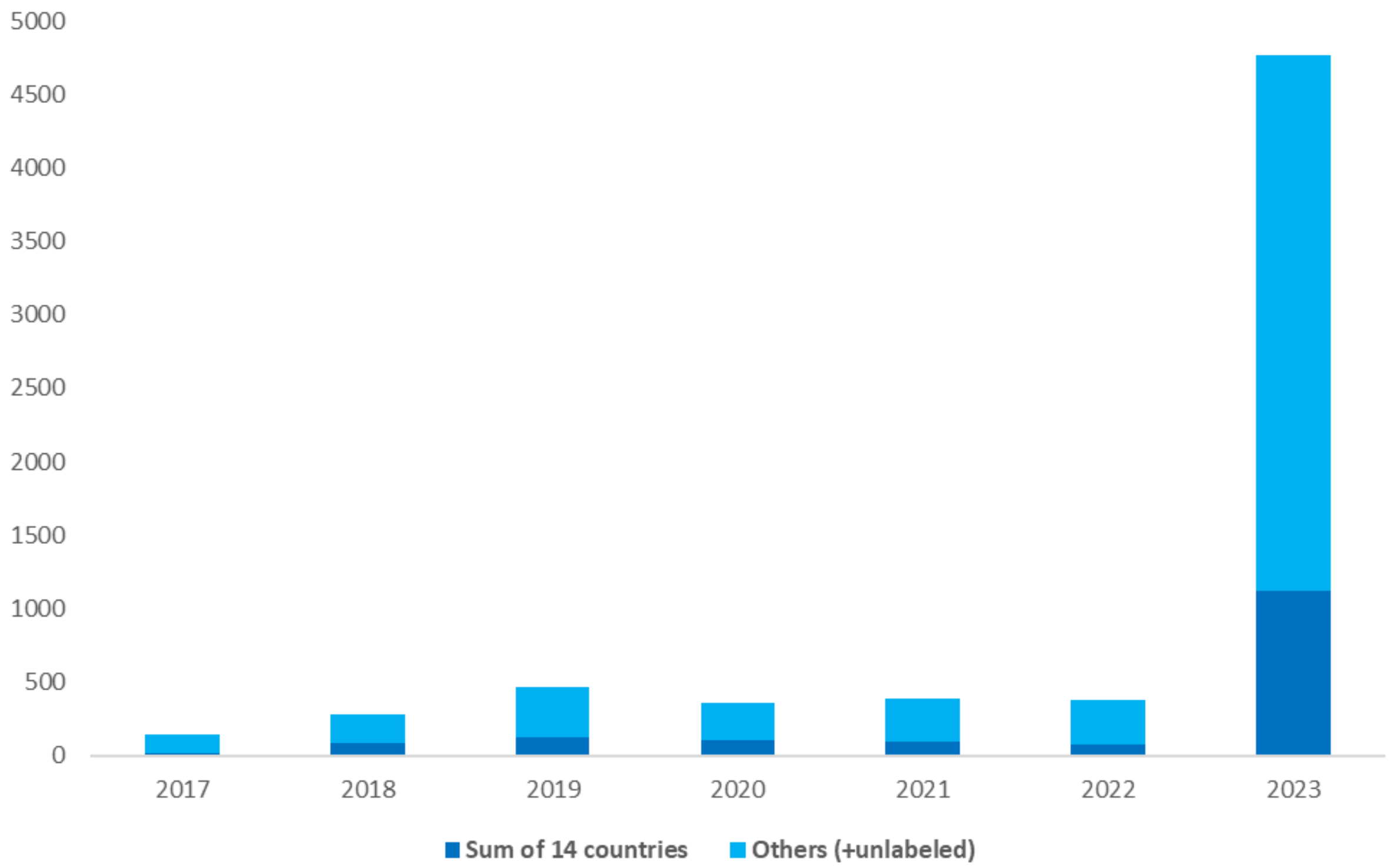

*Figure 7* ***Changes in the Number of AI Incidents from 2017 to 2023***

Data source*: OECD AIM*

Since generative AI technology like ChatGPT entered the scene in late 2022, its rapid development has been accompanied by growing concerns about potential risks. Data from the OECD AI Incidents Monitor reveals a significant surge in publicly reported AI risk incidents. By January 1, 2024, the database contained 7,198 such reports. The number of recorded incidents involving AI technology jumped significantly in 2023. Compared to 2022, there were 12.8 times more incidents, with 4,409 cases documented – representing a staggering 61% of all recorded incidents since the beginning. This trend held true across all 14 countries, with annual growth ranging from 4 to over 70 times. The sharp rise in AI incidents during 2023 highlights the urgent need for strong AI governance systems to catch up with the fast development of this technology.

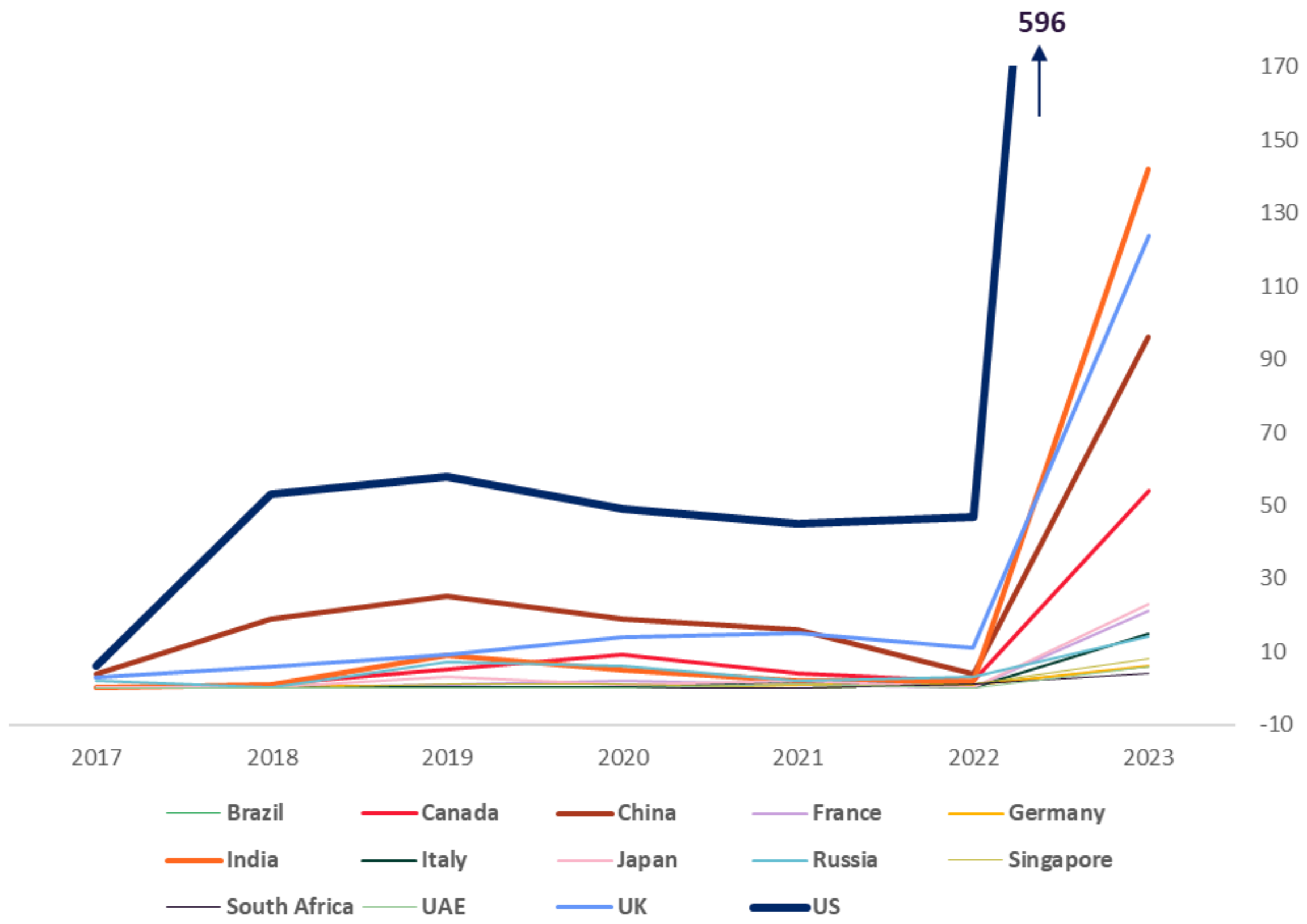

***Figure 8*** ***AI risk incidents of 14 countries (2017-2023)***
***(Data source: OECD AIM)***

---

## Observation 2.2: Among the 14 evaluated countries, especially the USA, face a significant proportion of documented AI risk incidents globally, highlighting the collective pressure on AI governance worldwide.

With 1,681 AI risk incidents (83% of identified origin) reported in 2023 originating from 14 countries, the assessed 14 countries hold significant responsibility in shaping effective global AI governance. Further analysis using data from other global AI risk incidents databases revealed that the United States accounted for 67% of all AI risk events involving the 14 assessed countries, far exceeding the contributions of the second-place China (9.3%) and the third-place United Kingdom (6.5%). The substantial gap in AI risk incidents between the US and other nations necessitates further investigation into the contributing factors, particularly considering its leading role in AI development.

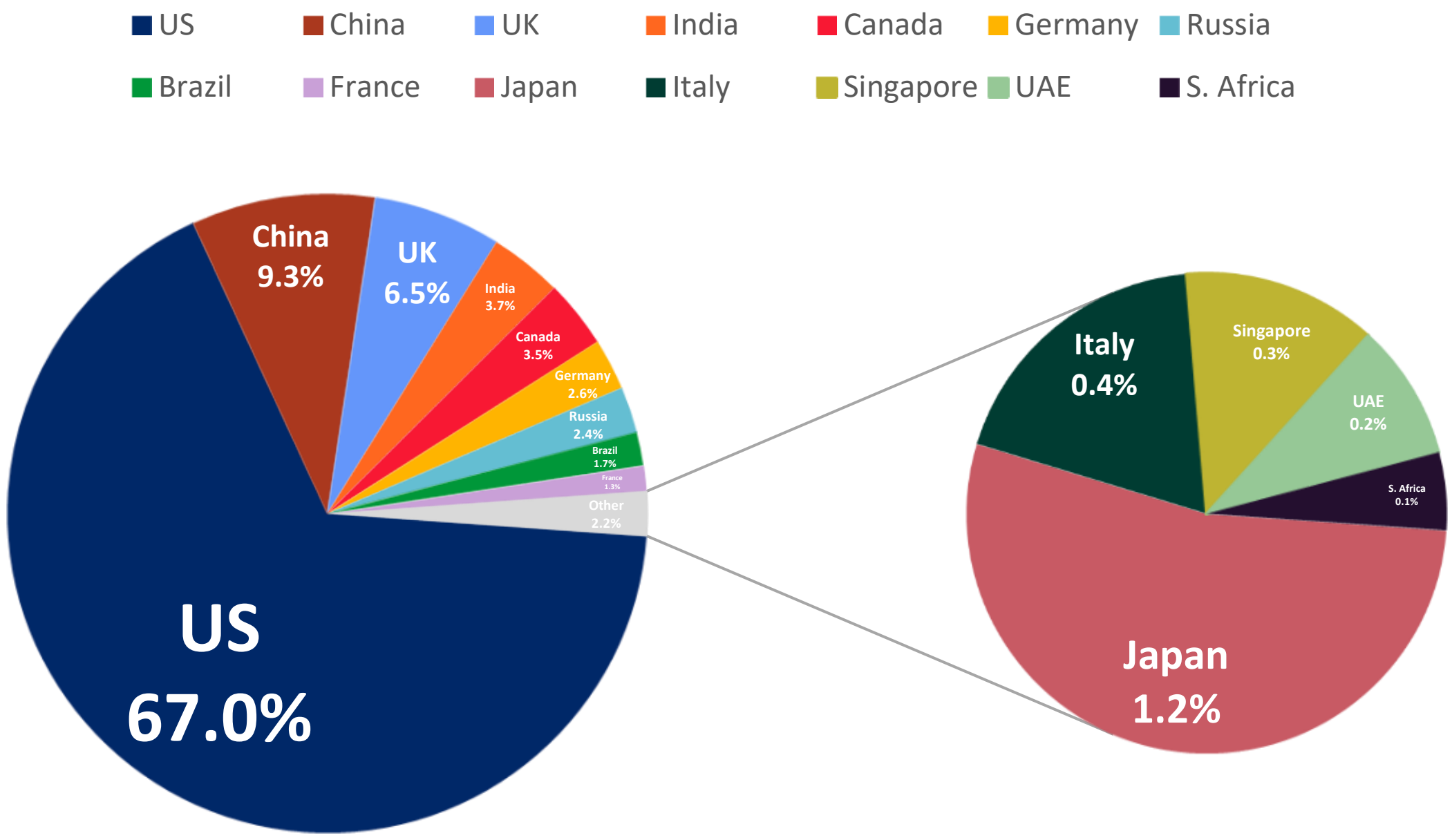

***Figure 9 Share of AI risk incidents in 14 Countries.***

---

## Observation 2.3: Although high-income countries often demonstrate a higher level of preparedness for AI governance, it's important to recognize the opportunity for all countries to excel in AI governance, regardless of their overall governance readiness.

To assess government general preparedness for the growing number of AI incidents, our evaluation examines each country's overall readiness to govern AI effectively. We combine indicators from two key aspects: 1) overall evaluation of the governance capability of a country, using the World Bank's Worldwide Governance Indicators (WGI) and the GovTech Maturity Index (GTMI), and 2) evaluation of a country's commitment to achieving Sustainable Development Goals (SDGs), using the SDG Development Index (SDGDI). While our analysis using WGI, GTMI, and SDGDI indices reveals that high-income countries generally exhibit better AI governance readiness compared to others, it's important to recognize the opportunity for all countries to excel in AI governance, regardless of their overall governance readiness. For example, although the US's performance in overall governance readiness is relatively low in high-income countries, it achieves the highest score in AGILE Index.

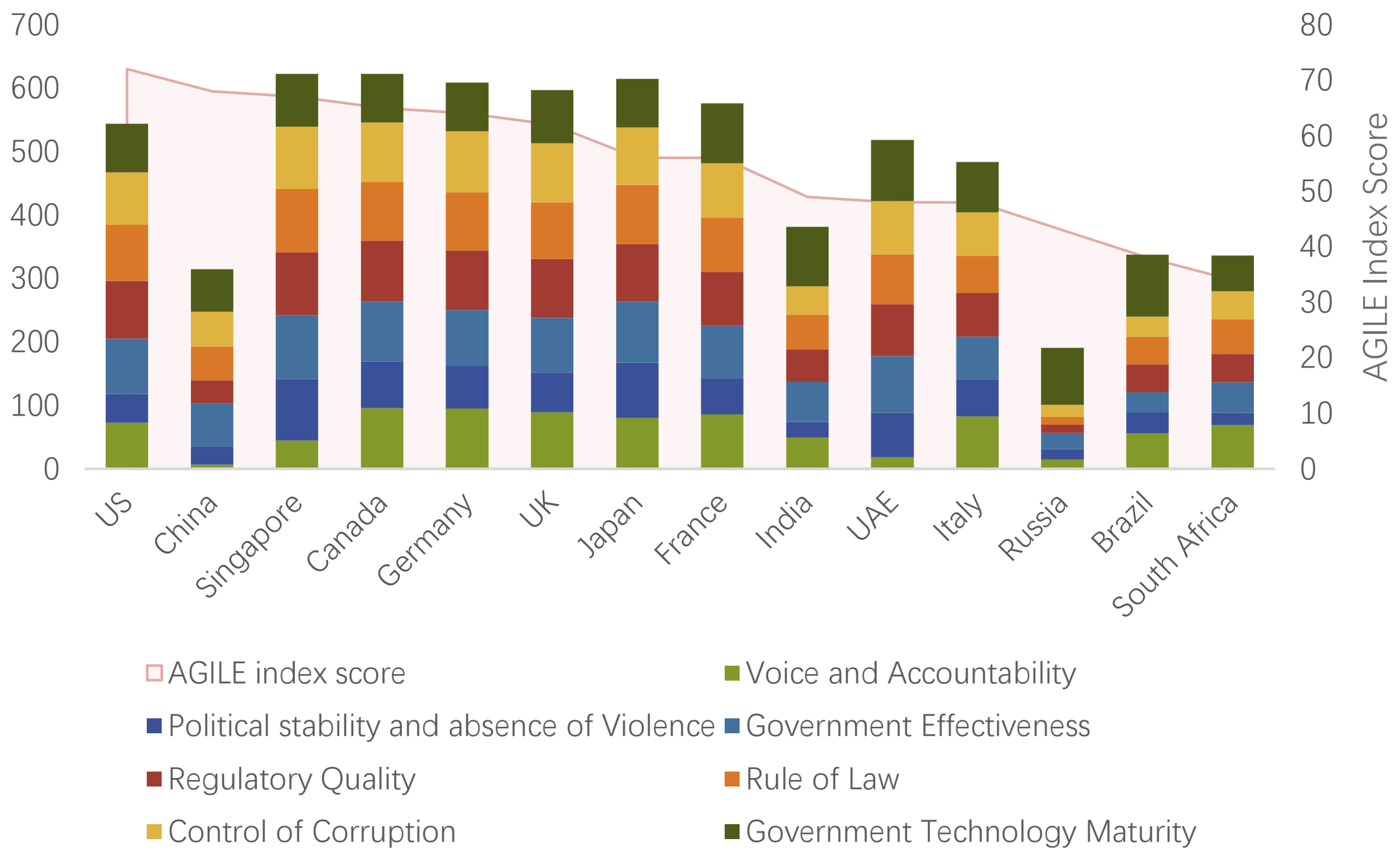

***Figure 10 Composition of Government overall readiness Scores.***

## 3.3. Pillar 3: AI Governance Instruments

### Pillar 3 overview: AGILE Index evaluates seven types of AI governance instruments.

AI governance encompasses a variety of instruments, each with a distinct function to ensure the responsible development and use of AI.

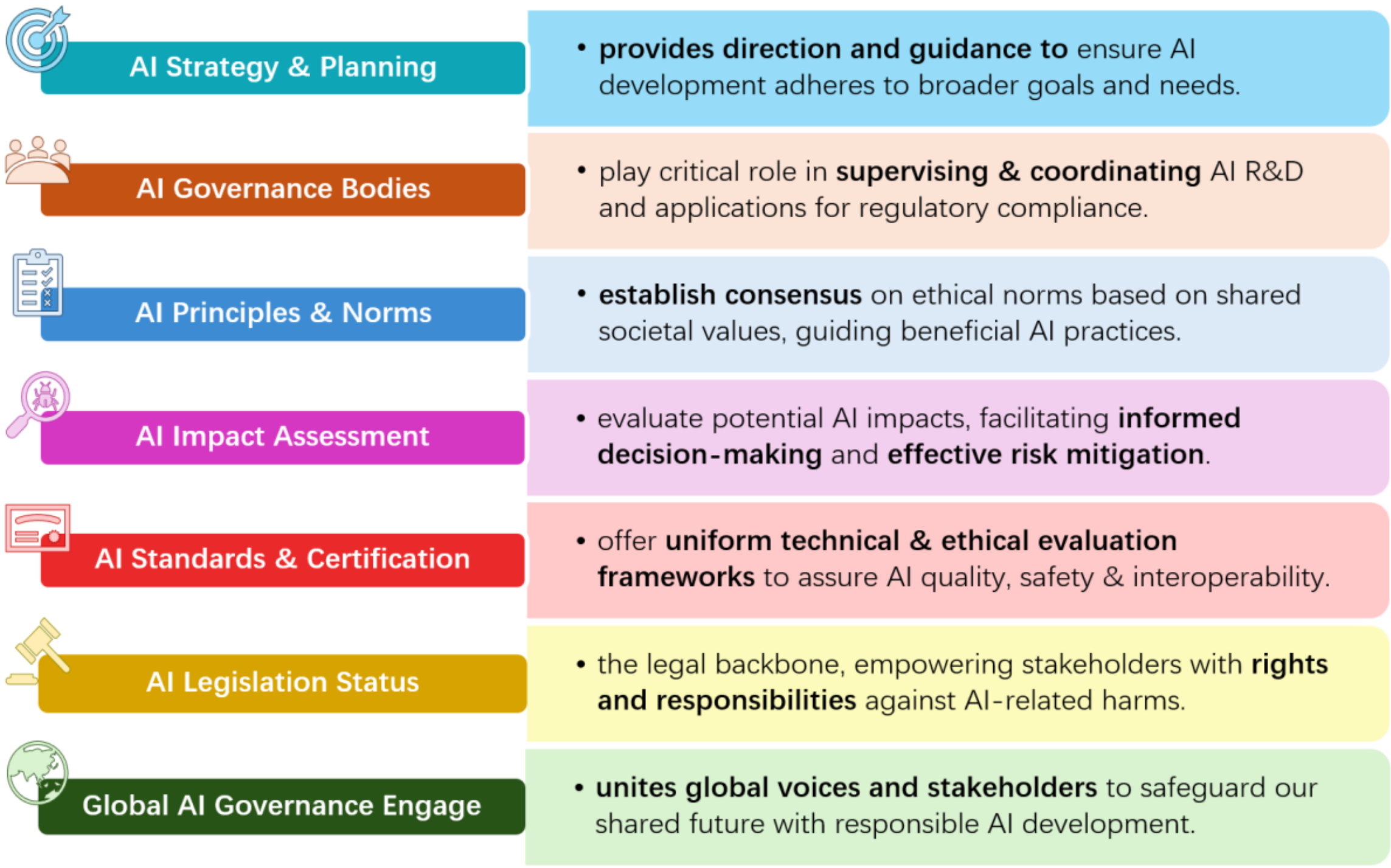

*Figure 11 A variety of AI governance instruments.*

### Observation 3.1: The 14 evaluated countries showed relatively strong performance in AI strategy, AI governance bodies, and participation in international AI governance.

The 14 evaluated countries showed relatively strong performance in AI strategy, AI governance bodies, and participation in international AI governance. However, there is room for improvement in areas such as AI standard certification, impact assessments, and legislation. Among these countries, most have published AI strategies. Ten countries have established AI principles and norms, six have introduced AI ethics assessment tools, five have implemented AI governance ethics standards, and four have enacted national laws pertaining AI.

We classify the instruments into 2 categories, one is the non-legislative category encompassing the five instruments of strategy, governance body, principle, assessment mechanism and standards, the other is the legislative category encompassing the three selected instruments of National AI laws, Data Protection laws for AI, Consumer Protection laws for AI. Based on our analysis, we observe that all high-income countries have released national AI strategies and formed AI governance bodies. France and Italy do not have nation-wise principles, one reason is that they have EU equivalents which can guide their practice as well. For example, EU has passed the Ethics Guidelines for Trustworthy AI, which are equally effective for France and Italy as members of EU.

| | AI **Strategies** Released | AI **Governance Bodies** | Government-issued AI **Principles** | AI **Impact Assessment** Mechanisms | AI **Standards** & Certifications |
|---|---|---|---|---|---|
| US | 2023 | 2018 | 2019 | 2023 | 2022 |
| Canada | 2017 | 2019 | 2023 | 2019 | 2023 |
| China | 2017 | 2019 | 2019 | - | 2023 |
| Germany | 2020 | 2019 | 2019 | - | 2022 |
| UK | 2021 | 2021 | 2018 | 2023 | - |
| Russia | 2019 | 2022 | 2022 | - | 2023 |
| Singapore | 2023 | 2018 | 2020 | 2022 | - |
| Japan | 2022 | 2017 | 2018 | - | - |
| UAE | 2017 | 2018 | 2019 | 2019 | - |
| France | 2021 | 2023 | - | 2022 | - |
| Italy | 2022 | 1996 | - | - | - |
| India | 2018 | - | - | - | - |
| Brazil | 2021 | - | - | - | - |
| S.Africa | - | - | - | - | - |

*Figure 12* ***The publication year of the 14 countries' non-legal AI governance instruments.***

Regarding the five non-legislative instruments, the United States and Canada have implemented all of these at a national level. For the legislative instruments, while no countries currently have comprehensive legislation covering AI, most are actively engaged in either drafting new AI-related laws or revising existing legislation to address AI-related incidents.

**Observation 3.2: Between 2020 and 2023, the governance instruments of AI has evolved from setting broad principles in the preceding five years to the development of tangible measures, including AI legislation, AI standards, and AI impact assessment tools.**

Generally, the majority of the 14 countries released their AI strategies, governance bodies, and principles between 2017 and 2020. However, from 2020 to 2023, there was a shift in focus towards AI legislation, AI standards, and AI impact assessment mechanisms.

In terms of publication timelines, we can categorize the development into three periods: before 2015, from 2015 to 2020, and from 2020 onwards. Most national governance bodies, strategies, and principles were established between 2015 and 2020, with some countries following suit in the subsequent three years. In contrast, the development of many standards and impact assessment tools began primarily after 2020, with numerous countries still in the process of formulating them. Regarding legislation on data protection and consumer protection, many were initially introduced before 2015 and have since been updated to address the challenges brought about by newer generations of AI technology.

Another notable point is the establishment of Italy's Data Protection Authority in 1996, known as Garante per la protezione dei dati personali. The Garante has proven to be a significant body in AI governance. For instance, on March 30, 2023, the Garante issued a temporary emergency order directing OpenAI LLC to cease using ChatGPT for processing personal data of individuals in Italy. This order was subsequently lifted at the end of April 2023.

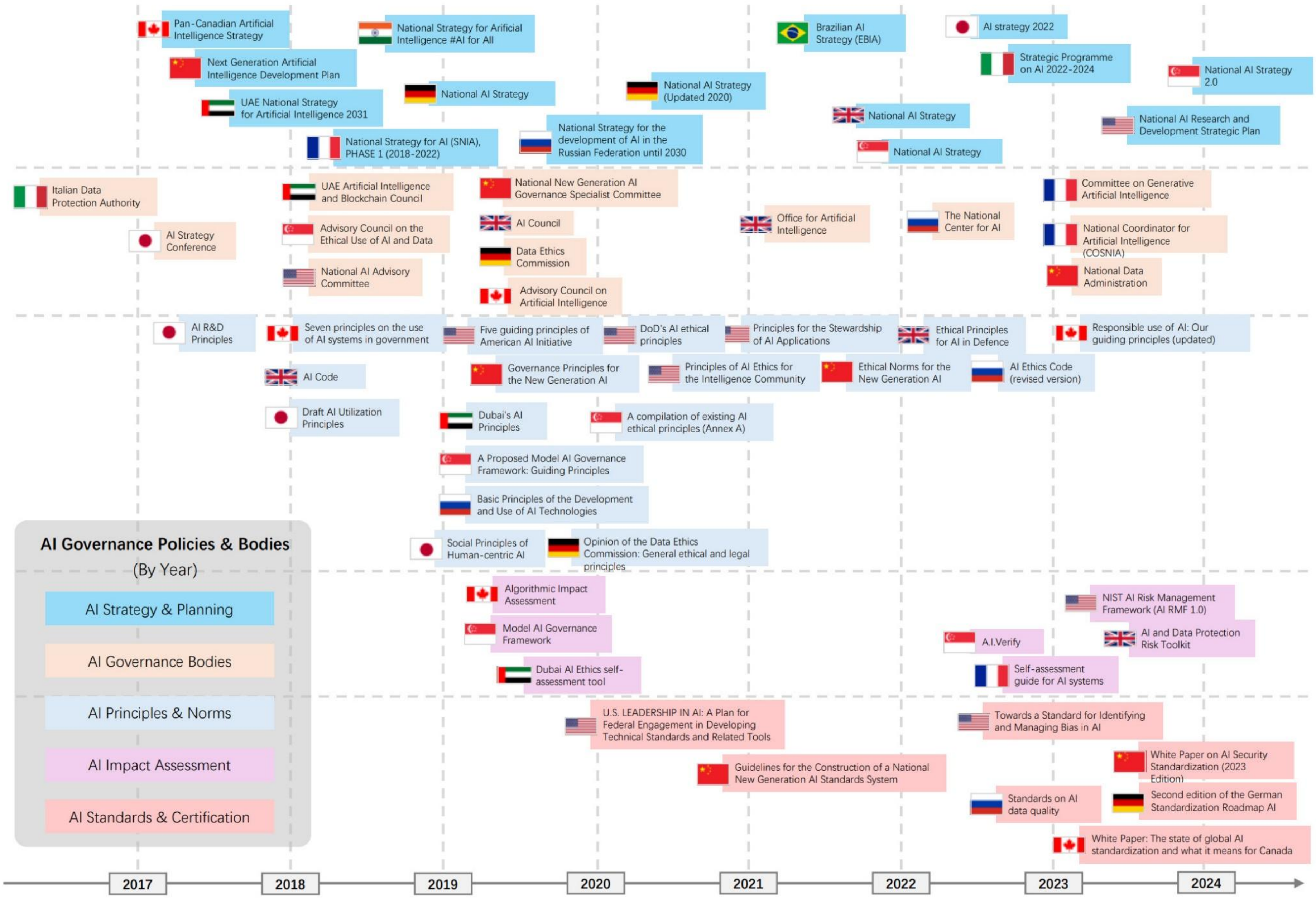
AI Governance Policies & Bodies
(By Year)
AI Strategy & Planning
AI Governance Bodies
AI Principles & Norms
AI Impact Assessment
AI Standards & Certification
Pan-Canadian Artificial Intelligence Strategy
Next Generation Artificial Intelligence Development Plan
UAE National Strategy for Artificial Intelligence 2031
National Strategy for Arificial Intelligence #AI for All
National Strategy for AI (SNIA), PHASE 1 (2018-2022)
National AI Strategy
National Strategy for the development of AI in the Russian Federation until 2030
National AI Strategy (Updated 2020)
Brazilian AI Strategy (EBIA)
National AI Strategy
National AI Strategy
AI strategy 2022
Strategic Programme on AI 2022-2024
National AI Strategy 2.0
National AI Research and Development Strategic Plan
Italian Data Protection Authority
AI Strategy Conference
UAE Artificial Intelligence and Blockchain Council
Advisory Council on the Ethical Use of AI and Data
National AI Advisory Committee
National New Generation AI Governance Specialist Committee
AI Council
Data Ethics Commission
Advisory Council on Artificial Intelligence
Office for Artificial Intelligence
The National Center for AI
Committee on Generative Artificial Intelligence
National Coordinator for Artificial Intelligence (COSNIA)
National Data Administration
AI R&D Principles
Seven principles on the use of AI systems in government
AI Code
Draft AI Utilization Principles
Five guiding principles of American AI Initiative
Governance Principles for the New Generation AI
Dubai's AI Principles
A Proposed Model AI Governance Framework: Guiding Principles
Basic Principles of the Development and Use of AI Technologies
Social Principles of Human-centric AI
DoD's AI ethical principles
A compilation of existing AI ethical principles (Annex A)
Opinion of the Data Ethics Commission: General ethical and legal principles
Principles for the Stewardship of AI Applications
Principles of AI Ethics for the Intelligence Community
Ethical Principles for AI in Defence
Ethical Norms for the New Generation AI
AI Ethics Code (revised version)
Responsible use of AI: Our guiding principles (updated)
Algorithmic Impact Assessment
Model AI Governance Framework
Dubai AI Ethics self-assessment tool
A.I.Verify
Self-assessment guide for AI systems
NIST AI Risk Management Framework (AI RMF 1.0)
AI and Data Protection Risk Toolkit
U.S. LEADERSHIP IN AI: A Plan for Federal Engagement in Developing Technical Standards and Related Tools
Guidelines for the Construction of a National New Generation AI Standards System
Towards a Standard for Identifying and Managing Bias in AI
Standards on AI data quality
White Paper on AI Security Standardization (2023 Edition)
Second edition of the German Standardization Roadmap AI
White Paper: The state of global AI standardization and what it means for Canada
2017
2018
2019
2020
2021
2022
2023
2024

## Observation 3.3: In the context of international participation in AI governance among the 14 evaluated countries, the United Kingdom, France, and Japan have demonstrated significant involvement.

| | Signed the **UNESCO**'s Recommendation on the Ethics of Artificial Intelligence? | Have experts in the UNESCO's Ad Hoc Expert Group (**AHEG**) for the Recommendation? | Have experts in the **UN's High-level Advisory Body** on Artificial Intelligence? | Adopted the **OECD/G20** AI principles? | A member of the Global Partnership on Artificial Intelligence (**GPAI**)? | Participated in the **Global AI Safety Summit** or co-signed the Bletchley Declaration? | One of the endorsing countries of the **REAIM 2023** Call to Action? |
|---|---|---|---|---|---|---|---|
| UK | ✓ | ✓ | ✓ | ✓ | ✓ | ✓ | ✓ |
| France | ✓ | ✓ | ✓ | ✓ | ✓ | ✓ | ✓ |
| Japan | ✓ | ✓ | ✓ | ✓ | ✓ | ✓ | ✓ |
| US | ✗ | ✓ | ✓ | ✓ | ✓ | ✓ | ✓ |
| China | ✓ | ✓ | ✓ | ✓ | ✗ | ✓ | ✓ |
| Germany | ✓ | ✗ | ✓ | ✓ | ✓ | ✓ | ✓ |
| Italy | ✓ | ✗ | ✓ | ✓ | ✓ | ✓ | ✓ |
| India | ✓ | ✓ | ✓ | ✓ | ✓ | ✓ | ✗ |
| Brazil | ✓ | ✓ | ✓ | ✓ | ✓ | ✓ | ✗ |
| Singapore | ✓ | ✗ | ✓ | ✓ | ✓ | ✓ | ✓ |
| Canada | ✓ | ✗ | ✗ | ✓ | ✓ | ✓ | ✓ |
| Russia | ✓ | ✓ | ✓ | ✓ | ✗ | ✗ | ✗ |
| S.Africa | ✓ | ✓ | ✓ | ✓ | ✗ | ✗ | ✗ |
| UAE | ✓ | ✓ | ✓ | ✗ | ✗ | ✓ | ✗ |

***Table 4** 14 Countries' participation in major global AI governance mechanisms*

Regarding international governance of artificial intelligence, the United Nations stands as the most inclusive governance mechanism worldwide. UNESCO's *Recommendations on the Ethics of Artificial Intelligence*, the first globally reached agreement on AI ethics, is the most broadly supported AI governance document. Among the 14 evaluated countries, all except the United States, which had already withdrawn from UNESCO at the time, signed the recommendation at the end of 2021. During its drafting, representatives from 10 of these countries, with the exception of Germany, Canada, Italy, and Singapore, were involved in the process. Additionally, in October 2023, the UN Secretary-General formed a High-Level Advisory Body on Artificial Intelligence consisting of 38 experts. In this group, experts from all the evaluated countries, except Canada, were selected.

Regarding international AI principles, the Organization for Economic Cooperation and Development (OECD) introduced the OECD AI Principles in 2019. Similarly, in June 2019, the Group of Twenty (G20) ratified the G20 Artificial Intelligence Principles, which promote a human-centered and responsible approach to development of AI. Of the 14 countries assessed, all except the United Arab Emirates signed the OECD/G20 AI Principles.

Among other influential international AI cooperation mechanisms, Canada and France established the Global Partnership on AI (GPAI) in 2020, hosted by the OECD. This organization, currently consisting of 28 member countries and the European Union, has set up four main working groups on Responsible AI, Data Governance, the Future of Work, and Innovation and Commercialization. It held a ministerial summit in New Delhi in December 2023, focusing on cooperation among member countries on these themes. Among the 14 evaluated countries, all except China, Russia, South Africa, and the United Arab Emirates are members of this organization.

In terms of globally influential AI initiatives and declarations formed in 2023, in November 2023, the UK government hosted the first AI Safety Summit to promote cooperation in AI safety, marking the first global summit in this field. Twenty-eight countries and the European Union jointly signed the Bletchley Declaration. Among the 14 evaluated countries, all except Russia and South Africa signed this declaration. Additionally, in February 2023, the Netherlands and South Korea co-hosted the first summit on Responsible Use of AI in the Military (REAIM) in The Hague. G7 countries, BRICS member China, and countries like Singapore participated in this initiative.

---

### Observation 3.4: AI legislation varies globally, with some countries adopting comprehensive state-level laws, while others integrate AI-specific amendments into existing frameworks or follow a more fragmented approach with state-specific initiatives and federal guidelines.

The top-down approach involves government-led initiatives to create broad, overarching regulations that directly address the nuances of AI technology. Contrastingly, other countries are adapting their existing legal frameworks to meet the evolving demands of AI. This approach, more evolutionary in nature, updates and extends current legislation to encompass the unique challenges and considerations posed by AI technologies. These varying strategies highlight the diverse responses to AI governance across the global landscape.

In the European Union, there's a move towards comprehensive, unified AI law, while the US shows a fragmented landscape with numerous state initiatives and federal guidelines. The UK's approach favors legal innovation, and France and Germany integrate AI within broader digital laws. Japan, Canada, and China are also actively creating specific AI governance legal frameworks. India, meanwhile, relies on existing laws for AI regulation. Other nations like South Africa, Brazil, Singapore, and the UAE focus on AI ethics frameworks without comprehensive state laws. This global picture reveals a mix of strategies, reflecting each country's governance style, technological advancement, and societal values towards AI.

*Figure 14* ***Four AI legislation approaches various countries are taking.***

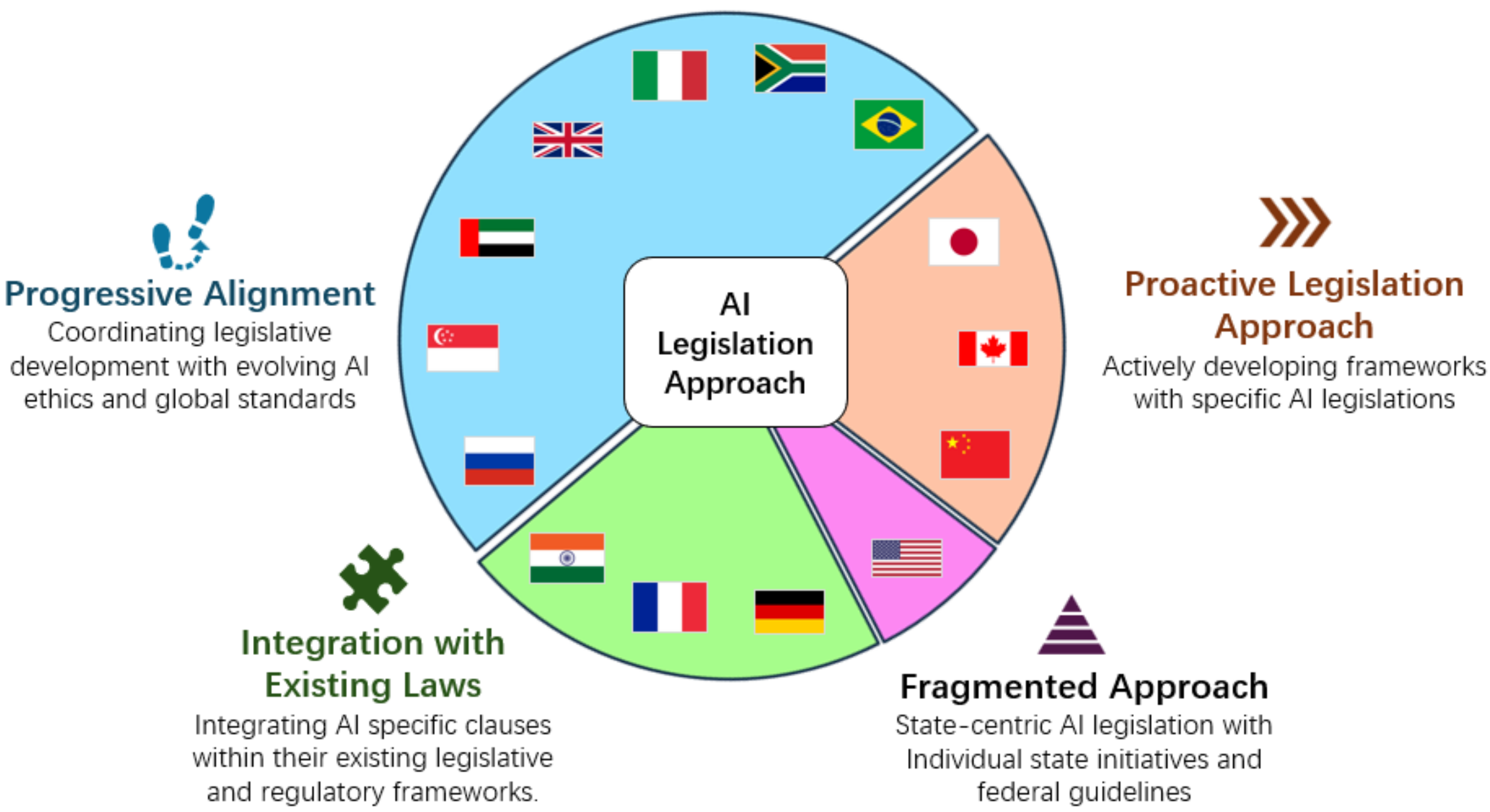

## 3.4 Pillar 4: AI Governance Effectiveness

### Observation 4.1: The public in BRICS countries generally express higher levels of trust in AI compared to their counterparts in high-income countries.

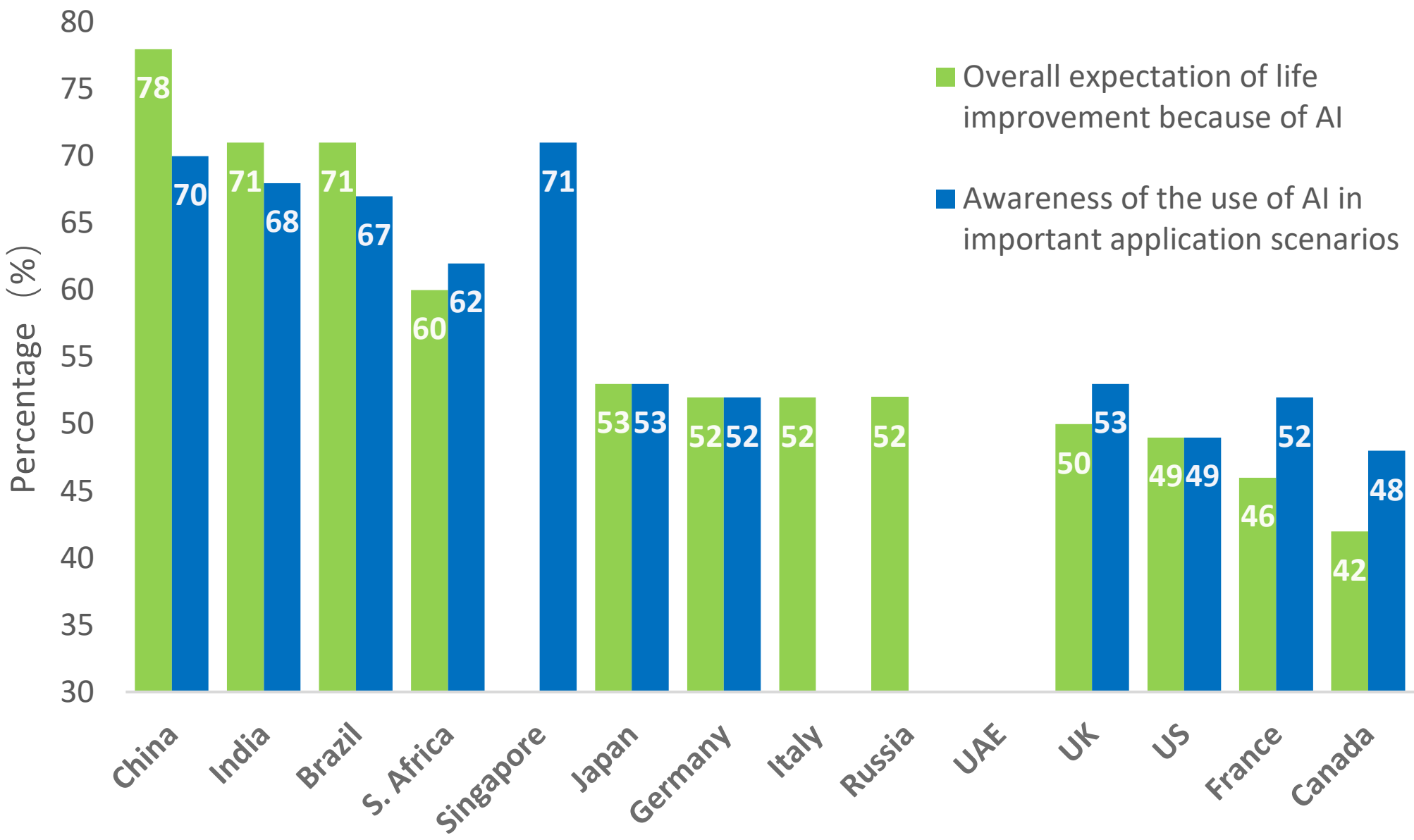

*Figure 15* ***Statistics on Public Trust and Awareness of AI in Different Countries*** *(Sources: IPSOS, KPMG)*

There is a strong positive correlation between public awareness and trust in AI, with a correlation coefficient of 0.84. This suggests that the more the public in a country is aware of AI applications in key scenarios, the more they trust AI. Among them, in China, India, Brazil, and South Africa, more than 60% of surveyed public expect AI to ultimately improve life rather than negatively impact it, which is significantly higher than other countries. In China, over three-quarters of respondents express trust in AI, the highest among the 14 countries. The KPMG survey[1] shows the percentage of scenarios in which people in various countries are aware of existing or upcoming AI applications and averages these percentages across countries. In this statistic, respondents from China, India, Brazil, South Africa, and Singapore are aware of AI in over 60% of scenarios on average. Singaporean respondents have the highest awareness ratio, exceeding 70%.

[1] Gillespie, N., Lockey, S., Curtis, C., Pool, J., & Akbari, A. (2023). Trust in Artificial Intelligence: A Global Study. The University of Queensland and KPMG Australia. doi:10.14264/00d3c94

**Observation 4.2: A stark gender imbalance permeates AI researchers across all 14 evaluated countries, with only about one in five researchers being female.**

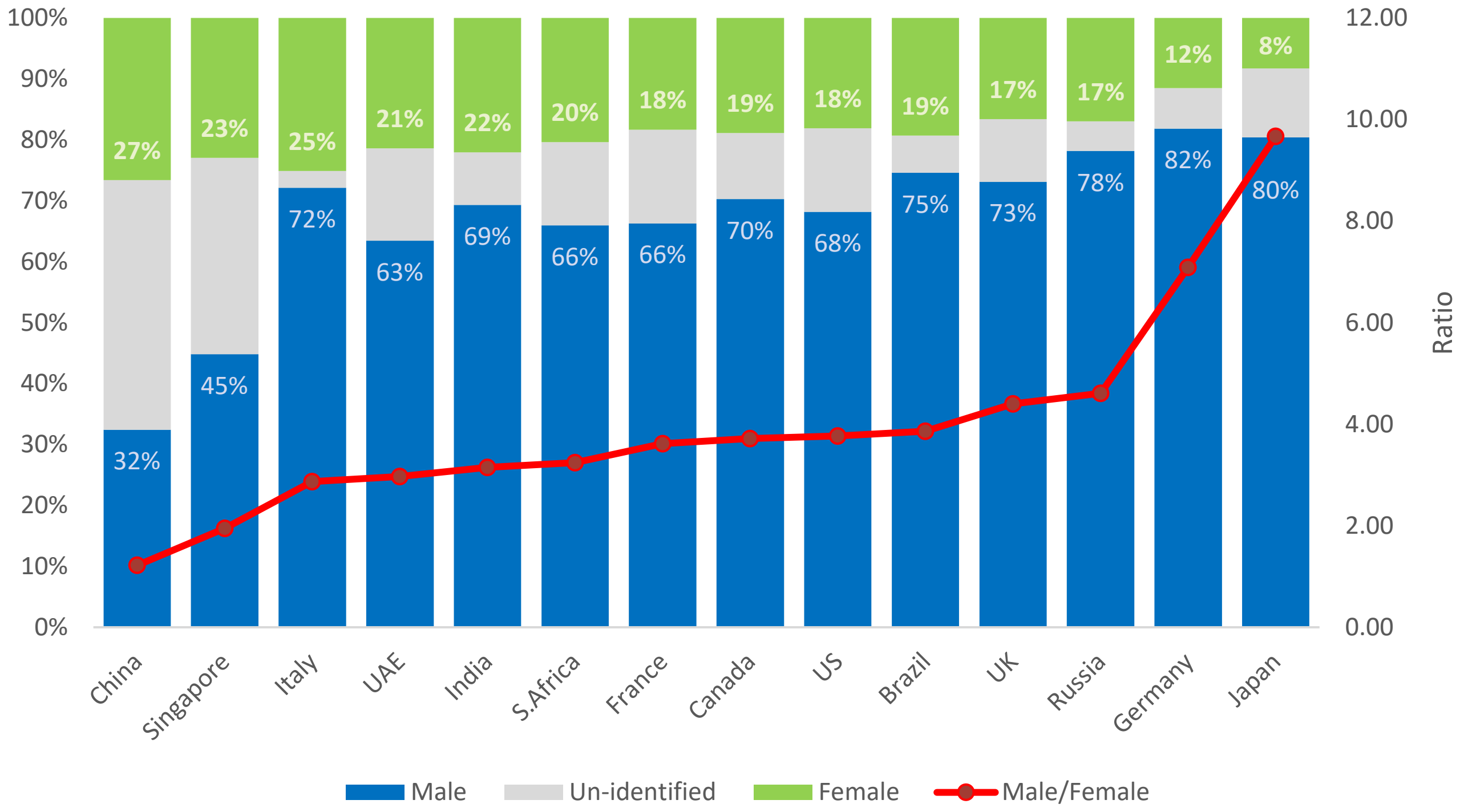

*Figure 16* ***AI literature author gender ratio and Male-Female ratio***

*（source： DBLP statistics）*

AGILE Index sees that China and Singapore have the lowest identified gender author ratios, at 1.2 and 2.0 respectively, ranking first and second. Italy follows closely with a ratio of 2.9. This indicates the most balanced male-to-female participation in AI literature in these three countries. In most other countries, the gender ratio is between 3 and 5. However, Germany and Japan have significantly higher gender ratios than other countries, at 7.1 and 9.7, indicating a need for increased female AI research participation in these countries.

## Observation 4.3: Further analysis on gender gap underscores a critical challenge: almost no countries excel at both AI gender inclusivity and broader societal gender equality, which calls for further research attention.

In comparison with the WEF Gender Gap report rating, which aims to score a country's general societal gender equality, we note countries like Germany, which have a small societal gender gap but a large gender ratio difference in literature, and countries like Japan with relatively large societal gender gaps and large gender ratios in literature, or countries like China with an observable societal gender gap and a more balanced gender ratio in literature. However, there are few countries that perform well in both aspects, indicating room for improvement in gender-ratio.

Further analysis reveals a weak negative correlation between the gender author ratio and the WEF Global Gender Gap score. Furthermore, we also observe a weak negative correlation between AI graduate ratio and the WEF Global Gender Gap score. The negative correlations between the gender ratio in AI and the overall societal gender gap suggests that if a country has a high performance in bridging the general gender gap, it tends to have a low score of the gender ratio in the field of AI**.** This counter-intuitive phenomenon worths reflection and further research. It would be especially interesting to see if the same negative correlation occurs in more countries. If this phenomenon is indeed occurring, then it means that all countries still have work to do in terms of gender equality.

***Figure 17*** ***Negative correlation between gender equality in AI graduate/author and whole society***

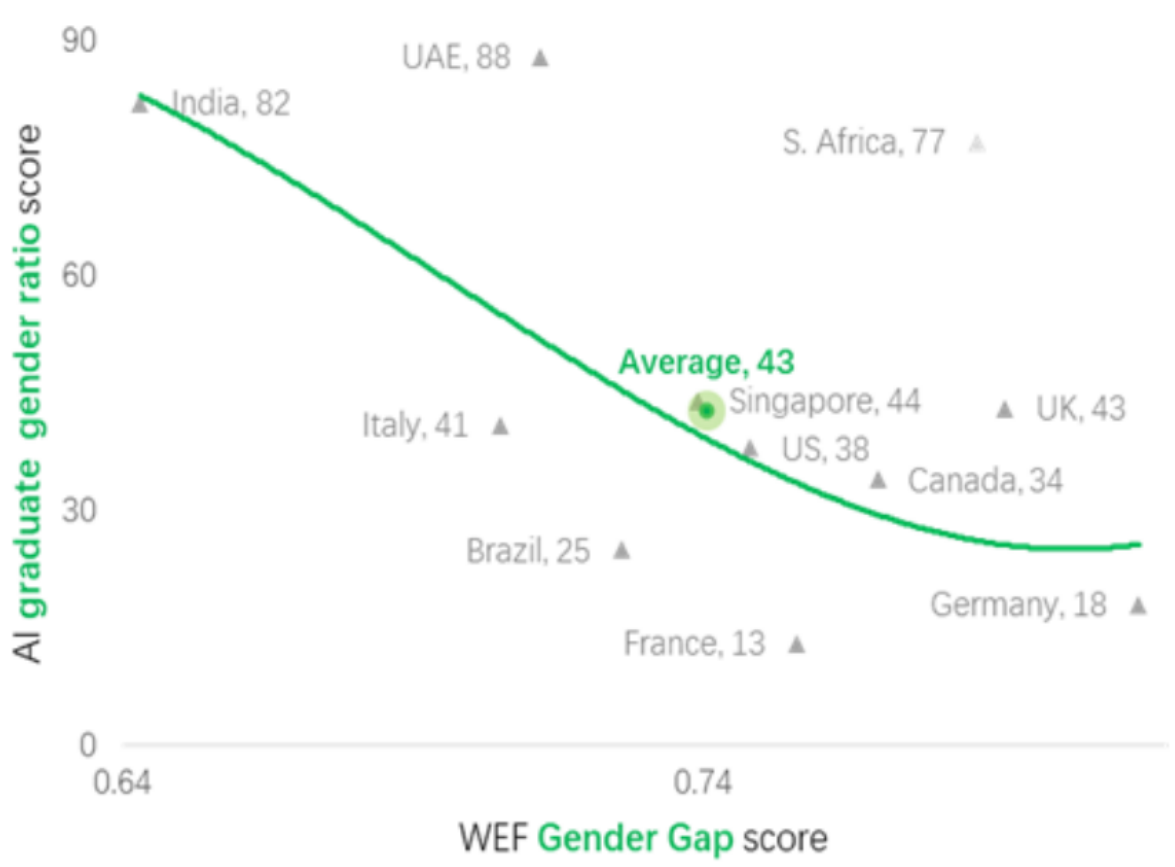

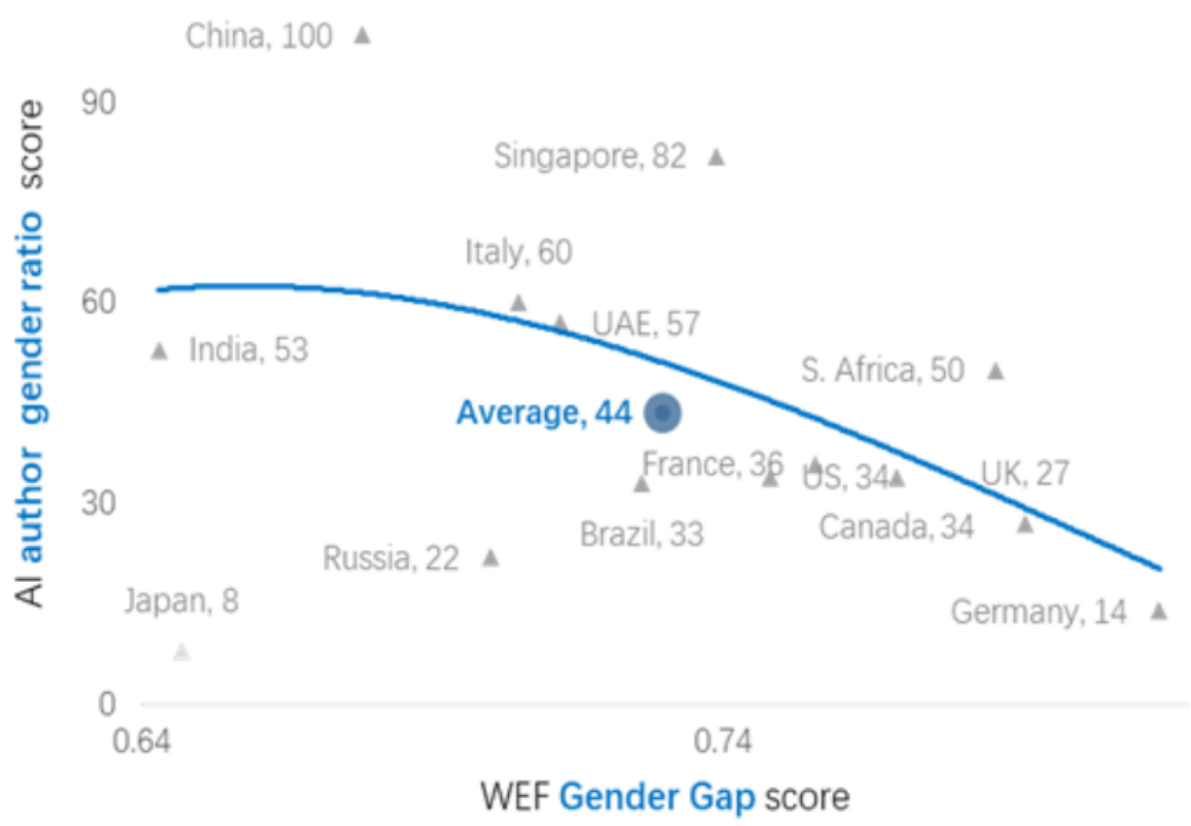

**Data Sources**: *Tortoise Media, World Economic Forum, DBLP*

### Observation 4.4: While all 14 countries actively engage in global developer communities, the United States, China, and India stand out for their substantial contributions and impact.

GitHub is an open-source code hosting platform with a multitude of open-source AI-related projects, including in fields like machine learning, natural language processing, and computer vision. We can see that India, despite having a very high number of total AI submissions on GitHub (about 49,000), has a low number of popular AI packages, accounting for only about 1.2%. The United States has the highest number of submissions to popular AI packages among the 14 countries, about 3,600, accounting for a significant percentage. Although China has fewer submissions to popular AI packages than the U.S., it has the highest percentage of popular packages, with the two ratios reaching about 39%.

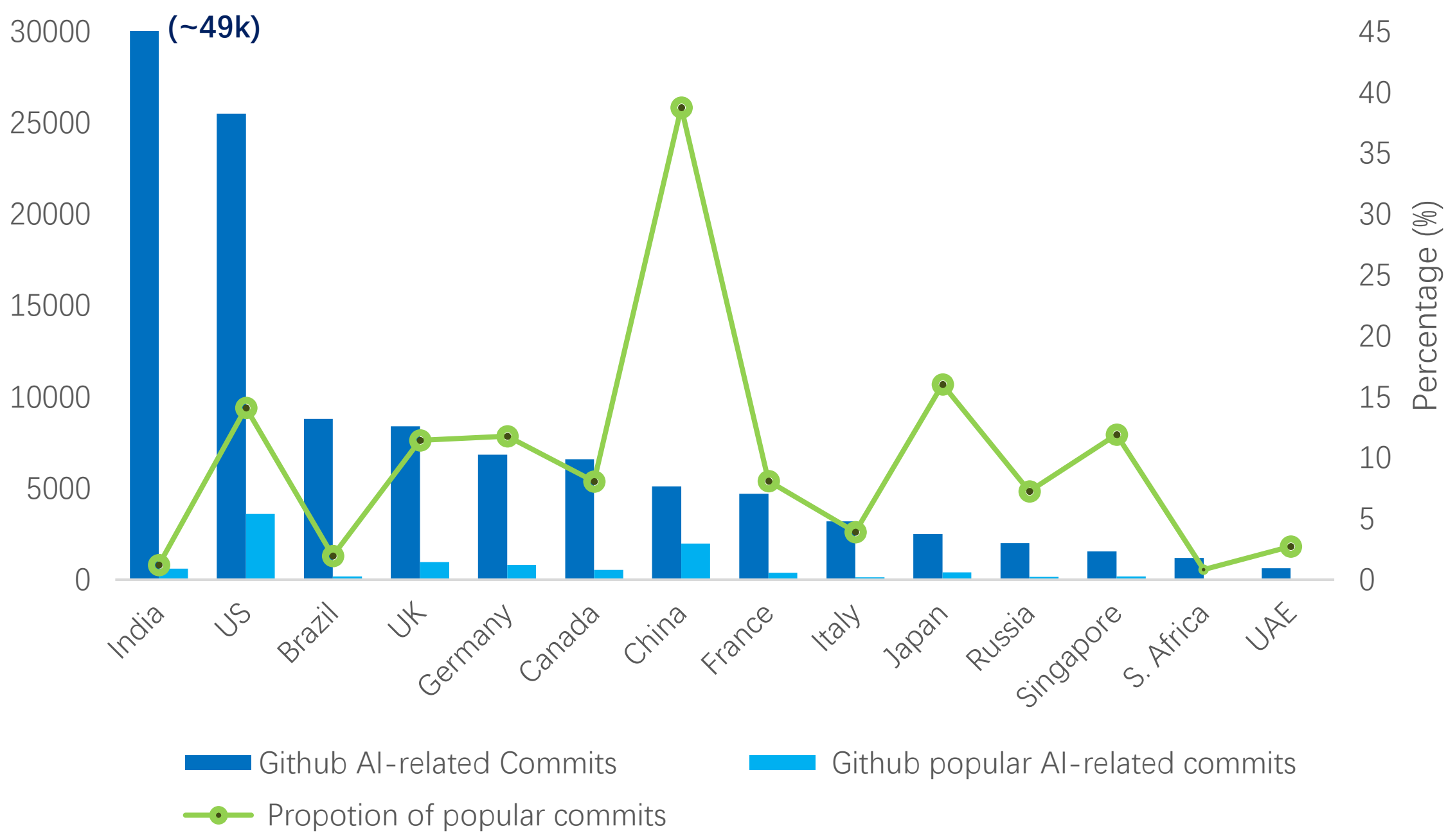

*Figure 18* ***GitHub AI-related commits and popularity by Country***

*(Based on Tortoise Media Data)*

AGILE Index also compiled contributions from different countries to the AI community Hugging Face, which is an open-source community dedicated to advancing natural language processing (NLP) technology and tools, such as pre-trained models, datasets, and tutorials. Overall, China and the United States belong to the first tier in terms of contributions, contributing over 300 models and datasets each to the Top 1000 influential ones. Meanwhile, Canada, France, India, the United Kingdom, Singapore, Brazil, and Japan belong to the second tier, contributing two-digit numbers of models and datasets.

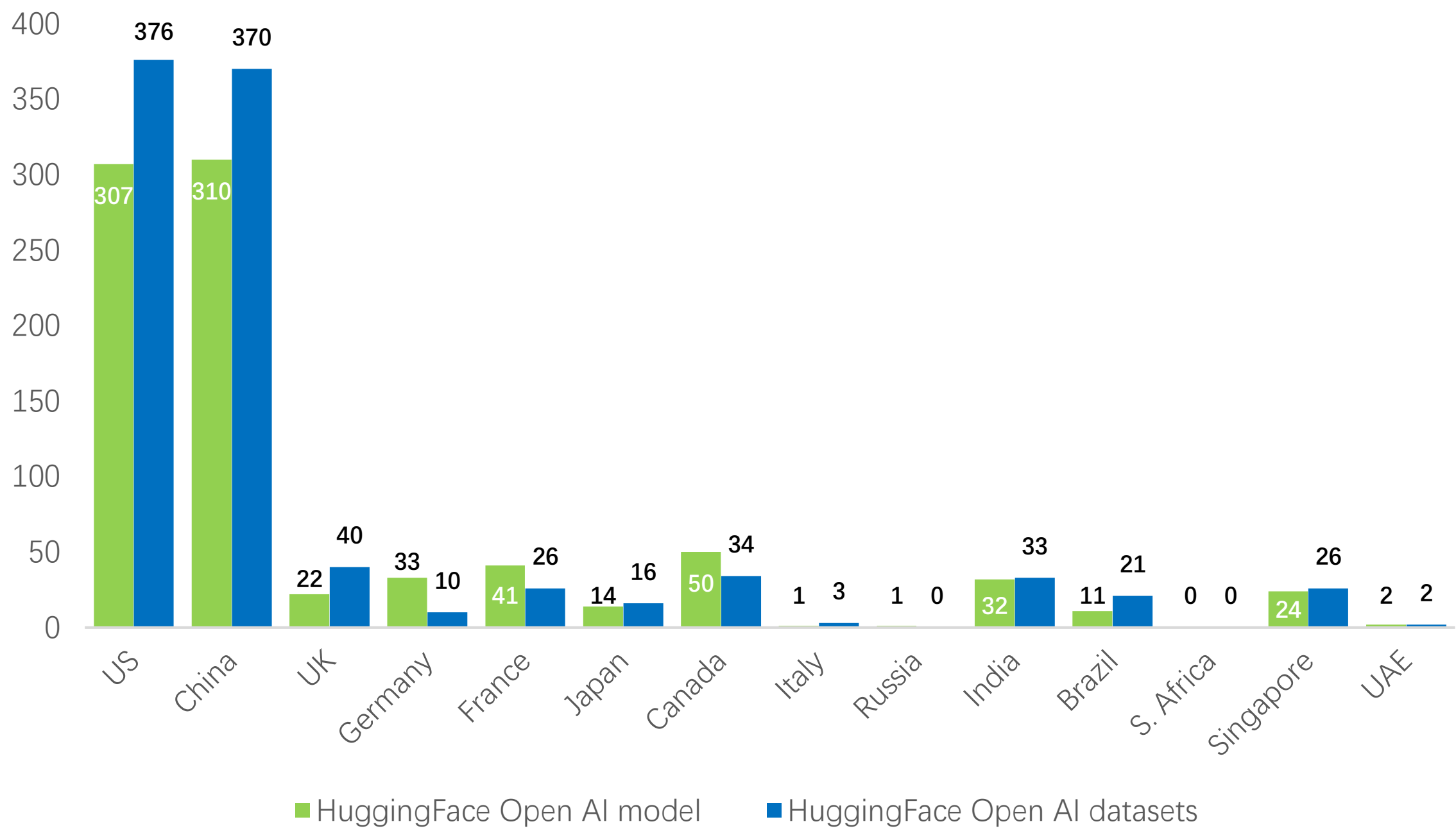

***Figure 19*** ***Number of Top 1000 Open Models and Datasets Released by Country***

*(Based on Hugging Face Statistics)*

**Observation 4.5: The proportion of publications related to AI governance is around 3%-4% of all AI-related publications in the 14 countries.**

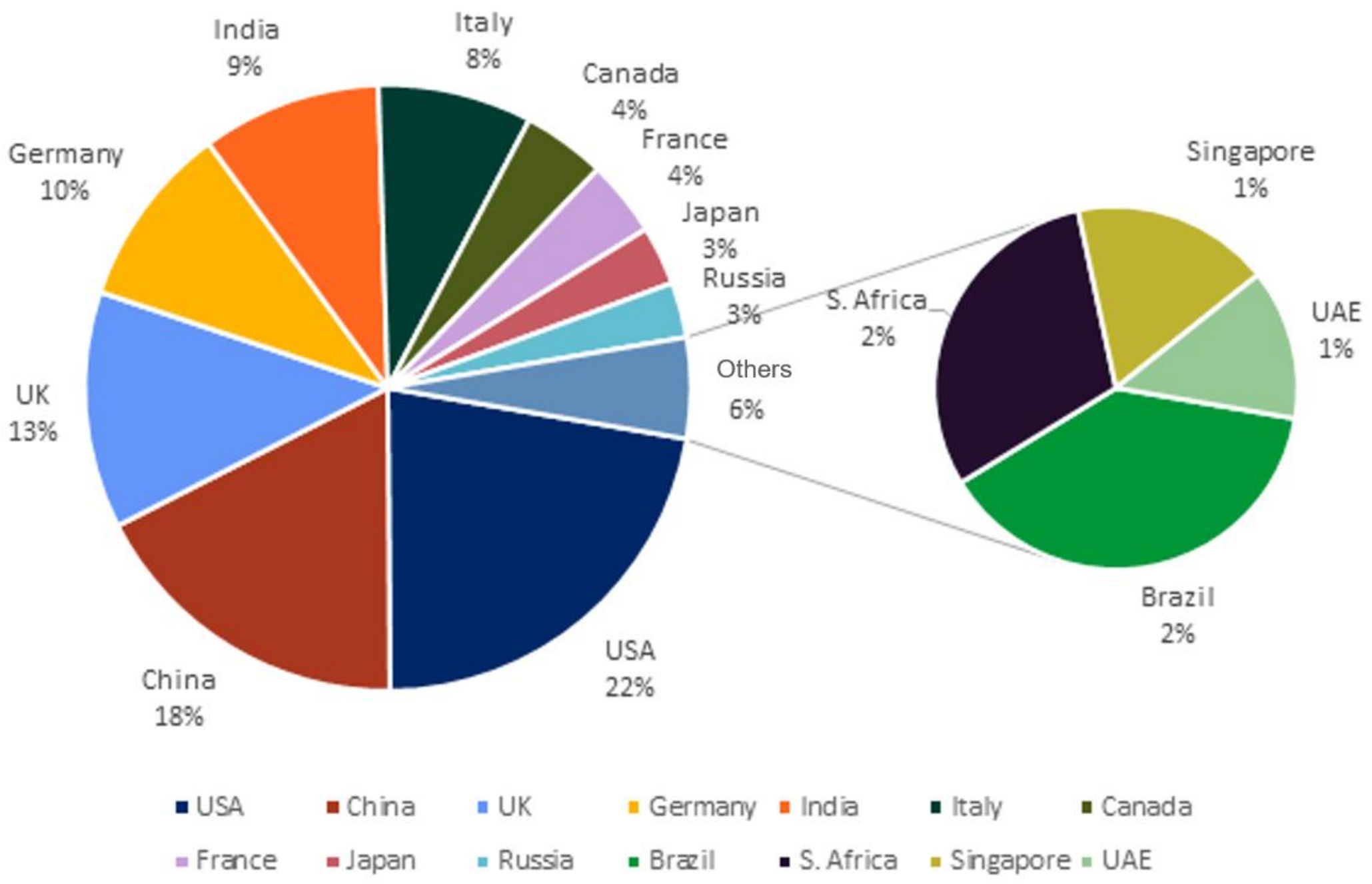

***Figure 20** Share of AI-governance related Publications by Country*

*(Based on DBLP Statistics)*

According to our statistics from the DBLP literature database, the 14 countries have a total of over 140,000 AI-governance-related publications. The proportion of publications related to AI governance is around 3%-4% of all AI-related publications. The total number of AI governance publications from the 14 countries was predominantly contributed by the US, China, and the UK, accounting for 22%, 18%, and 13%, respectively. Germany, the UK, India, and Italy also made significant contributions, with respective proportions of 10%, 9%, and 8%.

**Observation 4.6: The total volume of literature on AI governance has shown an exponential acceleration in recent years, with a growth rate reaching 45% in 2022.**

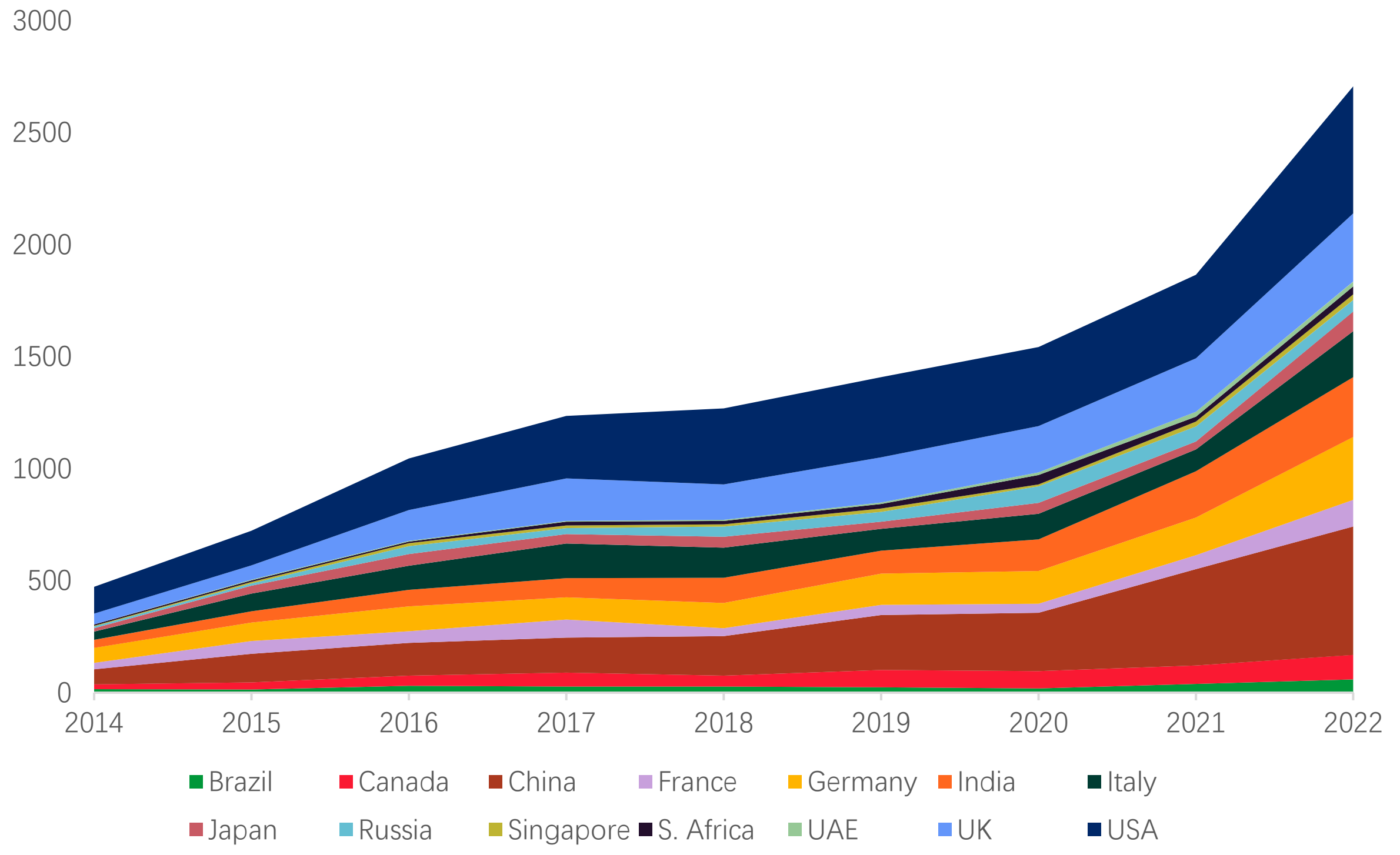

*Figure 21 AI-governance related papers focusing on social impact (2014-2022)*

*(Based on statistics from Springer, IEEE Xplore, and ACM digital library)*

Over 2014-2022, the total number of AI governance-related publications from these 14 countries has steadily increased, rising from 472 in 2014 to 2,707 in 2022. We have observed the largest increase in volume in 2022 with a 45% of growth rate.

Notably, the growth rate of AI governance publications from China was higher than other countries. In 2014, China's proportion of AI governance publications among the 14 countries was 14.2%, ranking second. In the same year, the US had a proportion of 25.2%, ranking first. However, by 2021, China reached 23.1%, becoming the leader in governance publications among the 14 countries and maintained the first place in 2022 with a contribution rate of 21.2%.

**Observation 4.7: Among the 14 countries, security, safety, and collaboration are consistently the most researched topics related to AI governance in AI literature, with long-term AI and accountability receiving significantly less research focus.**

We analysed the themes of AI literature in the 14 countries based on the key topics mentioned in AI principles[2]. Among 14 countries, security, safety, and collaboration were consistently the most researched topics related to AI governance, with long-term AI and accountability receiving significantly less research focus, both with less than 1% of literature on it.

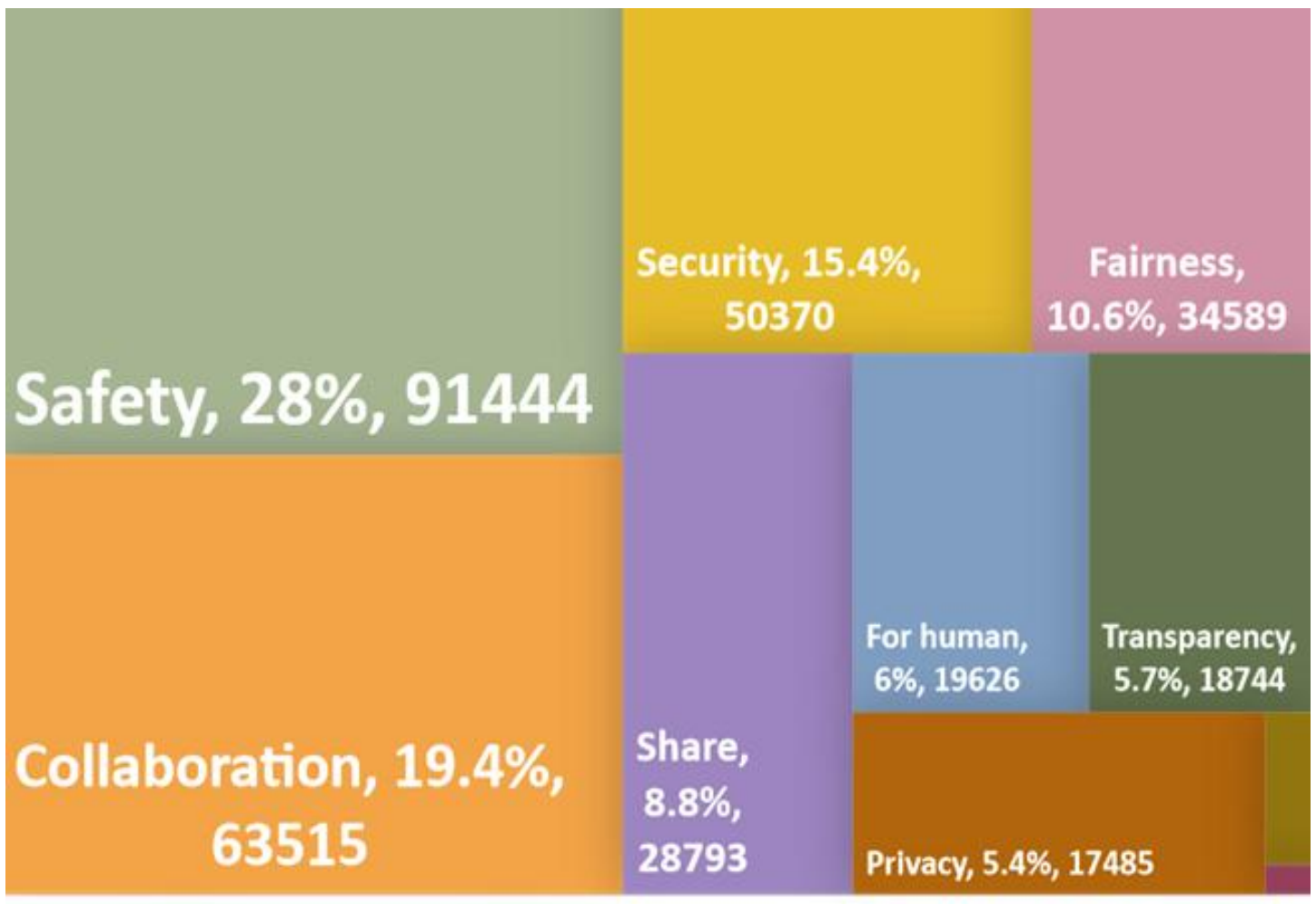

*Figure 22* ***AI Literature by Key Topics in AI Principles***

*(Based on DBLP Statistics)*

Further comparing the distribution of AI governance literature among countries, we find that China and Japan focus slightly more on the matter of collaboration relative to other countries. while France and Russia are slightly more concerned with safety. Germany and Italy show slightly more interest in transparency issues than other countries. Brazil, Russia, and South Africa focus more on the for-human theme.

[2] Here, the topics of AI principles used for analysis are from the Linking AI Principles platform (https://www.linking-ai-principles.org/). We specifically analyze 10 of the 11 key topics derived from AI principles, and for the topic of sustainability we examined it separately from the perspective of the 17 Sustainable Development Goals.

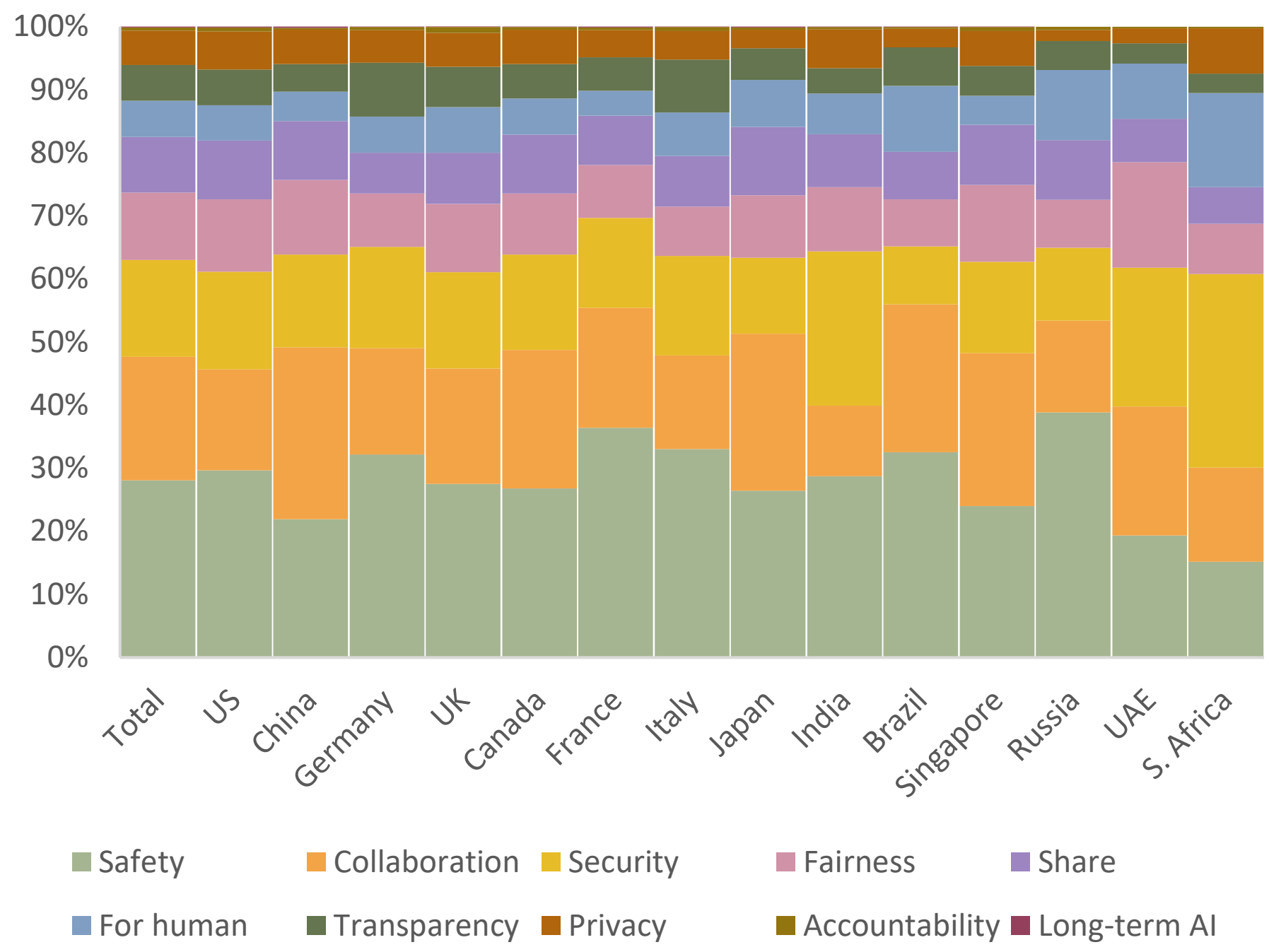

*Figure 23* ***Proportion of AI Literature on the Ten Key Topics of AI Principles in Different Countries***

*(Based on DBLP Statistics)*

---

**Observation 4.8: There are visible collaborations on AI governance among all countries, indicating that AI governance is globally connected and indivisible.**

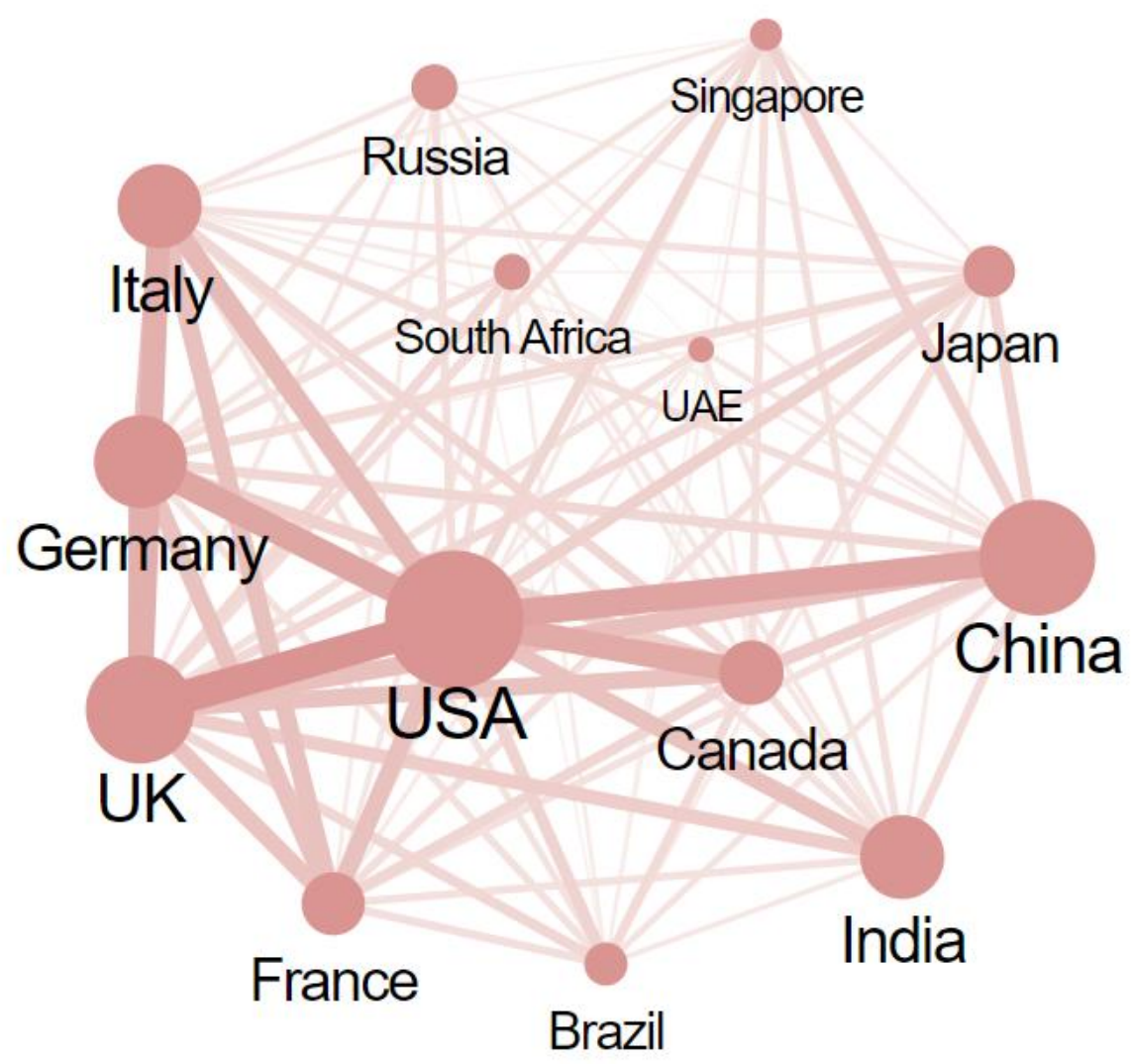

*Figure 24* ***Collaborations in AI Governance Literature from the 14 countries***

*(Based on statistics from Springer, IEEE Xplore, and ACM digital library)*

Based on analysis of collaborations in AI governance-related papers, visible collaborations can be found among all countries. Most collaborations occur between the US, the UK, and other countries, contributing about 21% and 16% of cooperation respectively, while regional cooperation within European, North American, and Eastern Asian countries are also strong. Noticeably, China-UK-US and UK-US-EU (France, Italy, Germany) collaboration network stands out from the rest, as they are also the most active hubs of governance research. The collaborations on AI governance among different countries indicates that AI governance is globally connected and indivisible.

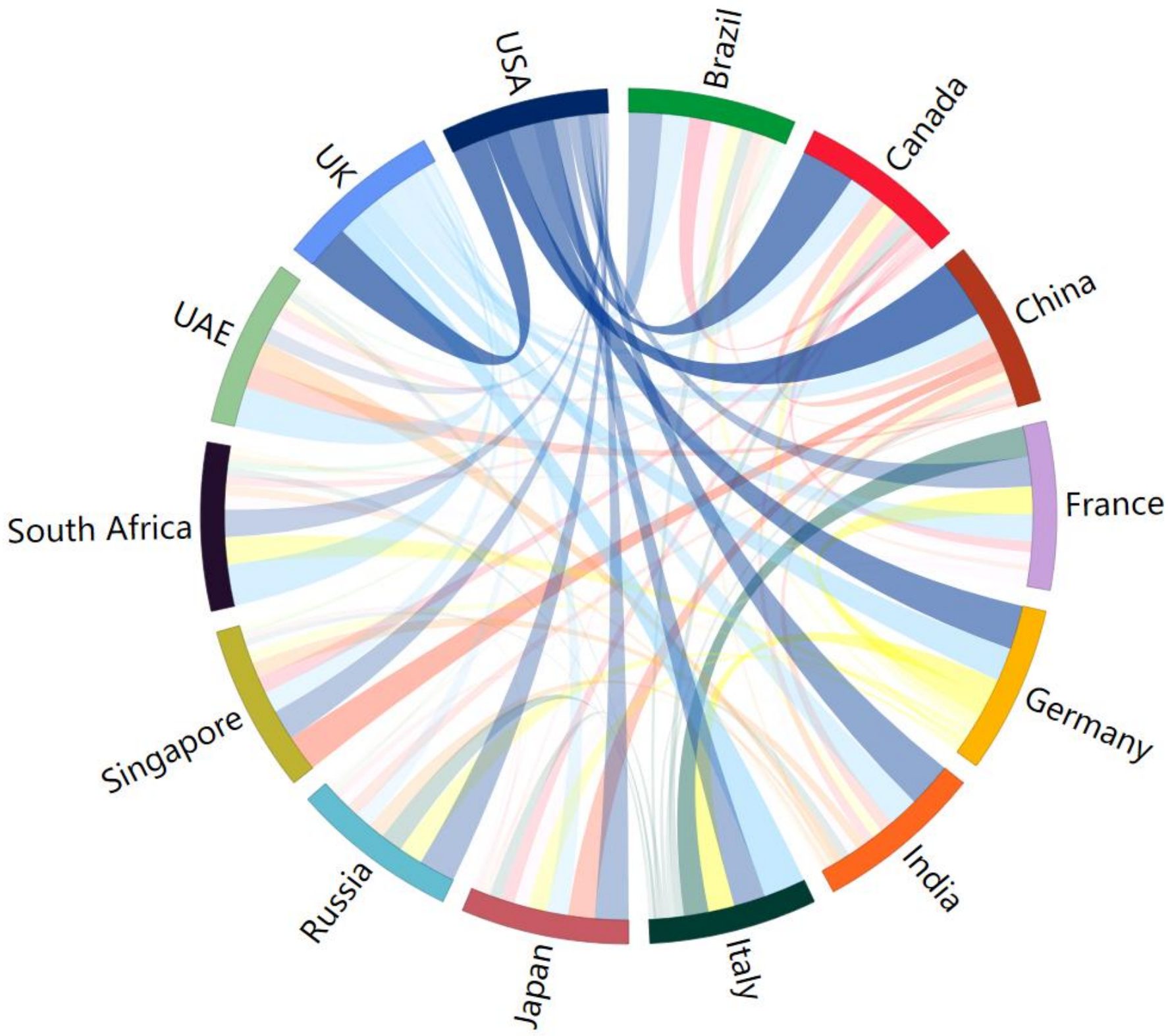

***Figure 25*** ***Relative Number of Collaborations in AI Governance Literature by Country***

*(Based on statistics from Springer, IEEE Xplore, and ACM Digital Library)*

## Observation 4.9: China and the United States together contribute more than half of the papers in nearly all AI for SDGs research directions.

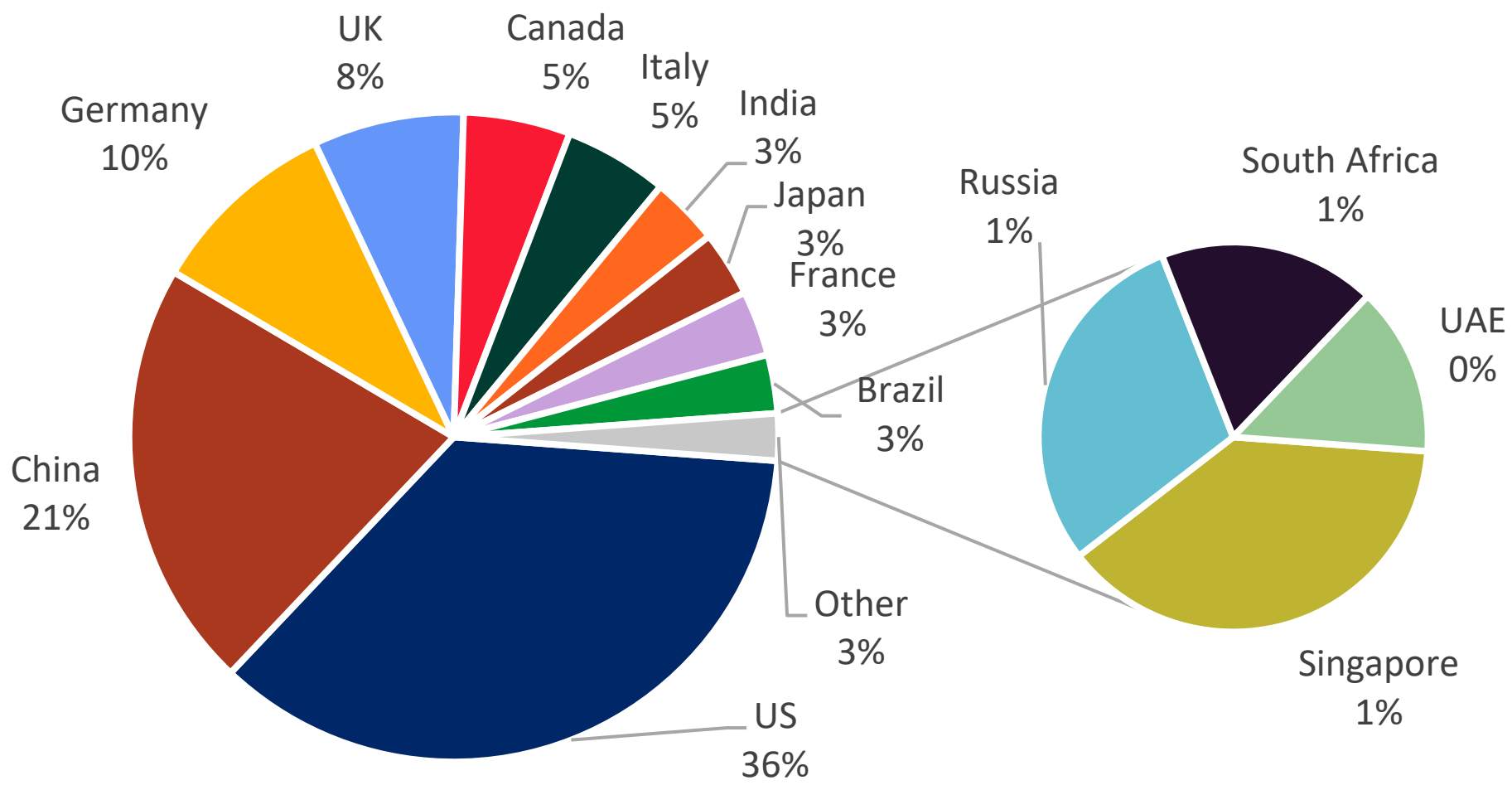

*Figure 26* ***Percentage of Total Number of AI Literature on 17 SDGs***

*(Based on DBLP Statistics)*

The number of papers published by a country using AI technology to help achieve sustainable development goals indicates the effort made by that country in using AI for Good. Overall, the United States and China are in the first tier in terms of the number of AI for SDGs papers published, accounting for 34.7% and 23.2% of the AI for SDGs literature from all 14 countries, respectively. They in total also contribute more than half of the papers in all AI for SDGs research directions, except on SDG2 (Zero Hunger) and SDG11 (Sustainable Cities and Communities) where India and Italy also play an important role.

**Observation 4.10: SDG3 (Good health and well-being), SDG9 (Industry, innovation, and infrastructure) and SDG4 (Quality Education) are the three most popular AI for SDGs research topics for all 14 countries.**

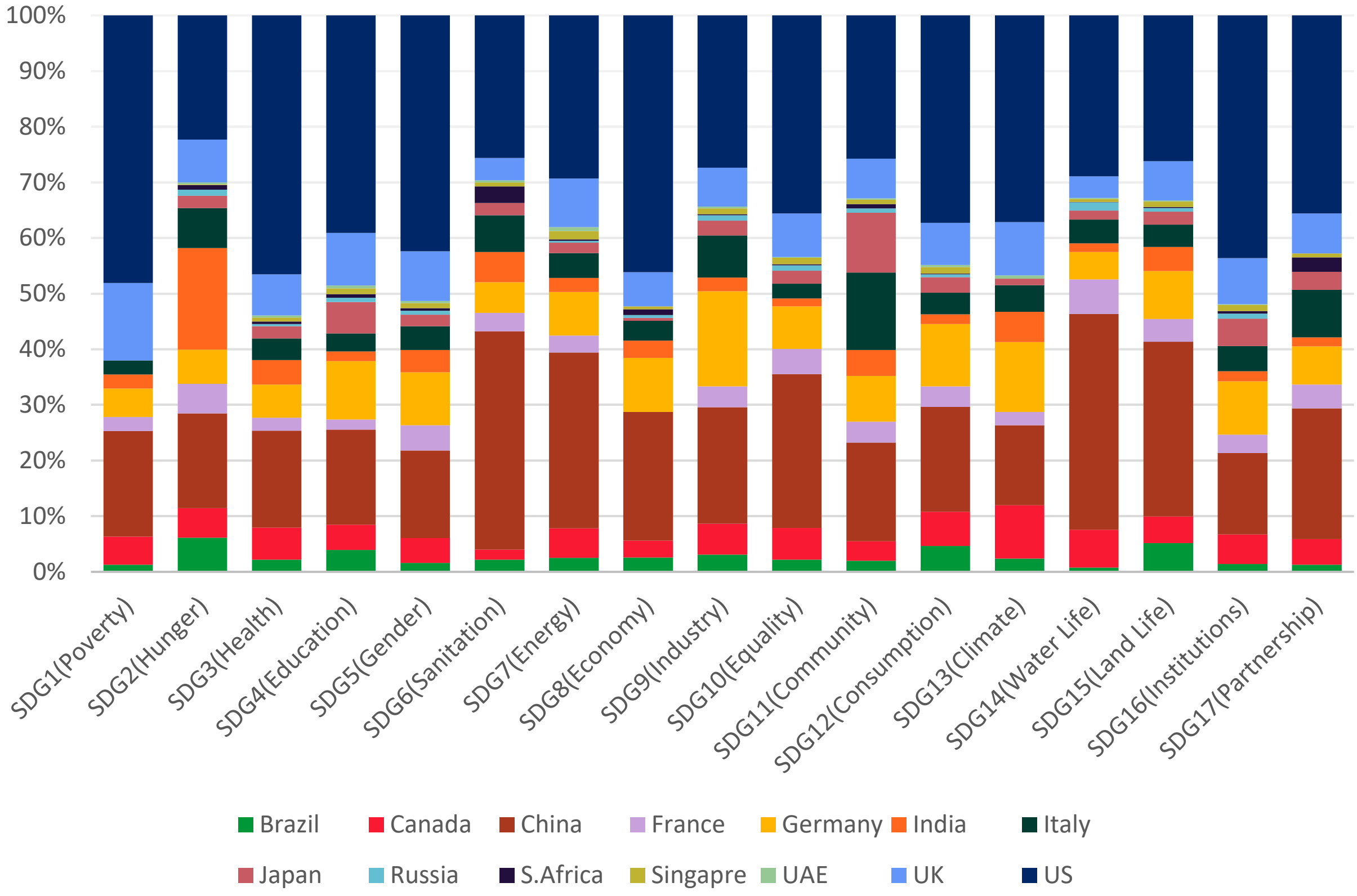

*Figure 27 The Percentage of AI Literatures Published by each Country on 17 SDGs*
*(Based on DBLP Statistics)*

All countries' focus on AI for SDGs literature is consistent, with SDG3 (Good health and well-being), SDG9 (Industry, innovation and infrastructure) and SDG4 (Quality Education) be three of the most popular topics for all 14 countries. For all countries, papers on these three SDGs account for more than half of all AI for SDGs relevant papers. Japan and Italy have relatively researched a lot on SDG7 (Affordable and clean Energy), which alongside with SDG14 (Life below water) and SDG14 (Life on Land), have also attracted observable amount of attention for almost all countries.

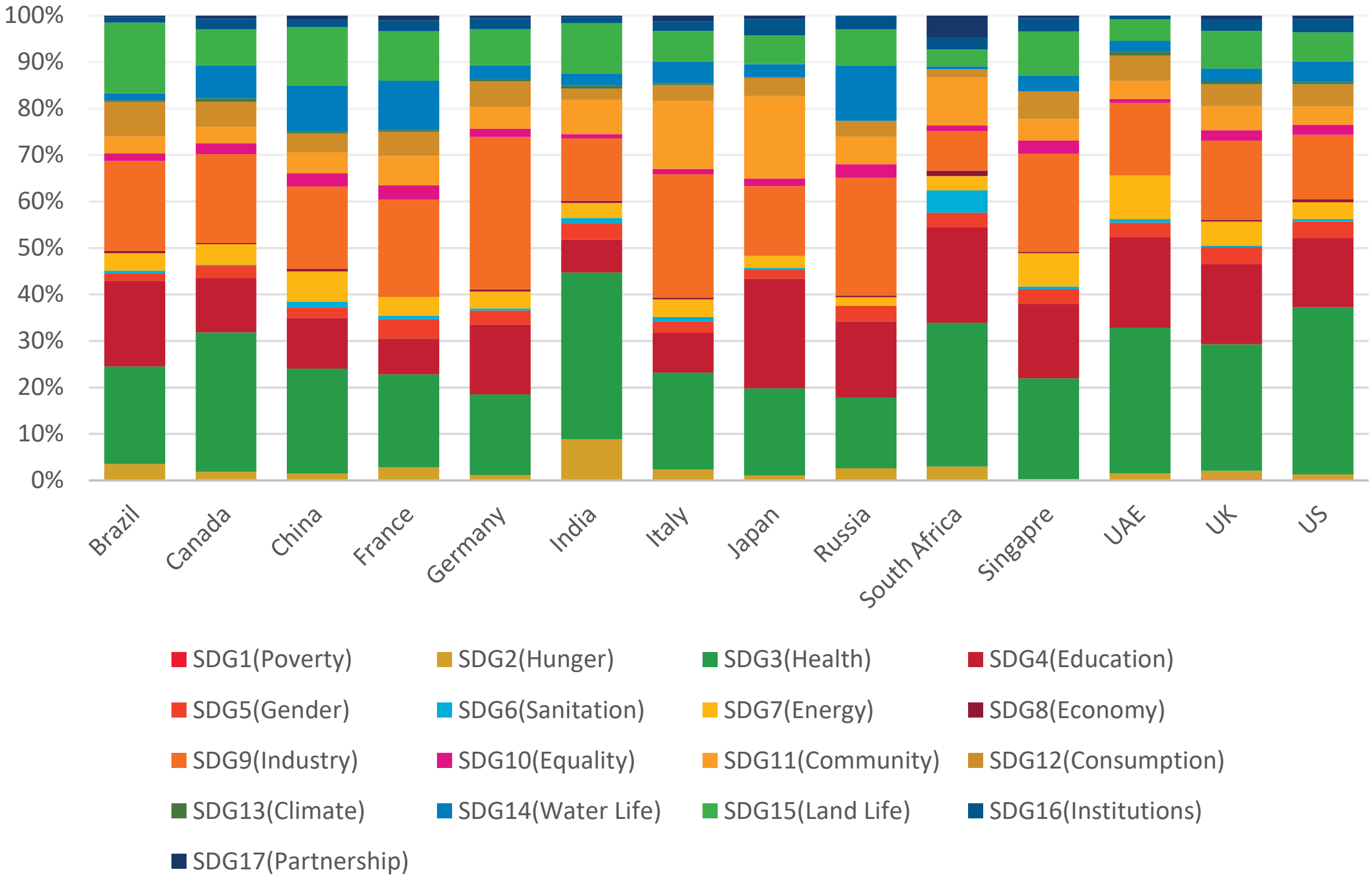

*Figure 28* ***The Percentage of AI Literature on each SDGs Published by the 14 Countries***

*(Based on DBLP Statistics)*

---

**Observation 4.11: In AI for SDGs application, while SDG3, SDG9, and SDG4 remain popular, there are also notable number of projects for SDG11 (Sustainable Cities and Communities), SDG12 (responsible consumption and production) and SDG13 (Climate actions).**

The AGILE Index also assesses the disparity between AI for SDGs literature and applications across the 14 countries. The United States continues to lead, contributing 43% of the total documented AI for SDGs use cases and significantly influencing all SDGs. In contrast, China exhibits a noticeably smaller share in AI for SDGs applications compared to literature. Conversely, India stands out with a higher proportion of SDGs application cases, particularly in the realms of SDG1 (No Poverty), SDG5 (Gender Equality), and SDG6 (Clean Water and Sanitation), in comparison to its literature contributions.

The distribution of application cases varies significantly among countries when compared to research papers. Diverse interests in AI4SDGs projects are evident across all nations. While SDG3 (Good Health and Well-being), SDG9 (Industry, Innovation, and Infrastructure), and SDG4 (Quality Education) remain popular for most countries, there is a notable increase in projects related to SDG11 (Sustainable Cities and Communities), SDG12 (Responsible Consumption and Production), and SDG13 (Climate Action).

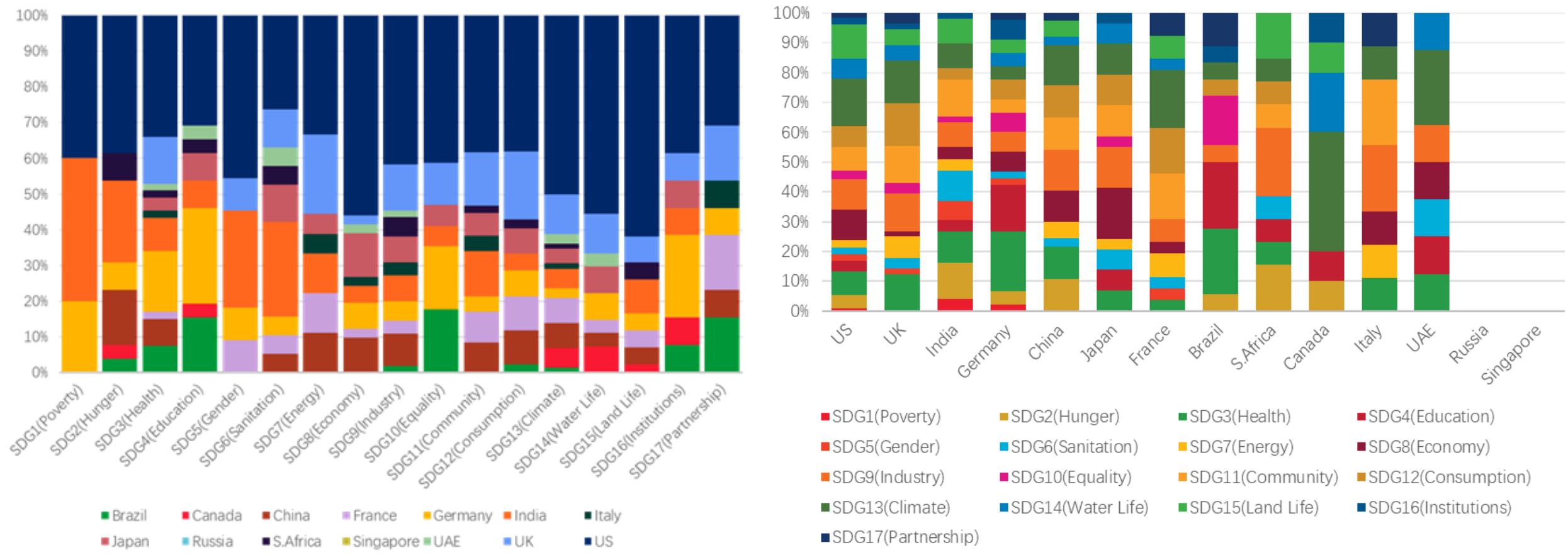

***Figure 29** AI for SDGs use case compositions: Countries and SDGs' Overview*

*(Based on DBLP Statistics)*

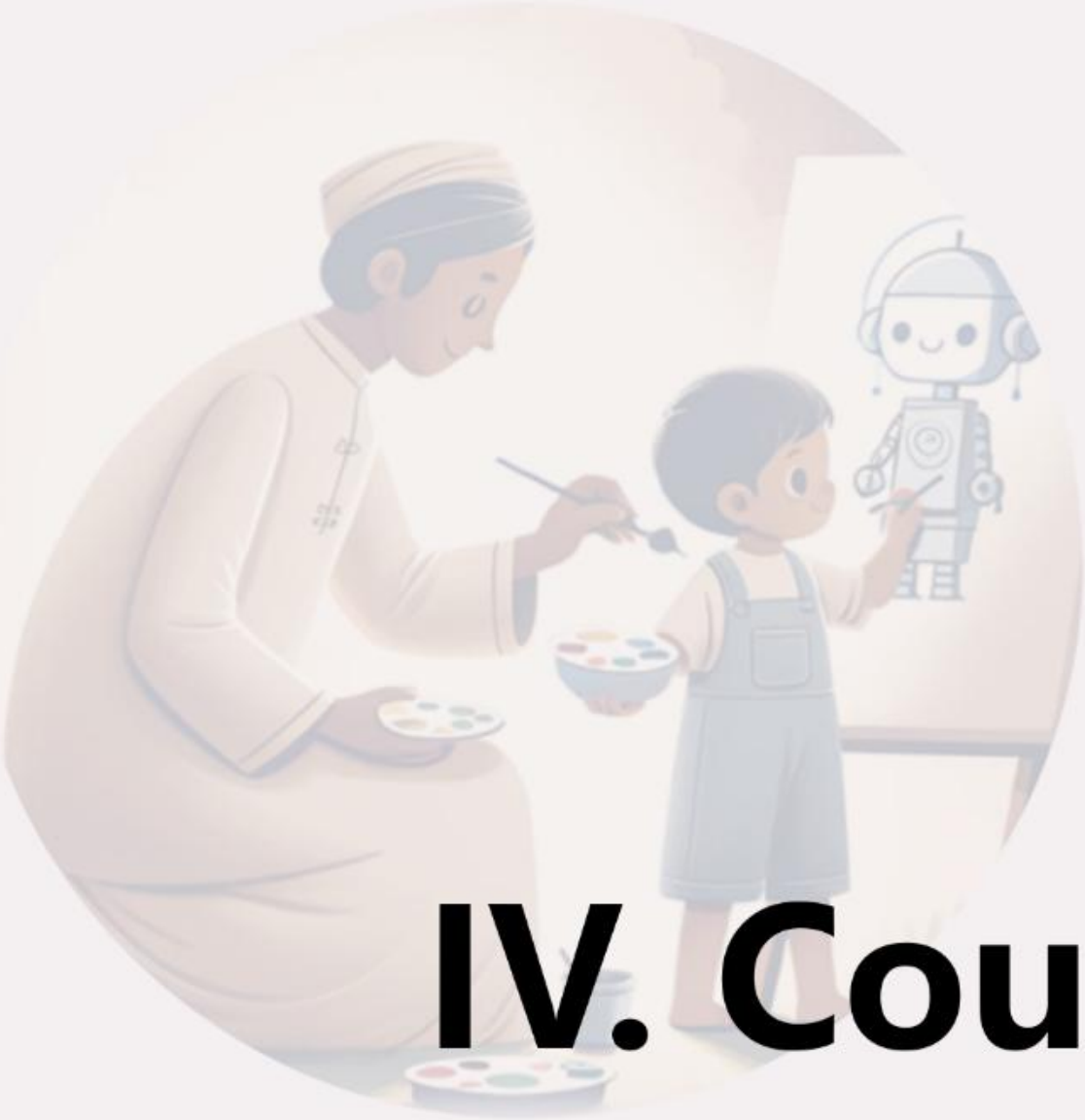

# IV. Country Profile

# Brazil

| AGILE index ranking | population | GDP per capita | Country Group |
|---|---|---|---|
| 13/14 | 216million | 8,870$ | middle-income |

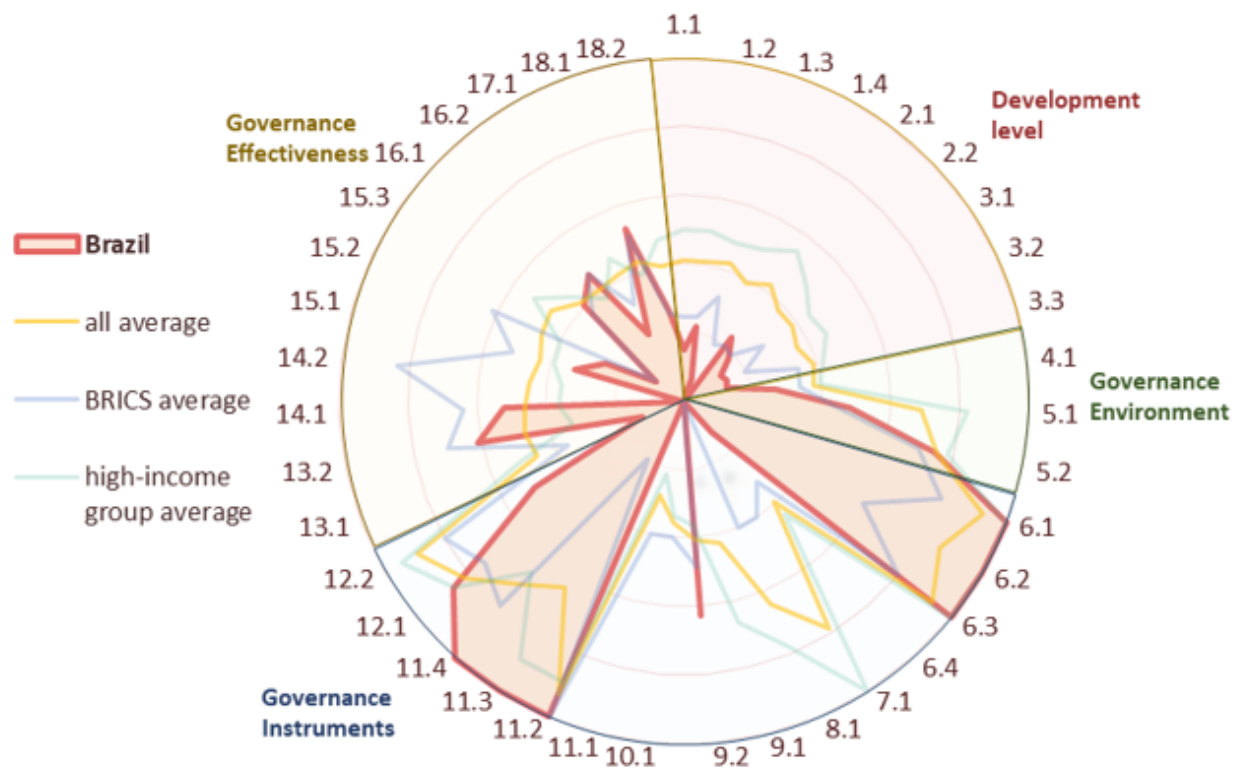

## P1 Development Level

|  |  |  |  |
|---|---|---|---|
| D1.1. AI Publications | (Number) | 15 | 78402 |
| D1.2.AI Professionals | (Number) | 22 | 48413 |
| D1.3. AI Patents | (Number) | 7 | 25 |
| D1.4. AI Systems | (Number) | 0 | 0 |
| D2.1. Colocation Data Centers | (Number) | 23 | 97 |
| D2.2. Supercomputer FLOP/s | (pFLOP/s) | 18 | 106 |
| D3.1. AI Companies' Funding | (billion$) | 13 | 1 |
| D3.2. AI Startups | (Number) | 14 | 198 |
| D3.3. Listed AI Companies | (Number) | 13 | 1 |

## P2 Governance Environment

|  |  |  |  |
|---|---|---|---|
| D4.1. AI Risk Incidents | (Number) | 27 | 115 |
| D5.1. Overall Governance Level | (WGI&GTMI) | 48 | 48 |
| D5.2. SGDs Progress | (SDGDI) | 74 | 74 |

## P3 Governance Instruments

|  |  |  |  |
|---|---|---|---|
| D6.1. AI Strategy Release Status | (Year of publish) | 100 | 2021 |
| D6.2. Measurable Goals in AI Strategy | | 100 | Yes |
| D6.3. Training Inclusion in AI Strategy | | 100 | Yes |
| D6.4. AI Budget | (billion$) | 12 | No |
| D7.1. AI Governance Bodies Establishment | (Year of publish) | 0 | No |
| D8.1. AI Principles Issued by Government | (Year of publish) | 0 | No |
| D9.1. AI Impact Assessment Mechanism | (Year of publish) | 0 | No |
| D9.2. Regulatory Sandboxes for AI | (Year of publish) | 63 | 2 |
| D10.1. AI Standards & Certification | (Year of publish) | 0 | No |
| D11.1. National Laws pertaining AI | (Year of publish) | 0 | No |
| D11.2. AI-Specific Data Protection Laws | (Year of publish) | 100 | 2019 |
| D11.3. AI Consumer Protection Legislation | (Year of publish) | 100 | 1990 |
| D11.4. Ongoing AI Legislation Process | (Year of publish) | 100 | 2023 |
| D12.1. International AI Governance Participation | (participation rate) | 86 | 6/7 |
| D12.2. ISO AI Standardization Participation | (Level) | 50 | Midium |

## P4 Governance Effectiveness

|  |  |  |  |
|---|---|---|---|
| D13.1. Public AI Skill Proficiency | (PISA score) | 13 | 384 |
| D13.2. Public Awareness of AI Impact | (scenario %) | 61 | 67 |
| D14.1. Positive Public Attitude towards AI | (positive %) | 52 | 71 |
| D14.2. Enterprise Attitude towards AI | (adoption %) | N/A | N/A |
| D15.1. Gender Ratio in AI literature | (male/female) | 33 | 3.86 |
| D15.2. Gender Ratio in AI Graduates | (female %) | 25 | 15 |
| D15.3. AI for Disadvantaged Groups | (internet %) | 10 | 58 |
| D16.1. Open AI Models & Datasets | (published numbers) | 40 | 11 |
| D16.2. AI Developer Community | (Github numbers) | 46 | 171 |
| D17.1. AI Governance Literature Volume | (Number) | 22 | 2565 |
| D18.1. AI & SDG Literature Volume | (Number) | 53 | 2090 |
| D18.2. AI for SDGs Cases | (Number) | 27 | 54 |

# Canada

| AGILE index ranking | population | GDP per capita | Country Group |
|---|---|---|---|
| 4/14 | 39million | 55,200$ | high-income |

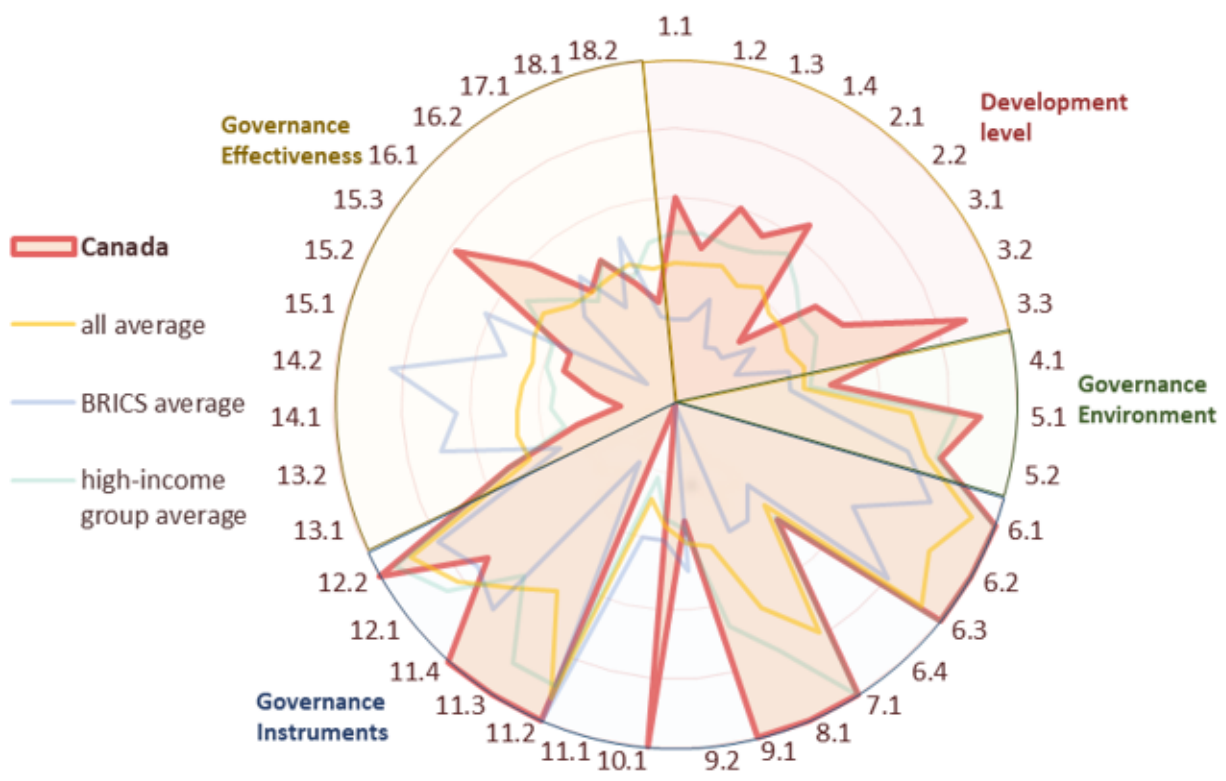

## P1 Development Level

|  |  |  |  |  |
|---|---|---|---|---|
| D1.1. AI Publications | 60 | 205201 | (Number) | ↑ |
| D1.2.AI Professionals | 46 | 122717 | (Number) | |
| D1.3. AI Patents | 60 | 1455 | (Number) | ↑ |
| D1.4. AI Systems | 55 | 45 | (Number) | |
| D2.1. Colocation Data Centers | 65 | 200 | (Number) | ↑ |
| D2.2. Supercomputer FLOP/s | 26 | 72 | (pFLOP/s) | ↓ |
| D3.1. AI Companies' Funding | 50 | 11 | (billion$) | |
| D3.2. AI Startups | 54 | 594 | (Number) | |
| D3.3. Listed AI Companies | 88 | 60 | (Number) | ↑ |

## P2 Governance Environment

|  |  |  |  |
|---|---|---|---|
| D4.1. AI Risk Incidents | 46 | 238 | (Number) |
| D5.1. Overall Governance Level | 89 | 89 | (WGI&GTMI) |
| D5.2. SGDs Progress | 79 | 79 | (SDGDI) |

## P3 Governance Instruments

|  |  |  |  |  |
|---|---|---|---|---|
| D6.1. AI Strategy Release Status | 100 | 2017 | (Year of publish) | |
| D6.2. Measurable Goals in AI Strategy | 100 | Yes | | |
| D6.3. Training Inclusion in AI Strategy | 100 | Yes | | |
| D6.4. AI Budget | 45 | 2.80 | (billion$) | |
| D7.1. AI Governance Bodies Establishment | 100 | 2019 | (Year of publish) | |
| D8.1. AI Principles Issued by Government | 100 | 2023 | (Year of publish) | ↑ |
| D9.1. AI Impact Assessment Mechanism | 100 | 2019 | (Year of publish) | ↑ |
| D9.2. Regulatory Sandboxes for AI | 34 | 1 | (Year of publish) | |
| D10.1. AI Standards & Certification | 100 | 2023 | (Year of publish) | ↑ |
| D11.1. National Laws pertaining AI | 0 | No | (Year of publish) | ↓ |
| D11.2. AI-Specific Data Protection Laws | 100 | 2019 | (Year of publish) | |
| D11.3. AI Consumer Protection Legislation | 100 | 1985 | (Year of publish) | ↑ |
| D11.4. Ongoing AI Legislation Process | 100 | 2022 | (Year of publish) | |
| D12.1. International AI Governance Participation | 71 | 5/7 | (participation rate) | |
| D12.2. ISO AI Standardization Participation | 100 | High | (Level) | |

## P4 Governance Effectiveness

|  |  |  |  |  |
|---|---|---|---|---|
| D13.1. Public AI Skill Proficiency | 52 | 512 | (PISA score) | |
| D13.2. Public Awareness of AI Impact | 29 | 48 | (scenario %) | |
| D14.1. Positive Public Attitude towards AI | 16 | 42 | (positive %) | |
| D14.2. Enterprise Attitude towards AI | 24 | 28 | (adoption %) | ↓ |
| D15.1. Gender Ratio in AI literature | 34 | 3.72 | (male/female) | |
| D15.2. Gender Ratio in AI Graduates | 34 | 30 | (female %) | |
| D15.3. AI for Disadvantaged Groups | 78 | 93 | (internet %) | ↑ |
| D16.1. Open AI Models & Datasets | 58 | 50 | (published numbers) | |
| D16.2. AI Developer Community | 41 | 532 | (Github numbers) | ↓ |
| D17.1. AI Governance Literature Volume | 47 | 7399 | (Number) | |
| D18.1. AI & SDG Literature Volume | 37 | 3976 | (Number) | |
| D18.2. AI for SDGs Cases | 30 | 30 | (Number) | ↓ |

# China

| AGILE index ranking | population | GDP per capita | Country Group |
|---|---|---|---|
| 2/14 | 1.43billion | 12,600$ | middle-income |

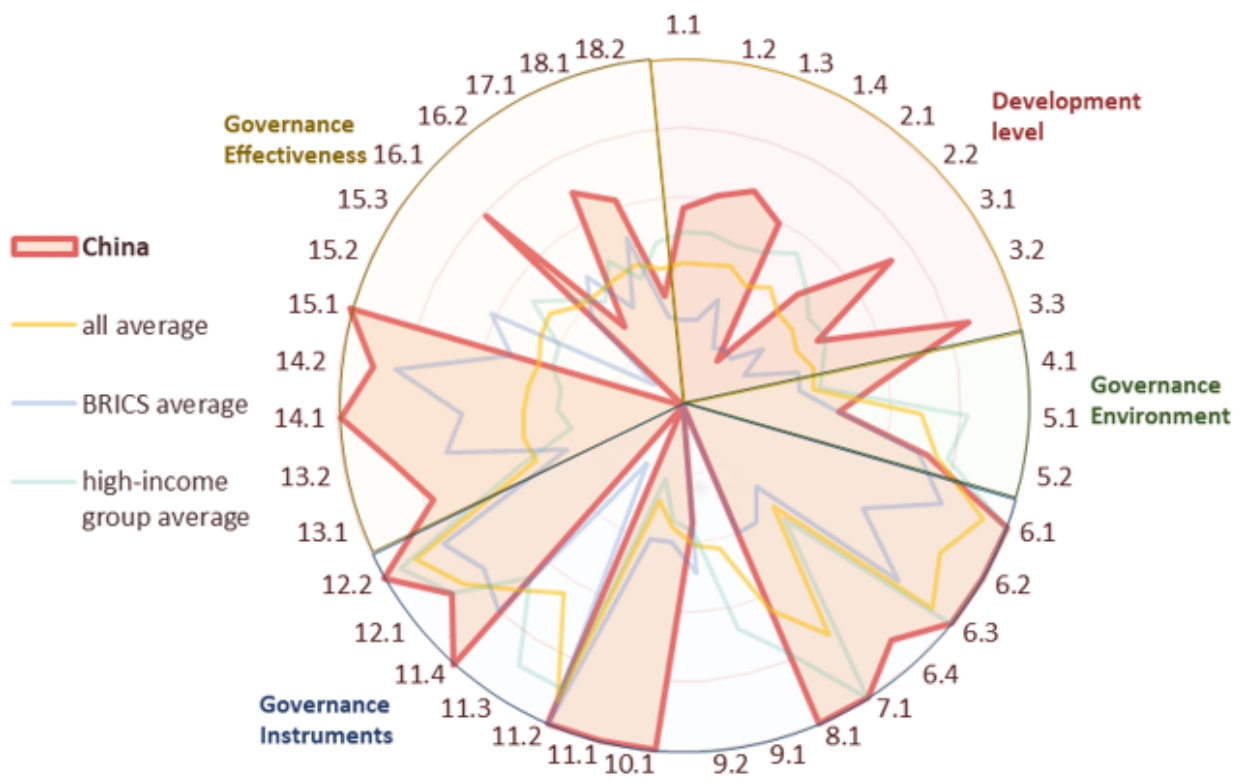

## P1 Development Level

|  |  |  |  |  |
|---|---|---|---|---|
| D1.1. AI Publications | (Number) | 57 | 957616 | |
| D1.2.AI Professionals | (Number) | 61 | 843167 | ↑ |
| D1.3. AI Patents | (Number) | 65 | 29160 | ↑ |
| D1.4. AI Systems | (Number) | 59 | 56 | |
| D2.1. Colocation Data Centers | (Number) | 16 | 92 | ↓ |
| D2.2. Supercomputer FLOP/s | (pFLOP/s) | 46 | 771 | ↑ |
| D3.1. AI Companies' Funding | (billion$) | 73 | 109 | ↑ |
| D3.2. AI Startups | (Number) | 43 | 1398 | |
| D3.3. Listed AI Companies | (Number) | 86 | 161 | ↑ |

## P2 Governance Environment

|  |  |  |  |  |
|---|---|---|---|---|
| D4.1. AI Risk Incidents | (Number) | 58 | 640 | |
| D5.1. Overall Governance Level | (WGI&GTMI) | 45 | 45 | |
| D5.2. SGDs Progress | (SDGDI) | 72 | 72 | |

## P3 Governance Instruments

|  |  |  |  |  |
|---|---|---|---|---|
| D6.1. AI Strategy Release Status | (Year of publish) | 100 | 2017 | |
| D6.2. Measurable Goals in AI Strategy | | 100 | Yes | |
| D6.3. Training Inclusion in AI Strategy | | 100 | Yes | |
| D6.4. AI Budget | (billion$) | 91 | 32.50 | ↑ |
| D7.1. AI Governance Bodies Establishment | (Year of publish) | 100 | 2019 | |
| D8.1. AI Principles Issued by Government | (Year of publish) | 100 | 2019 | ↑ |
| D9.1. AI Impact Assessment Mechanism | (Year of publish) | 0 | No | ↓ |
| D9.2. Regulatory Sandboxes for AI | (Year of publish) | 34 | 1 | |
| D10.1. AI Standards & Certification | (Year of publish) | 100 | 2023 | ↑ |
| D11.1. National Laws pertaining AI | (Year of publish) | 100 | 2021 | ↑ |
| D11.2. AI-Specific Data Protection Laws | (Year of publish) | 100 | 2021 | |
| D11.3. AI Consumer Protection Legislation | (Year of publish) | 0 | No | ↓ |
| D11.4. Ongoing AI Legislation Process | (Year of publish) | 100 | 2021 | |
| D12.1. International AI Governance Participation | (participation rate) | 86 | 6/7 | |
| D12.2. ISO AI Standardization Participation | (Level) | 100 | High | |

## P4 Governance Effectiveness

|  |  |  |  |  |
|---|---|---|---|---|
| D13.1. Public AI Skill Proficiency | (PISA score) | 77 | 591 | ↑ |
| D13.2. Public Awareness of AI Impact | (scenario %) | 85 | 70 | ↑ |
| D14.1. Positive Public Attitude towards AI | (positive %) | 99 | 78 | ↑ |
| D14.2. Enterprise Attitude towards AI | (adoption %) | 90 | 58 | ↑ |
| D15.1. Gender Ratio in AI literature | (male/female) | 100 | 1.22 | ↑ |
| D15.2. Gender Ratio in AI Graduates | (female %) | N/A | N/A | |
| D15.3. AI for Disadvantaged Groups | (internet %) | N/A | N/A | |
| D16.1. Open AI Models & Datasets | (published numbers) | 79 | 310 | ↑ |
| D16.2. AI Developer Community | (Github numbers) | 28 | 1978 | ↓ |
| D17.1. AI Governance Literature Volume | (Number) | 69 | 32436 | |
| D18.1. AI & SDG Literature Volume | (Number) | 62 | 18506 | |
| D18.2. AI for SDGs Cases | (Number) | 32 | 111 | |

# France

| AGILE index ranking | population | GDP per capita | Country Group |
|---|---|---|---|
| 8/14 | 64million | 43,000$ | high-income |

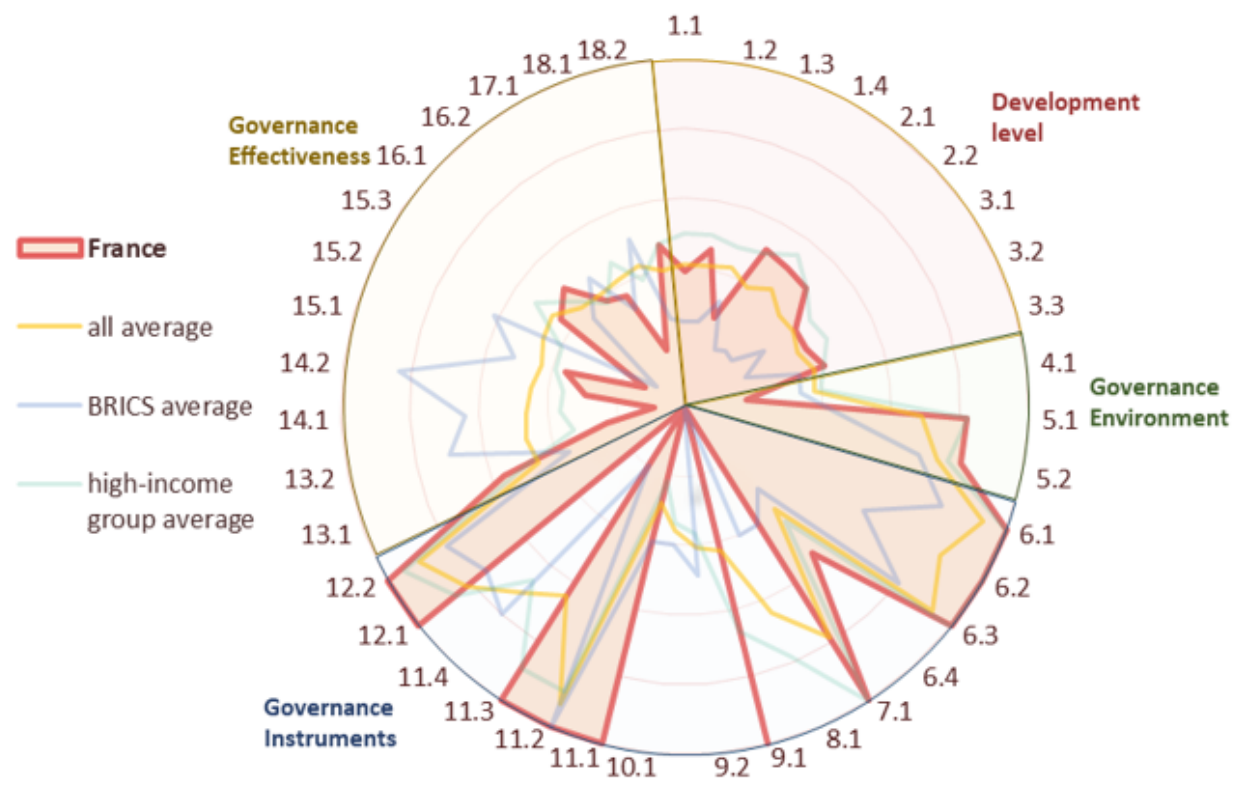

## P1 Development Level

|  |  |  |  |
|---|---|---|---|
| D1.1. AI Publications | 39 | 182837 | (Number) |
| D1.2.AI Professionals | 46 | 113144 | (Number) |
| D1.3. AI Patents | 27 | 565 | (Number) |
| D1.4. AI Systems | 51 | 16 | (Number) |
| D2.1. Colocation Data Centers | 50 | 178 | (Number) |
| D2.2. Supercomputer FLOP/s | 49 | 250 | (pFLOP/s) |
| D3.1. AI Companies' Funding | 39 | 7.7 | (billion$) |
| D3.2. AI Startups | 39 | 404 | (Number) |
| D3.3. Listed AI Companies | 42 | 13 | (Number) |

## P2 Governance Environment

|  |  |  |  |
|---|---|---|---|
| D4.1. AI Risk Incidents | 18 | 88 | (Number) |
| D5.1. Overall Governance Level | 82 | 82 | (WGI&GTMI) |
| D5.2. SGDs Progress | 82 | 82 | (SDGDI) |

## P3 Governance Instruments

|  |  |  |  |
|---|---|---|---|
| D6.1. AI Strategy Release Status | 100 | 2021 | (Year of publish) |
| D6.2. Measurable Goals in AI Strategy | 100 | Yes | |
| D6.3. Training Inclusion in AI Strategy | 100 | Yes | |
| D6.4. AI Budget | 56 | 4.50 | (billion$) |
| D7.1. AI Governance Bodies Establishment | 100 | 2023 | (Year of publish) |
| D8.1. AI Principles Issued by Government | 0 | No | (Year of publish) |
| D9.1. AI Impact Assessment Mechanism | 100 | 2022 | (Year of publish) |
| D9.2. Regulatory Sandboxes for AI | 0 | No | (Year of publish) |
| D10.1. AI Standards & Certification | 0 | No | (Year of publish) |
| D11.1. National Laws pertaining AI | 100 | 2016 | (Year of publish) |
| D11.2. AI-Specific Data Protection Laws | 100 | 2015 | (Year of publish) |
| D11.3. AI Consumer Protection Legislation | 100 | 2015 | (Year of publish) |
| D11.4. Ongoing AI Legislation Process | 0 | No | (Year of publish) |
| D12.1. International AI Governance Participation | 100 | 7/7 | (participation rate) |
| D12.2. ISO AI Standardization Participation | 100 | High | (Level) |

## P4 Governance Effectiveness

|  |  |  |  |
|---|---|---|---|
| D13.1. Public AI Skill Proficiency | 56 | 495 | (PISA score) |
| D13.2. Public Awareness of AI Impact | 23 | 52 | (scenario %) |
| D14.1. Positive Public Attitude towards AI | 9 | 46 | (positive %) |
| D14.2. Enterprise Attitude towards AI | 29 | 31 | (adoption %) |
| D15.1. Gender Ratio in AI literature | 36 | 3.62 | (male/female) |
| D15.2. Gender Ratio in AI Graduates | 13 | 14 | (female %) |
| D15.3. AI for Disadvantaged Groups | 44 | 78 | (internet %) |
| D16.1. Open AI Models & Datasets | 49 | 41 | (published numbers) |
| D16.2. AI Developer Community | 38 | 381 | (Github numbers) |
| D17.1. AI Governance Literature Volume | 36 | 6234 | (Number) |
| D18.1. AI & SDG Literature Volume | 17 | 2712 | (Number) |
| D18.2. AI for SDGs Cases | 47 | 78 | (Number) |

# Germany

| AGILE index ranking | population | GDP per capita | Country Group |
|---|---|---|---|
| 5/14 | 83million | 48,900$ | high-income |

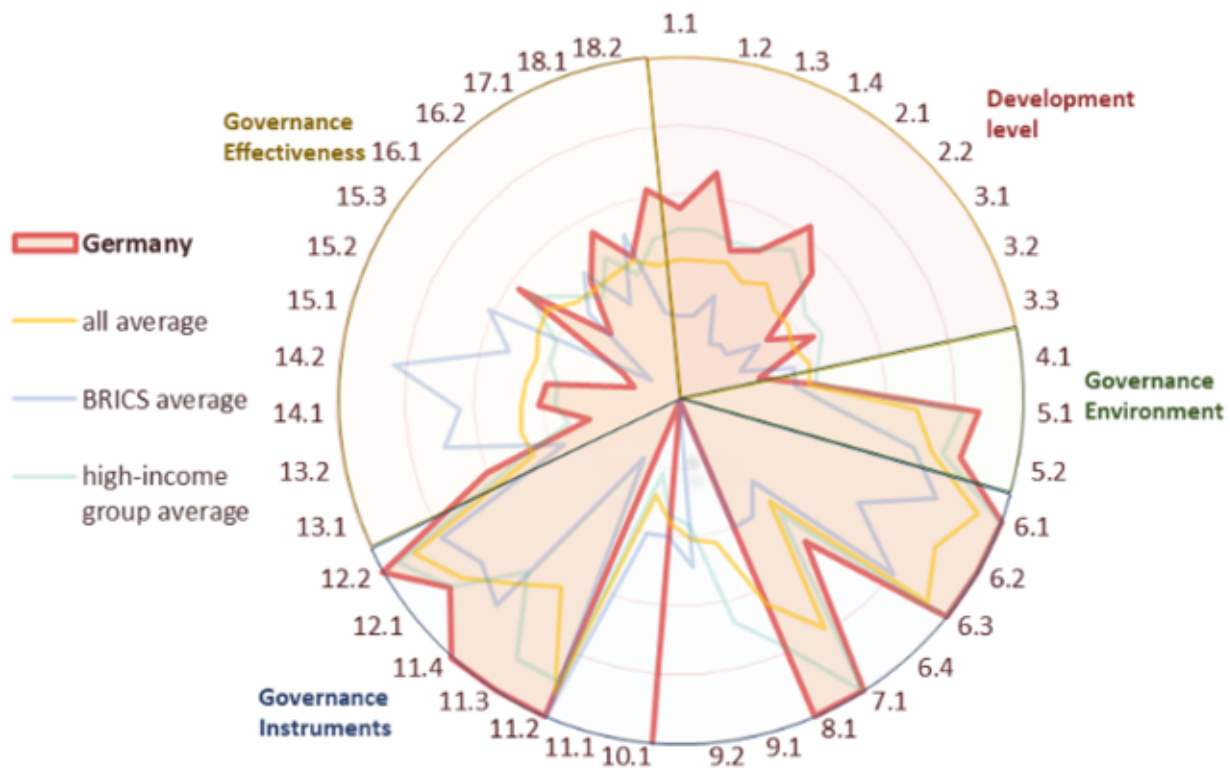

## P1 Development Level

|  |  |  |  |  |
|---|---|---|---|---|
| D1.1. AI Publications | 56 | 416706 | (Number) | |
| D1.2.AI Professionals | 67 | 276368 | (Number) | ↑ |
| D1.3. AI Patents | 46 | 2392 | (Number) | |
| D1.4. AI Systems | 49 | 18 | (Number) | |
| D2.1. Colocation Data Centers | 63 | 266 | (Number) | ↑ |
| D2.2. Supercomputer FLOP/s | 53 | 391 | (pFLOP/s) | |
| D3.1. AI Companies' Funding | 31 | 7.3 | (billion$) | |
| D3.2. AI Startups | 43 | 522 | (Number) | |
| D3.3. Listed AI Companies | 24 | 6 | (Number) | ↓ |

## P2 Governance Environment

|  |  |  |  |  |
|---|---|---|---|---|
| D4.1. AI Risk Incidents | 39 | 179 | (Number) | |
| D5.1. Overall Governance Level | 87 | 87 | (WGI&GTMI) | |
| D5.2. SGDs Progress | 83 | 83 | (SDGDI) | |

## P3 Governance Instruments

|  |  |  |  |  |
|---|---|---|---|---|
| D6.1. AI Strategy Release Status | 100 | 2020 | (Year of publish) | |
| D6.2. Measurable Goals in AI Strategy | 100 | Yes | | |
| D6.3. Training Inclusion in AI Strategy | 100 | Yes | | |
| D6.4. AI Budget | 55 | 5.80 | (billion$) | |
| D7.1. AI Governance Bodies Establishment | 100 | 2019 | (Year of publish) | |
| D8.1. AI Principles Issued by Government | 100 | 2019 | (Year of publish) | ↑ |
| D9.1. AI Impact Assessment Mechanism | 0 | No | (Year of publish) | ↓ |
| D9.2. Regulatory Sandboxes for AI | 0 | No | (Year of publish) | |
| D10.1. AI Standards & Certification | 100 | 2023 | (Year of publish) | ↑ |
| D11.1. National Laws pertaining AI | 0 | No | (Year of publish) | ↓ |
| D11.2. AI-Specific Data Protection Laws | 100 | 2017 | (Year of publish) | |
| D11.3. AI Consumer Protection Legislation | 100 | 2010 | (Year of publish) | ↑ |
| D11.4. Ongoing AI Legislation Process | 100 | 2020 | (Year of publish) | |
| D12.1. International AI Governance Participation | 86 | 6/7 | (participation rate) | |
| D12.2. ISO AI Standardization Participation | 100 | High | (Level) | |

## P4 Governance Effectiveness

|  |  |  |  |  |
|---|---|---|---|---|
| D13.1. Public AI Skill Proficiency | 60 | 500 | (PISA score) | |
| D13.2. Public Awareness of AI Impact | 27 | 52 | (scenario %) | ↓ |
| D14.1. Positive Public Attitude towards AI | 41 | 52 | (positive %) | |
| D14.2. Enterprise Attitude towards AI | 39 | 34 | (adoption %) | |
| D15.1. Gender Ratio in AI literature | 14 | 7.09 | (male/female) | ↓ |
| D15.2. Gender Ratio in AI Graduates | 18 | 19 | (female %) | ↓ |
| D15.3. AI for Disadvantaged Groups | 57 | 85 | (internet %) | |
| D16.1. Open AI Models & Datasets | 28 | 33 | (published numbers) | ↓ |
| D16.2. AI Developer Community | 43 | 807 | (Github numbers) | |
| D17.1. AI Governance Literature Volume | 55 | 13992 | (Number) | |
| D18.1. AI & SDG Literature Volume | 44 | 6871 | (Number) | |
| D18.2. AI for SDGs Cases | 62 | 135 | (Number) | |

# India

| AGILE index ranking | population | GDP per capita | Country Group |
|---|---|---|---|
| 11/14 | 1.43billion | 2,370$ | middle-income |

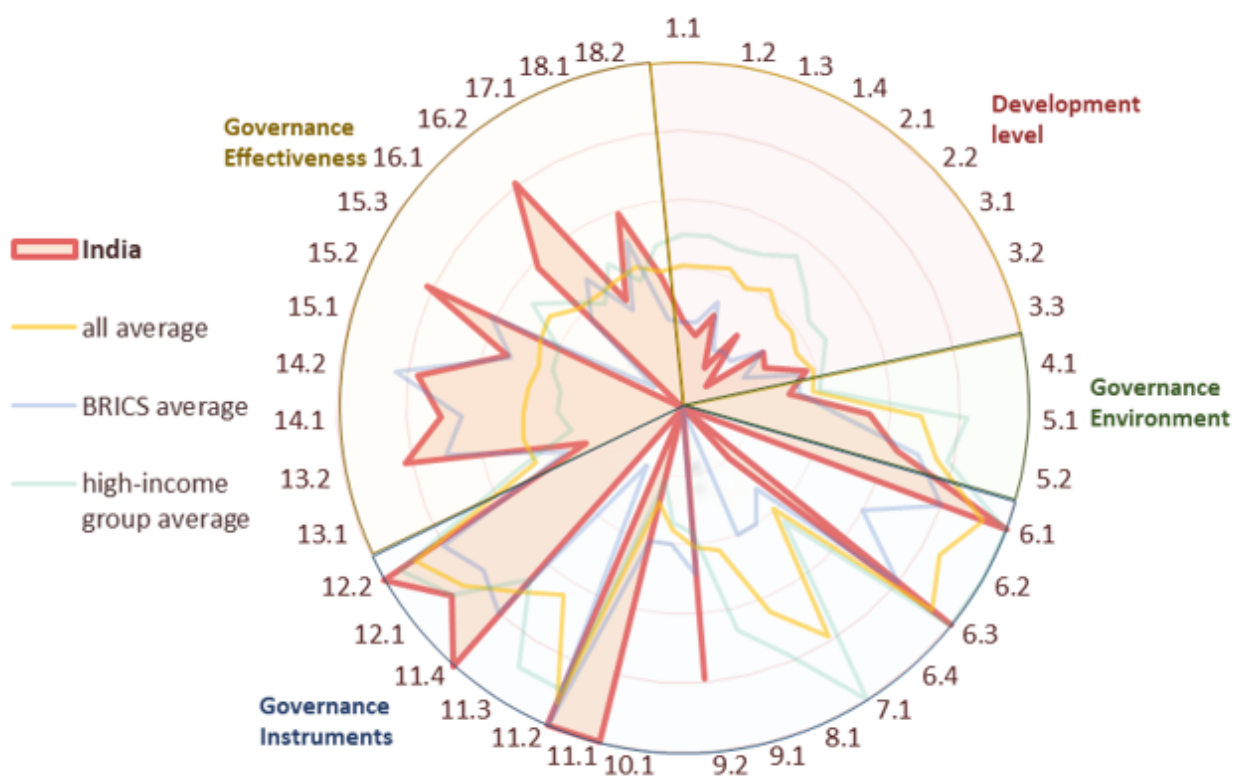

## P1 Development Level

|  |  |  |  |
|---|---|---|---|
| D1.1. AI Publications | (Number) | 25 | 130407 |
| D1.2.AI Professionals | (Number) | 21 | 75829 |
| D1.3. AI Patents | (Number) | 28 | 1211 |
| D1.4. AI Systems | (Number) | 13 | 2 |
| D2.1. Colocation Data Centers | (Number) | 26 | 165 |
| D2.2. Supercomputer FLOP/s | (pFLOP/s) | 9 | 25 |
| D3.1. AI Companies' Funding | (billion$) | 28 | 5.6 |
| D3.2. AI Startups | (Number) | 26 | 624 |
| D3.3. Listed AI Companies | (Number) | 37 | 12 |

## P2 Governance Environment

|  |  |  |  |
|---|---|---|---|
| D4.1. AI Risk Incidents | (Number) | 31 | 253 |
| D5.1. Overall Governance Level | (WGI&GTMI) | 54 | 54 |
| D5.2. SGDs Progress | (SDGDI) | 63 | 63 |

## P3 Governance Instruments

|  |  |  |  |
|---|---|---|---|
| D6.1. AI Strategy Release Status | (Year of publish) | 100 | 2018 |
| D6.2. Measurable Goals in AI Strategy | | 0 | No |
| D6.3. Training Inclusion in AI Strategy | | 100 | Yes |
| D6.4. AI Budget | (billion$) | 20 | 0.80 |
| D7.1. AI Governance Bodies Establishment | (Year of publish) | 0 | No |
| D8.1. AI Principles Issued by Government | (Year of publish) | 0 | No |
| D9.1. AI Impact Assessment Mechanism | (Year of publish) | 0 | No |
| D9.2. Regulatory Sandboxes for AI | (Year of publish) | 79 | 3 |
| D10.1. AI Standards & Certification | (Year of publish) | 0 | No |
| D11.1. National Laws pertaining AI | (Year of publish) | 100 | 2000 |
| D11.2. AI-Specific Data Protection Laws | (Year of publish) | 100 | 2023 |
| D11.3. AI Consumer Protection Legislation | (Year of publish) | 0 | No |
| D11.4. Ongoing AI Legislation Process | (Year of publish) | 100 | 2023 |
| D12.1. International AI Governance Participation | (participation rate) | 86 | 6/7 |
| D12.2. ISO AI Standardization Participation | (Level) | 100 | High |

## P4 Governance Effectiveness

|  |  |  |  |
|---|---|---|---|
| D13.1. Public AI Skill Proficiency | (PISA score) | 30 | N/A |
| D13.2. Public Awareness of AI Impact | (scenario %) | 82 | 68 |
| D14.1. Positive Public Attitude towards AI | (positive %) | 70 | 71 |
| D14.2. Enterprise Attitude towards AI | (adoption %) | 77 | 57 |
| D15.1. Gender Ratio in AI literature | (male/female) | 53 | 3.15 |
| D15.2. Gender Ratio in AI Graduates | (female %) | 82 | 46 |
| D15.3. AI for Disadvantaged Groups | (internet %) | N/A | N/A |
| D16.1. Open AI Models & Datasets | (published numbers) | 58 | 32 |
| D16.2. AI Developer Community | (Github numbers) | 81 | 596 |
| D17.1. AI Governance Literature Volume | (Number) | 35 | 4222 |
| D18.1. AI & SDG Literature Volume | (Number) | 59 | 3491 |
| D18.2. AI for SDGs Cases | (Number) | 38 | 147 |

# Italy

| AGILE index ranking | population | GDP per capita | Country Group |
|---|---|---|---|
| 10/14 | 59million | 34,100$ | high-income |

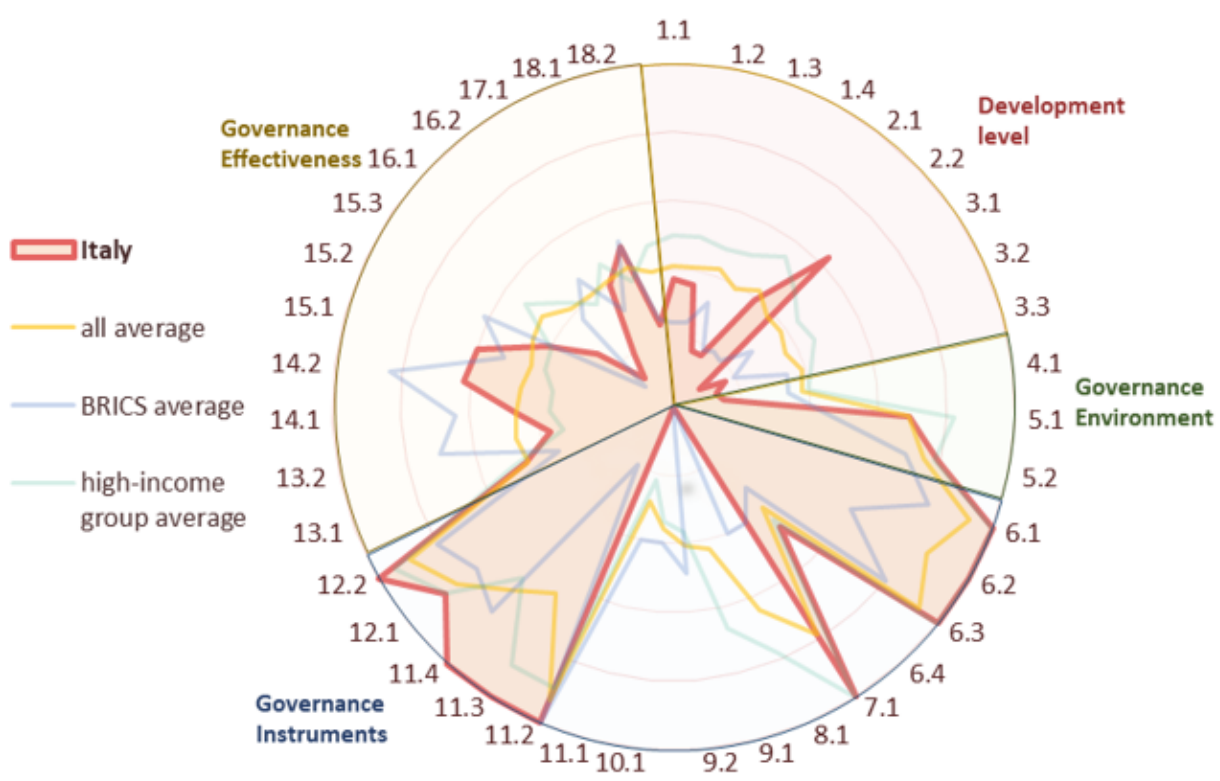

## P1 Development Level

|  |  |  |  |  |
|---|---|---|---|---|
| D1.1. AI Publications | 37 | 176252 | (Number) | |
| D1.2.AI Professionals | 36 | 121605 | (Number) | |
| D1.3. AI Patents | 17 | 252 | (Number) | ↓ |
| D1.4. AI Systems | 17 | 3 | (Number) | ↓ |
| D2.1. Colocation Data Centers | 39 | 97 | (Number) | |
| D2.2. Supercomputer FLOP/s | 63 | 464 | (pFLOP/s) | ↑ |
| D3.1. AI Companies' Funding | 9 | 0.5 | (billion$) | ↓ |
| D3.2. AI Startups | 17 | 128 | (Number) | ↓ |
| D3.3. Listed AI Companies | 13 | 1 | (Number) | ↓ |

## P2 Governance Environment

|  |  |  |  |  |
|---|---|---|---|---|
| D4.1. AI Risk Incidents | 15 | 29 | (Number) | ↑ |
| D5.1. Overall Governance Level | 69 | 69 | (WGI&GTMI) | |
| D5.2. SGDs Progress | 79 | 79 | (SDGDI) | |

## P3 Governance Instruments

|  |  |  |  |  |
|---|---|---|---|---|
| D6.1. AI Strategy Release Status | 100 | 2022 | (Year of publish) | |
| D6.2. Measurable Goals in AI Strategy | 100 | Yes | | |
| D6.3. Training Inclusion in AI Strategy | 100 | Yes | | |
| D6.4. AI Budget | 47 | 2.90 | (billion$) | |
| D7.1. AI Governance Bodies Establishment | 100 | 1996 | (Year of publish) | |
| D8.1. AI Principles Issued by Government | 0 | No | (Year of publish) | |
| D9.1. AI Impact Assessment Mechanism | 0 | No | (Year of publish) | ↓ |
| D9.2. Regulatory Sandboxes for AI | 0 | No | (Year of publish) | |
| D10.1. AI Standards & Certification | 0 | No | (Year of publish) | ↓ |
| D11.1. National Laws pertaining AI | 0 | No | (Year of publish) | ↓ |
| D11.2. AI-Specific Data Protection Laws | 100 | 2016 | (Year of publish) | |
| D11.3. AI Consumer Protection Legislation | 100 | 2005 | (Year of publish) | ↑ |
| D11.4. Ongoing AI Legislation Process | 100 | 2021 | (Year of publish) | |
| D12.1. International AI Governance Participation | 86 | 6/7 | (participation rate) | |
| D12.2. ISO AI Standardization Participation | 100 | High | (Level) | |

## P4 Governance Effectiveness

|  |  |  |  |  |
|---|---|---|---|---|
| D13.1. Public AI Skill Proficiency | 46 | 487 | (PISA score) | |
| D13.2. Public Awareness of AI Impact | 37 | N/A | (scenario %) | |
| D14.1. Positive Public Attitude towards AI | 46 | 52 | (positive %) | |
| D14.2. Enterprise Attitude towards AI | 62 | 42 | (adoption %) | ↑ |
| D15.1. Gender Ratio in AI literature | 60 | 2.87 | (male/female) | |
| D15.2. Gender Ratio in AI Graduates | 41 | 16 | (female %) | |
| D15.3. AI for Disadvantaged Groups | 27 | 71 | (internet %) | ↓ |
| D16.1. Open AI Models & Datasets | 12 | 1 | (published numbers) | ↓ |
| D16.2. AI Developer Community | 19 | 125 | (Github numbers) | ↓ |
| D17.1. AI Governance Literature Volume | 40 | 6033 | (Number) | |
| D18.1. AI & SDG Literature Volume | 49 | 3974 | (Number) | |
| D18.2. AI for SDGs Cases | 24 | 27 | (Number) | ↓ |

# Japan

| AGILE index ranking | population | GDP per capita | Country Group |
|---|---|---|---|
| 7/14 | 123million | 34,300$ | high-income |

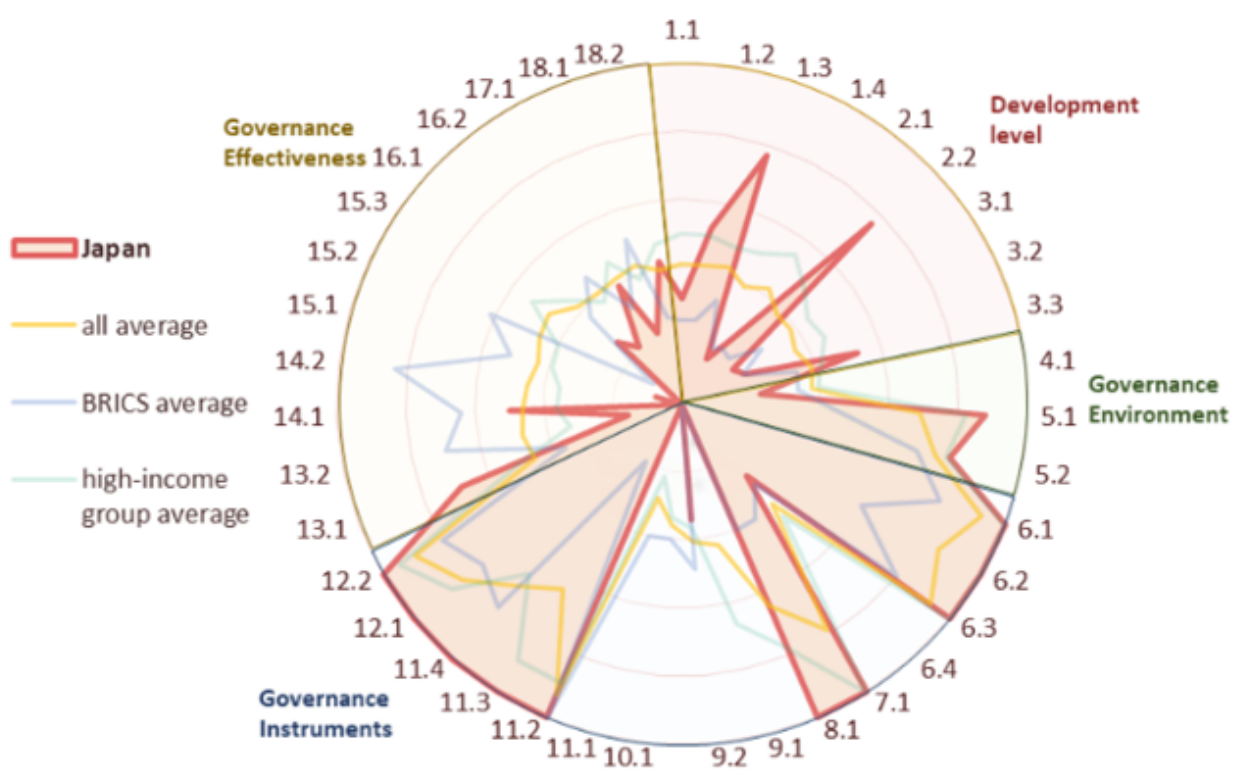

## P1 Development Level

|  |  |  |  |
|---|---|---|---|
| D1.1. AI Publications | 31 | 154795 | (Number) |
| D1.2.AI Professionals | 52 | 122626 | (Number) |
| D1.3. AI Patents | 77 | 4881 | (Number) |
| D1.4. AI Systems | 15 | 7 | (Number) |
| D2.1. Colocation Data Centers | 23 | 68 | (Number) |
| D2.2. Supercomputer FLOP/s | 76 | 866 | (pFLOP/s) |
| D3.1. AI Companies' Funding | 18 | 3 | (billion$) |
| D3.2. AI Startups | 20 | 250 | (Number) |
| D3.3. Listed AI Companies | 53 | 26 | (Number) |

## P2 Governance Environment

|  |  |  |  |
|---|---|---|---|
| D4.1. AI Risk Incidents | 23 | 82 | (Number) |
| D5.1. Overall Governance Level | 88 | 88 | (WGI&GTMI) |
| D5.2. SGDs Progress | 79 | 79 | (SDGDI) |

## P3 Governance Instruments

|  |  |  |  |
|---|---|---|---|
| D6.1. AI Strategy Release Status | 100 | 2022 | (Year of publish) |
| D6.2. Measurable Goals in AI Strategy | 100 | Yes | |
| D6.3. Training Inclusion in AI Strategy | 100 | Yes | |
| D6.4. AI Budget | 28 | 1.60 | (billion$) |
| D7.1. AI Governance Bodies Establishment | 100 | 2017 | (Year of publish) |
| D8.1. AI Principles Issued by Government | 100 | 2018 | (Year of publish) |
| D9.1. AI Impact Assessment Mechanism | 0 | No | (Year of publish) |
| D9.2. Regulatory Sandboxes for AI | 34 | 1 | (Year of publish) |
| D10.1. AI Standards & Certification | 0 | No | (Year of publish) |
| D11.1. National Laws pertaining AI | 0 | No | (Year of publish) |
| D11.2. AI-Specific Data Protection Laws | 100 | 2003 | (Year of publish) |
| D11.3. AI Consumer Protection Legislation | 100 | 1994 | (Year of publish) |
| D11.4. Ongoing AI Legislation Process | 100 | 2020 | (Year of publish) |
| D12.1. International AI Governance Participation | 100 | 7/7 | (participation rate) |
| D12.2. ISO AI Standardization Participation | 100 | High | (Level) |

## P4 Governance Effectiveness

|  |  |  |  |
|---|---|---|---|
| D13.1. Public AI Skill Proficiency | 68 | 527 | (PISA score) |
| D13.2. Public Awareness of AI Impact | 16 | 53 | (scenario %) |
| D14.1. Positive Public Attitude towards AI | 50 | 53 | (positive %) |
| D14.2. Enterprise Attitude towards AI | N/A | N/A | (adoption %) |
| D15.1. Gender Ratio in AI literature | 8 | 9.67 | (male/female) |
| D15.2. Gender Ratio in AI Graduates | N/A | N/A | (female %) |
| D15.3. AI for Disadvantaged Groups | N/A | N/A | (internet %) |
| D16.1. Open AI Models & Datasets | 26 | 14 | (published numbers) |
| D16.2. AI Developer Community | 21 | 401 | (Github numbers) |
| D17.1. AI Governance Literature Volume | 39 | 5542 | (Number) |
| D18.1. AI & SDG Literature Volume | 22 | 2530 | (Number) |
| D18.2. AI for SDGs Cases | 42 | 87 | (Number) |

# Russia

| AGILE index ranking | population | GDP per capita | Country Group |
|---|---|---|---|
| 12 /14 | 144million | 15,500$ | middle-income |

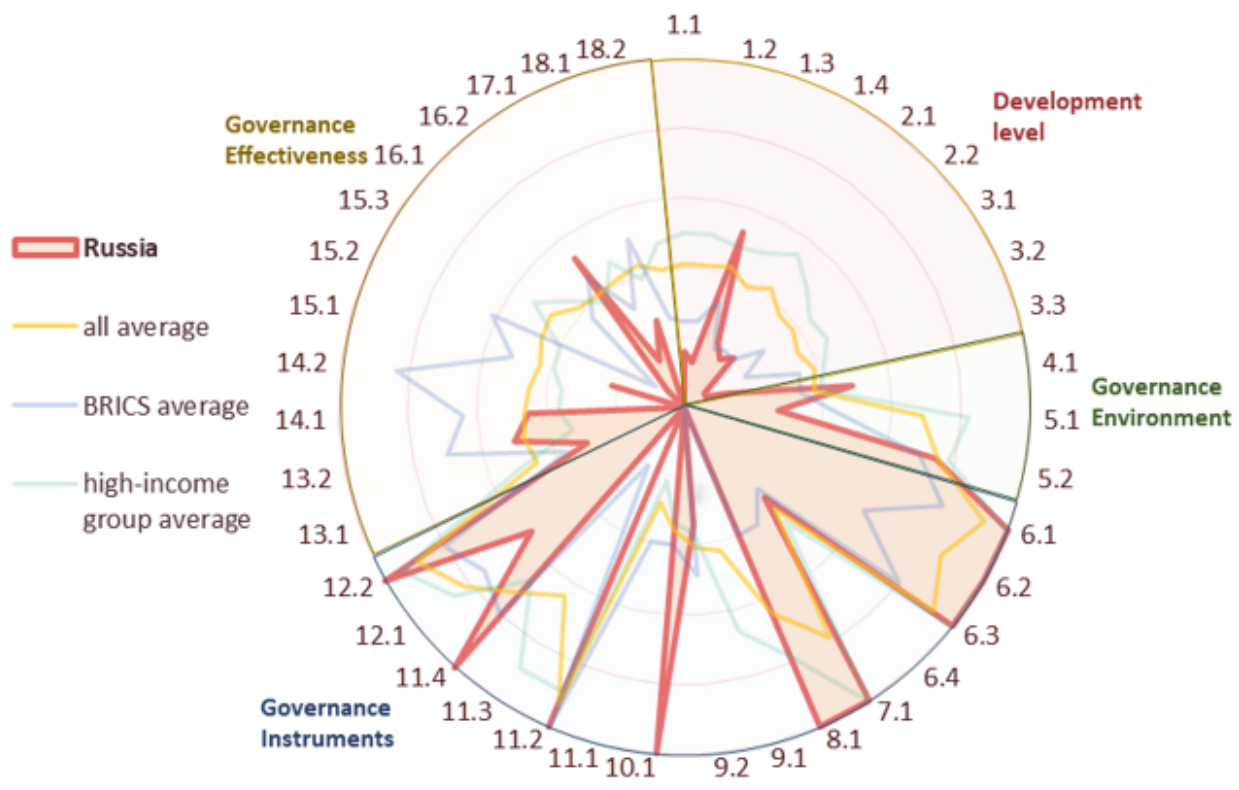

## P1 Development Level

|  |  |  |  |
|---|---|---|---|
| D1.1. AI Publications | 16 | 22956 | (Number) |
| D1.2.AI Professionals | 13 | 11859 | (Number) |
| D1.3. AI Patents | 53 | 2964 | (Number) |
| D1.4. AI Systems | 21 | 1 | (Number) |
| D2.1. Colocation Data Centers | 17 | 59 | (Number) |
| D2.2. Supercomputer FLOP/s | 20 | 102 | (pFLOP/s) |
| D3.1. AI Companies' Funding | 7 | 0.3 | (billion$) |
| D3.2. AI Startups | 7 | 64 | (Number) |
| D3.3. Listed AI Companies | 13 | 1 | (Number) |

## P2 Governance Environment

|  |  |  |  |
|---|---|---|---|
| D4.1. AI Risk Incidents | 49 | 162 | (Number) |
| D5.1. Overall Governance Level | 27 | 27 | (WGI&GTMI) |
| D5.2. SGDs Progress | 74 | 74 | (SDGDI) |

## P3 Governance Instruments

|  |  |  |  |
|---|---|---|---|
| D6.1. AI Strategy Release Status | 100 | 2019 | (Year of publish) |
| D6.2. Measurable Goals in AI Strategy | 100 | Yes | |
| D6.3. Training Inclusion in AI Strategy | 100 | Yes | |
| D6.4. AI Budget | 35 | 1.50 | (billion$) |
| D7.1. AI Governance Bodies Establishment | 100 | 2022 | (Year of publish) |
| D8.1. AI Principles Issued by Government | 100 | 2022 | (Year of publish) |
| D9.1. AI Impact Assessment Mechanism | 0 | No | (Year of publish) |
| D9.2. Regulatory Sandboxes for AI | 34 | No | (Year of publish) |
| D10.1. AI Standards & Certification | 100 | 2023 | (Year of publish) |
| D11.1. National Laws pertaining AI | 0 | No | (Year of publish) |
| D11.2. AI-Specific Data Protection Laws | 100 | 2006 | (Year of publish) |
| D11.3. AI Consumer Protection Legislation | 0 | No | (Year of publish) |
| D11.4. Ongoing AI Legislation Process | 100 | 2020 | (Year of publish) |
| D12.1. International AI Governance Participation | 57 | 4/7 | (participation rate) |
| D12.2. ISO AI Standardization Participation | 100 | High | (Level) |

## P4 Governance Effectiveness

|  |  |  |  |
|---|---|---|---|
| D13.1. Public AI Skill Proficiency | 30 | 488 | (PISA score) |
| D13.2. Public Awareness of AI Impact | 50 | N/A | (scenario %) |
| D14.1. Positive Public Attitude towards AI | 45 | 52 | (positive %) |
| D14.2. Enterprise Attitude towards AI | N/A | N/A | (adoption %) |
| D15.1. Gender Ratio in AI literature | 22 | 4.61 | (male/female) |
| D15.2. Gender Ratio in AI Graduates | N/A | N/A | (female %) |
| D15.3. AI for Disadvantaged Groups | N/A | N/A | (internet %) |
| D16.1. Open AI Models & Datasets | 7 | 1 | (published numbers) |
| D16.2. AI Developer Community | 53 | 145 | (Github numbers) |
| D17.1. AI Governance Literature Volume | 15 | 528 | (Number) |
| D18.1. AI & SDG Literature Volume | 26 | 474 | (Number) |
| D18.2. AI for SDGs Cases | 0 | 0 | (Number) |

# South Africa

| AGILE index ranking | population | GDP per capita | Country Group |
|---|---|---|---|
| 14/14 | 60million | 6700$ | middle-income |

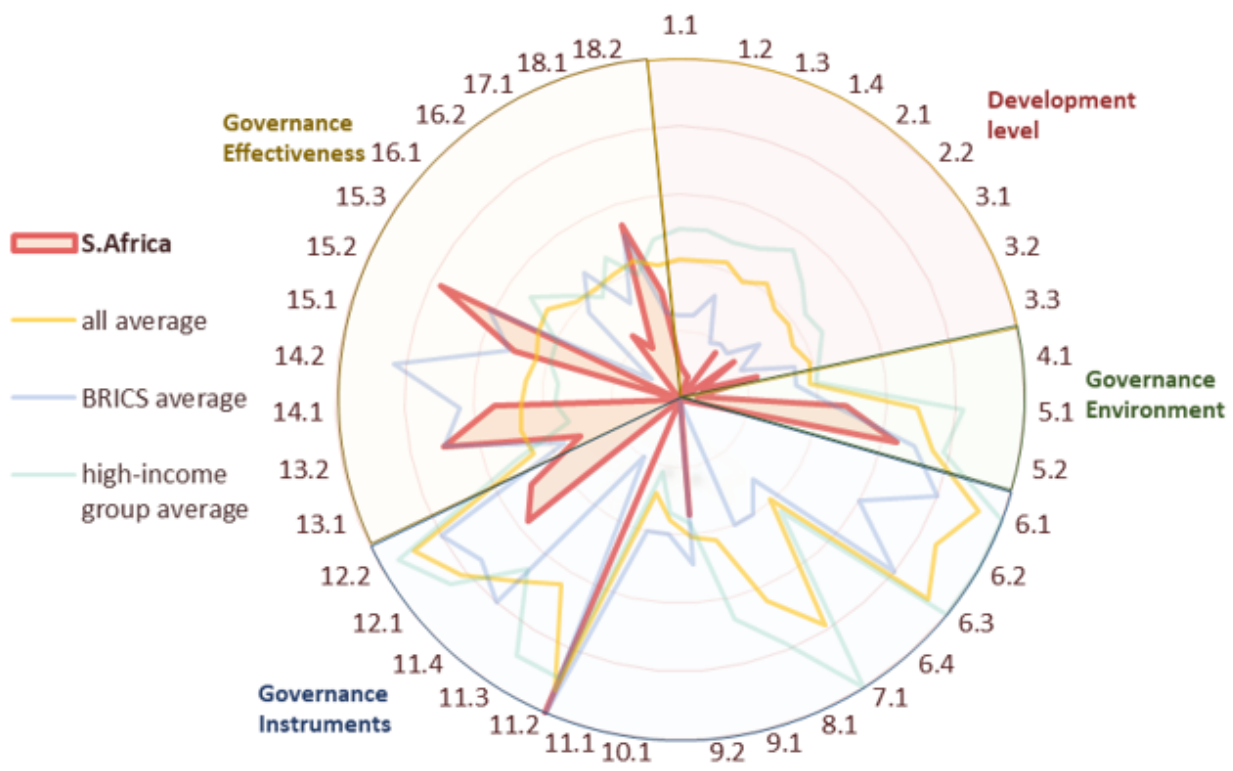

## P1 Development Level

|  |  |  |  |  |
|---|---|---|---|---|
| D1.1. AI Publications | (Number) | 10 | 10810 | ↓ |
| D1.2.AI Professionals | (Number) | 8 | 4334 | ↓ |
| D1.3. AI Patents | (Number) | 7 | 25 | ↓ |
| D1.4. AI Systems | (Number) | 0 | 0 | ↓ |
| D2.1. Colocation Data Centers | (Number) | 17 | 32 | ↓ |
| D2.2. Supercomputer FLOP/s | (pFLOP/s) | 0 | 0 | ↓ |
| D3.1. AI Companies' Funding | (billion$) | 19 | 0.6 | ↓ |
| D3.2. AI Startups | (Number) | 7 | 32 | ↓ |
| D3.3. Listed AI Companies | (Number) | 23 | 1 | ↓ |

## P2 Governance Environment

|  |  |  |  |  |
|---|---|---|---|---|
| D4.1. AI Risk Incidents | (Number) | 5 | 8 | ↑ |
| D5.1. Overall Governance Level | (WGI&GTMI) | 48 | 48 | |
| D5.2. SGDs Progress | (SDGDI) | 64 | 64 | |

## P3 Governance Instruments

|  |  |  |  |  |
|---|---|---|---|---|
| D6.1. AI Strategy Release Status | (Year of publish) | 0 | No | ↓ |
| D6.2. Measurable Goals in AI Strategy | | 0 | No | ↓ |
| D6.3. Training Inclusion in AI Strategy | | 0 | No | ↓ |
| D6.4. AI Budget | (billion$) | 0 | No | ↓ |
| D7.1. AI Governance Bodies Establishment | (Year of publish) | 0 | No | ↓ |
| D8.1. AI Principles Issued by Government | (Year of publish) | 0 | No | ↓ |
| D9.1. AI Impact Assessment Mechanism | (Year of publish) | 0 | No | ↓ |
| D9.2. Regulatory Sandboxes for AI | (Year of publish) | 34 | 1 | |
| D10.1. AI Standards & Certification | (Year of publish) | 0 | No | |
| D11.1. National Laws pertaining AI | (Year of publish) | 0 | No | |
| D11.2. AI-Specific Data Protection Laws | (Year of publish) | 100 | 2013 | |
| D11.3. AI Consumer Protection Legislation | (Year of publish) | 0 | No | ↓ |
| D11.4. Ongoing AI Legislation Process | (Year of publish) | 0 | No | ↓ |
| D12.1. International AI Governance Participation | (participation rate) | 57 | 4/7 | ↓ |
| D12.2. ISO AI Standardization Participation | (Level) | 50 | Midium | |

## P4 Governance Effectiveness

|  |  |  |  |  |
|---|---|---|---|---|
| D13.1. Public AI Skill Proficiency | (PISA score) | 31 | N/A | |
| D13.2. Public Awareness of AI Impact | (scenario %) | 70 | 62 | ↑ |
| D14.1. Positive Public Attitude towards AI | (positive %) | 54 | 60 | |
| D14.2. Enterprise Attitude towards AI | (adoption %) | N/A | N/A | |
| D15.1. Gender Ratio in AI literature | (male/female) | 50 | 3.24 | |
| D15.2. Gender Ratio in AI Graduates | (female %) | 77 | 38 | ↑ |
| D15.3. AI for Disadvantaged Groups | (internet %) | N/A | N/A | |
| D16.1. Open AI Models & Datasets | (published numbers) | 0 | 0 | |
| D16.2. AI Developer Community | (Github numbers) | 23 | 10 | ↓ |
| D17.1. AI Governance Literature Volume | (Number) | 17 | 362 | ↓ |
| D18.1. AI & SDG Literature Volume | (Number) | 54 | 331 | |
| D18.2. AI for SDGs Cases | (Number) | 32 | 39 | |

# Singapore

| AGILE index ranking | population | GDP per capita | Country Group |
|---|---|---|---|
| 3/14 | 6million | 77,600$ | high-income |

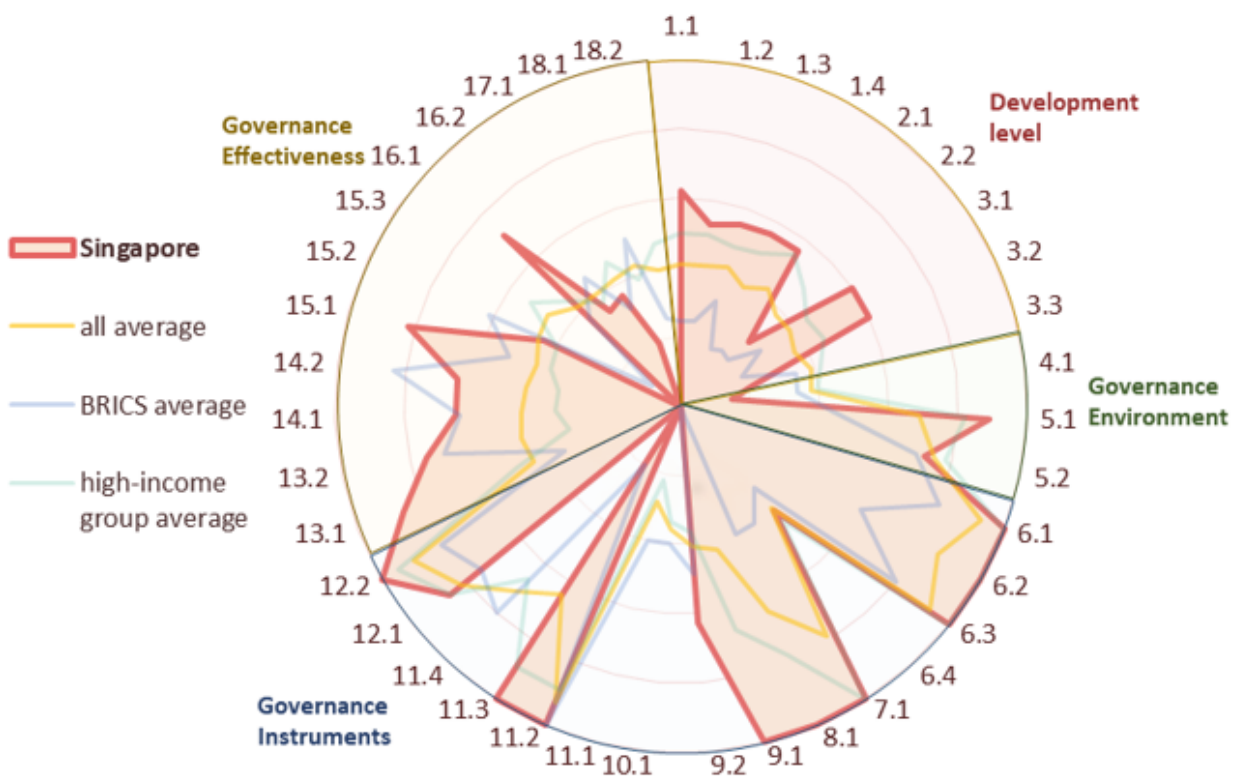

## P1 Development Level

|  |  |  |  |  |
|---|---|---|---|---|
| D1.1. AI Publications | 62 | 50307 | (Number) | ↑ |
| D1.2.AI Professionals | 53 | 28721 | (Number) | |
| D1.3. AI Patents | 55 | 266 | (Number) | |
| D1.4. AI Systems | 56 | 7 | (Number) | |
| D2.1. Colocation Data Centers | 56 | 45 | (Number) | |
| D2.2. Supercomputer FLOP/s | 27 | 16 | (pFLOP/s) | ↓ |
| D3.1. AI Companies' Funding | 60 | 4.8 | (billion$) | ↑ |
| D3.2. AI Startups | 60 | 271 | (Number) | ↑ |
| D3.3. Listed AI Companies | 22 | 1 | (Number) | ↓ |

## P2 Governance Environment

|  |  |  |  |  |
|---|---|---|---|---|
| D4.1. AI Risk Incidents | 15 | 20 | (Number) | ↑ |
| D5.1. Overall Governance Level | 89 | 89 | (WGI&GTMI) | |
| D5.2. SGDs Progress | 72 | 72 | (SDGDI) | |

## P3 Governance Instruments

|  |  |  |  |  |
|---|---|---|---|---|
| D6.1. AI Strategy Release Status | 100 | 2023 | (Year of publish) | |
| D6.2. Measurable Goals in AI Strategy | 100 | Yes | | |
| D6.3. Training Inclusion in AI Strategy | 100 | Yes | | |
| D6.4. AI Budget | 40 | 0.74 | (billion$) | |
| D7.1. AI Governance Bodies Establishment | 100 | 2018 | (Year of publish) | |
| D8.1. AI Principles Issued by Government | 100 | 2020 | (Year of publish) | ↑ |
| D9.1. AI Impact Assessment Mechanism | 100 | 2022 | (Year of publish) | ↑ |
| D9.2. Regulatory Sandboxes for AI | 63 | 2 | (Year of publish) | |
| D10.1. AI Standards & Certification | 0 | No | (Year of publish) | ↓ |
| D11.1. National Laws pertaining AI | 0 | No | (Year of publish) | ↓ |
| D11.2. AI-Specific Data Protection Laws | 100 | 2012 | (Year of publish) | |
| D11.3. AI Consumer Protection Legislation | 100 | 2019 | (Year of publish) | ↑ |
| D11.4. Ongoing AI Legislation Process | 0 | No | (Year of publish) | ↓ |
| D12.1. International AI Governance Participation | 86 | 6/7 | (participation rate) | |
| D12.2. ISO AI Standardization Participation | 100 | High | (Level) | |

## P4 Governance Effectiveness

|  |  |  |  |  |
|---|---|---|---|---|
| D13.1. Public AI Skill Proficiency | 86 | 569 | (PISA score) | ↑ |
| D13.2. Public Awareness of AI Impact | 75 | 71 | (scenario %) | ↑ |
| D14.1. Positive Public Attitude towards AI | 65 | N/A | (positive %) | ↑ |
| D14.2. Enterprise Attitude towards AI | 65 | 39 | (adoption %) | ↑ |
| D15.1. Gender Ratio in AI literature | 82 | 1.95 | (male/female) | ↑ |
| D15.2. Gender Ratio in AI Graduates | 44 | 32 | (female %) | |
| D15.3. AI for Disadvantaged Groups | N/A | N/A | (internet %) | |
| D16.1. Open AI Models & Datasets | 71 | 24 | (published numbers) | ↑ |
| D16.2. AI Developer Community | 34 | 185 | (Github numbers) | |
| D17.1. AI Governance Literature Volume | 36 | 1965 | (Number) | |
| D18.1. AI & SDG Literature Volume | 19 | 836 | (Number) | ↓ |
| D18.2. AI for SDGs Cases | 0 | 0 | (Number) | ↓ |

# United Arab Emirates

| AGILE index ranking | population | GDP per capita | Country Group |
|---|---|---|---|
| 9/14 | 9.5million | 53,300$ | high-income |

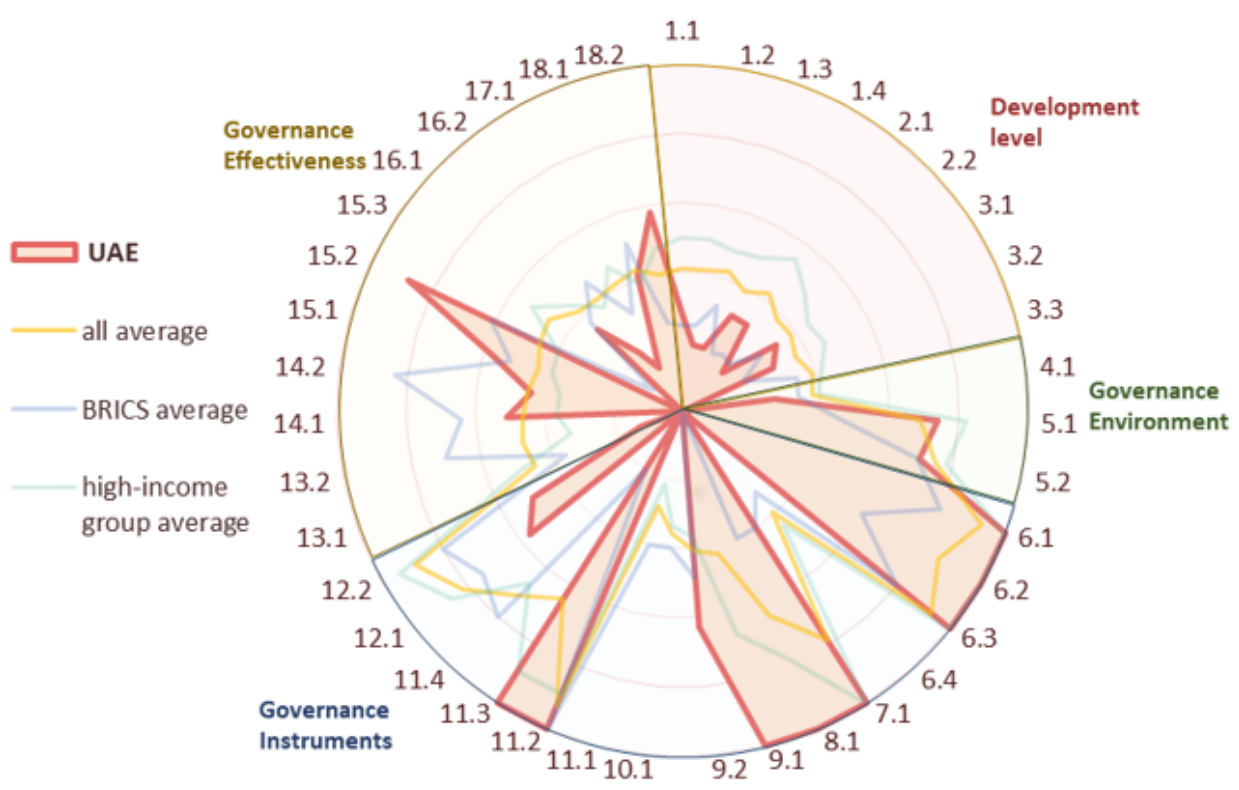

## P1 Development Level

|  |  |  |  |
|---|---|---|---|
| D1.1. AI Publications | 28 | 11401 | (Number) |
| D1.2.AI Professionals | 19 | 5764 | (Number) |
| D1.3. AI Patents | 19 | 41 | (Number) |
| D1.4. AI Systems | 31 | 3 | (Number) |
| D2.1. Colocation Data Centers | 31 | 19 | (Number) |
| D2.2. Supercomputer FLOP/s | 16 | 9 | (pFLOP/s) |
| D3.1. AI Companies' Funding | 33 | 0.9 | (billion$) |
| D3.2. AI Startups | 29 | 64 | (Number) |
| D3.3. Listed AI Companies | 0 | 0 | (Number) |

## P2 Governance Environment

|  |  |  |  |
|---|---|---|---|
| D4.1. AI Risk Incidents | 27 | 14 | (Number) |
| D5.1. Overall Governance Level | 74 | 74 | (WGI&GTMI) |
| D5.2. SGDs Progress | 70 | 70 | (SDGDI) |

## P3 Governance Instruments

|  |  |  |  |
|---|---|---|---|
| D6.1. AI Strategy Release Status | 100 | 2017 | (Year of publish) |
| D6.2. Measurable Goals in AI Strategy | 100 | Yes | |
| D6.3. Training Inclusion in AI Strategy | 100 | Yes | |
| D6.4. AI Budget | 0 | No | (billion$) |
| D7.1. AI Governance Bodies Establishment | 100 | 2018 | (Year of publish) |
| D8.1. AI Principles Issued by Government | 100 | 2019 | (Year of publish) |
| D9.1. AI Impact Assessment Mechanism | 100 | 2019 | (Year of publish) |
| D9.2. Regulatory Sandboxes for AI | 63 | 2 | (Year of publish) |
| D10.1. AI Standards & Certification | 0 | No | (Year of publish) |
| D11.1. National Laws pertaining AI | 0 | No | (Year of publish) |
| D11.2. AI-Specific Data Protection Laws | 100 | 2021 | (Year of publish) |
| D11.3. AI Consumer Protection Legislation | 100 | 2020 | (Year of publish) |
| D11.4. Ongoing AI Legislation Process | 0 | No | (Year of publish) |
| D12.1. International AI Governance Participation | 57 | 4/7 | (participation rate) |
| D12.2. ISO AI Standardization Participation | 50 | Midium | (Level) |

## P4 Governance Effectiveness

|  |  |  |  |
|---|---|---|---|
| D13.1. Public AI Skill Proficiency | 13 | 435 | (PISA score) |
| D13.2. Public Awareness of AI Impact | N/A | N/A | (scenario %) |
| D14.1. Positive Public Attitude towards AI | 51 | N/A | (positive %) |
| D14.2. Enterprise Attitude towards AI | 44 | 38 | (adoption %) |
| D15.1. Gender Ratio in AI literature | 57 | 2.97 | (male/female) |
| D15.2. Gender Ratio in AI Graduates | 88 | 55 | (female %) |
| D15.3. AI for Disadvantaged Groups | N/A | N/A | (internet %) |
| D16.1. Open AI Models & Datasets | 34 | 2 | (published numbers) |
| D16.2. AI Developer Community | 21 | 17 | (Github numbers) |
| D17.1. AI Governance Literature Volume | 14 | 377 | (Number) |
| D18.1. AI & SDG Literature Volume | 41 | 288 | (Number) |
| D18.2. AI for SDGs Cases | 58 | 24 | (Number) |

# United Kingdoms

| AGILE index ranking | population | GDP per capita | Country Group |
|---|---|---|---|
| 6/14 | 68million | 45,300$ | high-income |

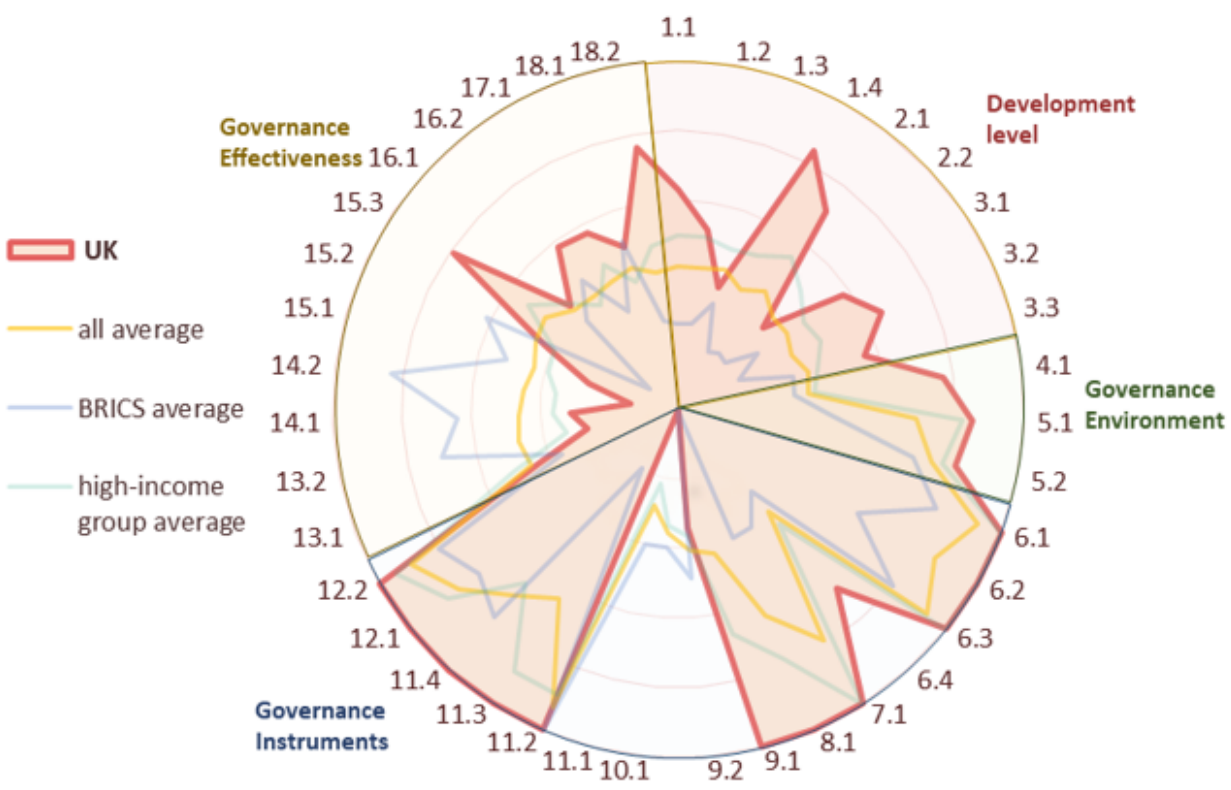

## P1 Development Level

|  |  |  |  |  |
|---|---|---|---|---|
| D1.1. AI Publications | (Number) | 63 | 308276 | ∧ |
| D1.2.AI Professionals | (Number) | 52 | 173647 | |
| D1.3. AI Patents | (Number) | 37 | 1129 | |
| D1.4. AI Systems | (Number) | 84 | 57 | ∧ |
| D2.1. Colocation Data Centers | (Number) | 71 | 286 | ∧ |
| D2.2. Supercomputer FLOP/s | (pFLOP/s) | 34 | 142 | ∨ |
| D3.1. AI Companies' Funding | (billion$) | 58 | 17 | |
| D3.2. AI Startups | (Number) | 65 | 1052 | ∧ |
| D3.3. Listed AI Companies | (Number) | 56 | 23 | |

## P2 Governance Environment

|  |  |  |  |  |
|---|---|---|---|---|
| D4.1. AI Risk Incidents | (Number) | 77 | 445 | ∨ |
| D5.1. Overall Governance Level | (WGI&GTMI) | 85 | 85 | |
| D5.2. SGDs Progress | (SDGDI) | 82 | 82 | |

## P3 Governance Instruments

|  |  |  |  |  |
|---|---|---|---|---|
| D6.1. AI Strategy Release Status | (Year of publish) | 100 | 2021 | |
| D6.2. Measurable Goals in AI Strategy | | 100 | Yes | |
| D6.3. Training Inclusion in AI Strategy | | 100 | Yes | |
| D6.4. AI Budget | (billion$) | 69 | 5.80 | |
| D7.1. AI Governance Bodies Establishment | (Year of publish) | 100 | 2021 | |
| D8.1. AI Principles Issued by Government | (Year of publish) | 100 | 2018 | ∧ |
| D9.1. AI Impact Assessment Mechanism | (Year of publish) | 100 | 2023 | ∧ |
| D9.2. Regulatory Sandboxes for AI | (Year of publish) | 34 | 1 | |
| D10.1. AI Standards & Certification | (Year of publish) | 0 | No | ∨ |
| D11.1. National Laws pertaining AI | (Year of publish) | 0 | No | ∨ |
| D11.2. AI-Specific Data Protection Laws | (Year of publish) | 100 | 2018 | |
| D11.3. AI Consumer Protection Legislation | (Year of publish) | 100 | 2015 | ∧ |
| D11.4. Ongoing AI Legislation Process | (Year of publish) | 100 | 2023 | |
| D12.1. International AI Governance Participation | (participation rate) | 100 | 7/7 | |
| D12.2. ISO AI Standardization Participation | (Level) | 100 | High | |

## P4 Governance Effectiveness

|  |  |  |  |  |
|---|---|---|---|---|
| D13.1. Public AI Skill Proficiency | (PISA score) | 43 | 502 | |
| D13.2. Public Awareness of AI Impact | (scenario %) | 27 | 53 | ∨ |
| D14.1. Positive Public Attitude towards AI | (positive %) | 31 | 50 | |
| D14.2. Enterprise Attitude towards AI | (adoption %) | 14 | 26 | ∨ |
| D15.1. Gender Ratio in AI literature | (male/female) | 27 | 4.4 | ∨ |
| D15.2. Gender Ratio in AI Graduates | (female %) | 43 | 19 | |
| D15.3. AI for Disadvantaged Groups | (internet %) | 79 | 92 | ∧ |
| D16.1. Open AI Models & Datasets | (published numbers) | 43 | 22 | |
| D16.2. AI Developer Community | (Github numbers) | 58 | 961 | |
| D17.1. AI Governance Literature Volume | (Number) | 57 | 10823 | |
| D18.1. AI & SDG Literature Volume | (Number) | 49 | 6009 | |
| D18.2. AI for SDGs Cases | (Number) | 76 | 168 | ∧ |

# United States

| AGILE index ranking | population | GDP per capita | Country Group |
|---|---|---|---|
| 1/14 | 339million | 73,900$ | high-income |

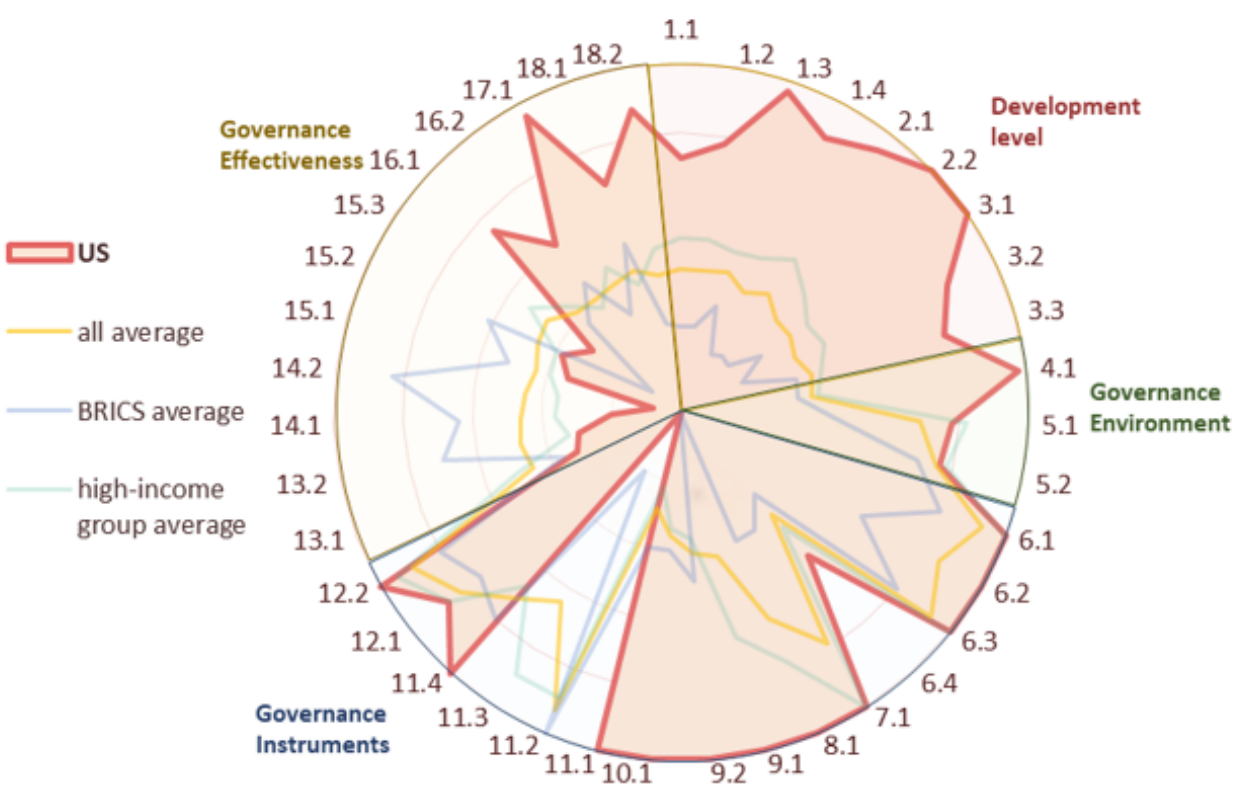

## P1 Development Level

|  |  |  |  |  |
|---|---|---|---|---|
| D1.1. AI Publications | (Number) | 73 | 1352481 | ↑ |
| D1.2.AI Professionals | (Number) | 78 | 1039461 | ↑ |
| D1.3. AI Patents | (Number) | 97 | 45913 | ↑ |
| D1.4. AI Systems | (Number) | 89 | 377 | ↑ |
| D2.1. Colocation Data Centers | (Number) | 94 | 2105 | ↑ |
| D2.2. Supercomputer FLOP/s | (pFLOP/s) | 100 | 5711 | ↑ |
| D3.1. AI Companies' Funding | (billion$) | 100 | 288 | ↑ |
| D3.2. AI Startups | (Number) | 85 | 5938 | ↑ |
| D3.3. Listed AI Companies | (Number) | 79 | 172 | ↑ |

## P2 Governance Environment

|  |  |  |  |  |
|---|---|---|---|---|
| D4.1. AI Risk Incidents | (Number) | 98 | 4615 | ↓ |
| D5.1. Overall Governance Level | (WGI&GTMI) | 78 | 78 | |
| D5.2. SGDs Progress | (SDGDI) | 76 | 76 | |

## P3 Governance Instruments

|  |  |  |  |  |
|---|---|---|---|---|
| D6.1. AI Strategy Release Status | (Year of publish) | 100 | 2023 | |
| D6.2. Measurable Goals in AI Strategy | | 100 | Yes | |
| D6.3. Training Inclusion in AI Strategy | | 100 | Yes | |
| D6.4. AI Budget | (billion$) | 55 | 12.10 | |
| D7.1. AI Governance Bodies Establishment | (Year of publish) | 100 | 2018 | |
| D8.1. AI Principles Issued by Government | (Year of publish) | 100 | 2019 | ↑ |
| D9.1. AI Impact Assessment Mechanism | (Year of publish) | 100 | 2023 | ↑ |
| D9.2. Regulatory Sandboxes for AI | (Year of publish) | 100 | 5 | ↑ |
| D10.1. AI Standards & Certification | (Year of publish) | 100 | 2022 | ↑ |
| D11.1. National Laws pertaining AI | (Year of publish) | 100 | 2023 | ↑ |
| D11.2. AI-Specific Data Protection Laws | (Year of publish) | 0 | No | ↓ |
| D11.3. AI Consumer Protection Legislation | (Year of publish) | 0 | No | ↓ |
| D11.4. Ongoing AI Legislation Process | (Year of publish) | 100 | 2022 | |
| D12.1. International AI Governance Participation | (participation rate) | 86 | 6/7 | |
| D12.2. ISO AI Standardization Participation | (Level) | 100 | High | |

## P4 Governance Effectiveness

|  |  |  |  |  |
|---|---|---|---|---|
| D13.1. Public AI Skill Proficiency | (PISA score) | 32 | 478 | |
| D13.2. Public Awareness of AI Impact | (scenario %) | 30 | 49 | ↓ |
| D14.1. Positive Public Attitude towards AI | (positive %) | 20 | 49 | ↓ |
| D14.2. Enterprise Attitude towards AI | (adoption %) | 8 | 25 | ↓ |
| D15.1. Gender Ratio in AI literature | (male/female) | 34 | 3.77 | |
| D15.2. Gender Ratio in AI Graduates | (female %) | 38 | 24 | |
| D15.3. AI for Disadvantaged Groups | (internet %) | 31 | 80 | |
| D16.1. Open AI Models & Datasets | (published numbers) | 75 | 307 | ↑ |
| D16.2. AI Developer Community | (Github numbers) | 60 | 3600 | |
| D17.1. AI Governance Literature Volume | (Number) | 96 | 53957 | ↑ |
| D18.1. AI & SDG Literature Volume | (Number) | 69 | 27685 | ↑ |
| D18.2. AI for SDGs Cases | (Number) | 88 | 681 | ↑ |

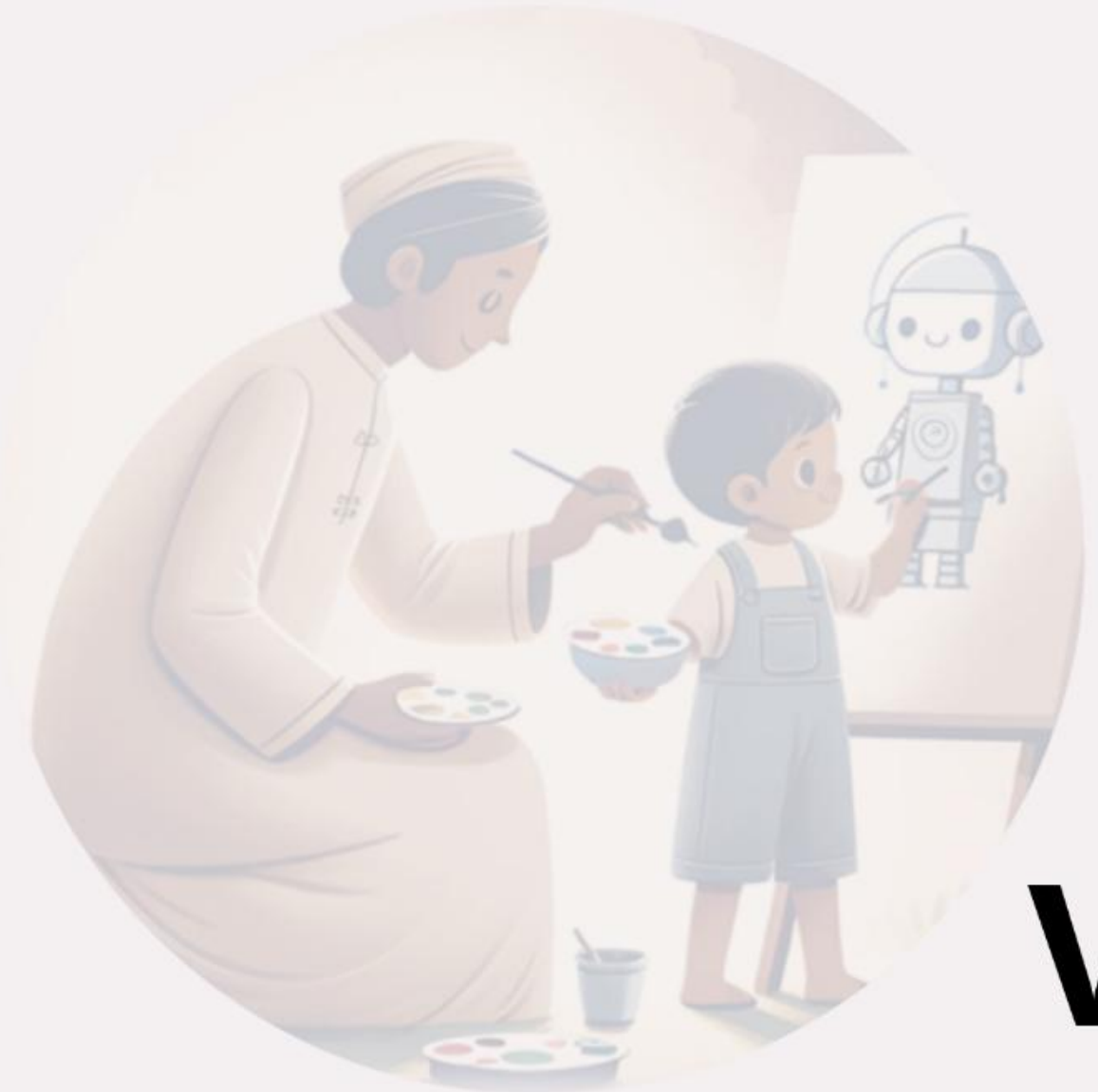

# V. Appendix

# 1. Dimension Details and Data Sources

*Table 5 AGILE Index Dimensions and Indicators (in detail)*

| Pillars | Dimensions | Content of Evaluation | Referencing Articles from UNESCO Rec.[3] | Indicators |
|---|---|---|---|---|
| **P1. Development Level** | D1. AI Research and Development Activity | Assessment of countries' level of activity in AI-related R&D | A83 | D1.1. Number of **publications in AI**-related journals/conferences & the per capita ratio |
| | | | | D1.2. Number of **professionals** in the field of AI & the per capita ratio |
| | | | | D1.3. Number of granted **AI patents** & the per capita ratio |
| | | | | D1.4. Number of **AI systems** developed & the GDP ratio |
| | D2. AI Infrastructure | Assessment of the level of deployment and access to AI technologies and digital ecosystem infrastructure in each country | A59, A80 | D2.1. Number of colocation **data centers** & the per capita ratio |
| | | | | D2.2. Non-distributed **supercomputers** floating point operations per second & the per capita ratio |
| | D3. AI Industry Scale | Assessment of the level of activity in AI-related industries in each country | A117 | D3.1. AI companies' total **funding** & the GDP ratio |
| | | | | D3.2. Number of **AI startups** & the GDP ratio |
| | | | | D3.3. Number of **AI companies listed** on stock exchanges & the GDP ratio |
| **P2. Governance Environment** | D4. AI Risk Exposure | Assessment of the level of exposure to AI-related ethical and safety risks in each country | A50 | D4.1. Number of **AI-related risk cases/incidents** & the GDP ratio |
| | D5. AI Governance Readiness | Assessment of countries' preparedness and implementation capabilities in AI governance | A54 | D5.1. Overall assessment of the **level of governance** in the country |
| | | | | D5.2. The overall process of achieving **sustainable development goals** in the country |
| **P3. Governance Instruments** | D6. AI Strategy & Planning | Assessment of the development of AI strategy/planning/roadmap in each country | A56, A71 | D6.1. Whether an **AI strategy** has been released in the country |
| | | | | D6.2. Whether the AI strategy has **measurable goals** |
| | | | | D6.3. Whether the AI strategy mentions **training or skills upgrading** |
| | | | | D6.4. **Budget scale** & the GDP ratio for AI-specific expenditure |
| | D7. AI Governance | Assessment of the establishment of | A58 | D7.1. Whether **AI governance bodies** have been established in the country |

[3] The references in this column indicate the supporting articles from the 'IV. Areas of policy action' section of UNESCO's *Recommendation on the Ethics of Artificial Intelligence* that most closely correspond to the evaluation content of the specific AGILE Index Dimension.

| | | | | |
|---|---|---|---|---|
| | Bodies | AI governance institutions or bodies in each country | | |
| | D8. AI Principles & Norms | Assessment of the development of AI governance principles and norms in each country | A48 | D8.1. Whether governments have issued **AI principles or norms** |
| | D9. AI Impact Assessment | Assessment of the development of AI impact assessment tools/frameworks in each country | A50 | D9.1. Whether governments have introduced **AI impact assessment mechanisms** |
| | | | | D9.2. Number of **regulatory sandboxes** for safety test of (financial) AI |
| | D10. AI Standards & Certification | Assessment of the establishment of AI standards/certification mechanisms in each country | A64 | D10.1. Whether governments have developed **standards and certification mechanisms** for AI |
| | D11. AI Legislation Status | Assessment of the enactment status of AI laws and related regulations in each country | A133 | D11.1. Whether countries have enacted a **national-level law** regarding AI |
| | | | | D11.2. Whether countries have implemented **data protection legislation** specifically addressing AI |
| | | | | D11.3. Whether countries have enacted **consumer protection legislation** specifically tailored to AI |
| | | | | D11.4. Whether countries are working on **AI legal instruments** which are at a later stage of enactment |
| | D12. Global AI Governance Engagement | Assessment of the degree of countries' participation in international AI governance | A80 | D12.1. The participation level in **international AI governance mechanisms** |
| | | | | D12.2. The participation level in **ISO AI standardization** |
| **P4. Governance Effectiveness** | D13. Public Understanding of AI | Assessment of the public's AI competence and AI risk awareness in each country | A101 | D13.1. The **AI-related skill proficiencies** of the public |
| | | | | D13.2. The level of the public's **awareness of AI's impact** |
| | D14. Public Trust in AI | Assessment of the degree of public trust in AI technologies and applications in each country | A39 | D14.1. The level of the **public's positive attitude** towards AI's development |
| | | | | D14.2. The level of **enterprises' positive attitudes** towards AI's adoption |
| | D15. AI Development Inclusivity | Assessment of the inclusiveness of AI R&D and applications in each country | A91, A105 | D15.1. Gender ratio of **AI literature authors** |
| | | | | D15.2. Gender ratio of **graduates in AI-related majors** |
| | | | | D15.3. The level of AI accessibility by **disadvantaged groups** |
| | D16. Data & Algorithm Openness | Assessment of the level of open source and openness of AI data | A75, A76 | D16.1. Number of impactful **open AI models and datasets** released |
| | | | | D16.2. The level of contributions in the **AI developer community** |
| | | and algorithms in each country | | |
| | D17. AI Governance Research Activity | Assessment of countries' activity of research in AI governance | A131 | D17.1. Total number & the proportion of **literature on AI governance** topics |
| | D18. AI for SDGs (AI4SDGs) Activity | Assessment of countries' level of research and application on AI for Sustainable Development Goals | A47 | D18.1. Total number & the proportion of **literature on AI and SDGs** topics |
| | | | | D18.2. Number of reported cases of **AI applications for SDGs** & the GDP ratio |

## a) P1. AI Development Level

### D1. AI Research and Development Activity

AI Research and Development Activity refers to the level of activity in artificial intelligence related research and development in various countries. According to UNESCO's *Recommendation on the Ethics of Artificial Intelligence*, Article 83, "Member States should encourage international cooperation and collaboration in the field of AI to bridge geo-technological lines." This recommendation aligns with Dimension 1, which involves assessing the level of AI development to facilitate comparative analysis of technological gaps among different countries and regions.

The Dimension 1 currently covers four indicators:

- **D1.1. Number of publications in AI-related journals/conferences & the per capita ratio**
  - Data Source: Based on statistical analysis of the DBLP Computer Science Bibliography literature database; the Tortoise Media's Global AI Index Indicators (Number of AI Articles, Number of Submissions to AI Conferences)
- **D1.2. Number of professionals in the field of AI & the per capita ratio**
  - Data Source: Based on statistical analysis of the DBLP Computer Science Bibliography literature database; the Tortoise Media's Global AI Index indicators (number of researchers).
- **D1.3. Number of granted AI patents & the per capita ratio**
  - Data Source: The Tortoise Media's Global AI Index indicators (numbers of granted ai patents by applicant, TGAII numbers of granted patents by inventors).
- **D1.4. Number of AI systems developed & the GDP ratio**
  - Data Source: The Tortoise Media's Global AI Index indicators (Number of Significant Machine Learning Systems). The number of important large language models in the LifeArchitect.AI database. The number of machine learning systems in the Epoch.AI database.

### D2. AI Infrastructure

AI Infrastructure refers to the foundational technology and digital ecosystem for artificial intelligence. According to UNESCO's *Recommendation on the Ethics of Artificial Intelligence*, Article 59, "Member States should foster the development of, and access to, a digital ecosystem for ethical and inclusive development of AI systems at the national level...Such an ecosystem includes, in particular, digital technologies and infrastructure..." and Article 80, "Member States should work through international organizations to provide platforms for international cooperation on AI for development, including...infrastructure, and facilitating multi-stakeholder collaboration...". These recommendations align with Dimension 2.

The Dimension 2 currently covers two indicators:

- **D2.1. Number of colocation data centers & the per capita ratio**
    - Data Source: The number of data centers in the Data Center Map database.
- **D2.2. Non-distributed supercomputers floating point operations per second & the per capita ratio**
    - Data Source: The TOP500 List of Supercomputer (Total cores, Rpeak, Rmax)

**D3. AI Industry Scale**

The AI Industry Scale refers to the activity of a country in artificial intelligence related industries. According to UNESCO's *Recommendation on the Ethics of Artificial Intelligence*, Article 117, "Member States should support collaboration agreements among governments, academic institutions, vocational education and training institutions, industry, workers' organizations and civil society to bridge the gap of skillset requirements to align training programmes and strategies with the implications of the future of work and the needs of industry, including small and medium enterprises," this recommendation is consistent with Dimension 3.

The D3 dimension currently covers three indicators:

- **D3.1. AI companies' total funding & the GDP ratio**
    - Data Source: The Tortoise Media's Global AI Index indicators (Total Funding of AI Companies).
- **D3.2. Number of AI startups & the GDP ratio**
    - Data Source: The Tortoise Media's Global AI Index indicators (Number of AI Startups).
- **D3.3. Number of AI companies listed on stock exchanges & the GDP ratio**
    - Data Source: The Tortoise Media's Global AI Index indicators (Number of AI Companies on Country's Stock Exchange).

## b) P2. AI Governance Environment

**D4. AI Risk Exposure**

AI Risk Exposure refers to the degree of exposure to ethical and safety risks and issues related to AI in various countries. The more issues there are, the higher the urgency for AI governance in that country. Therefore, this dimension has a negative impact on the background. According to UNESCO's *Recommendation on the Ethics of Artificial Intelligence*, Article 50, "Member States should introduce frameworks for impact assessments, such as ethical impact assessment, to identify and assess benefits, concerns and risks of AI systems, as well as appropriate risk prevention, mitigation and monitoring measures, among other assurance mechanisms," this recommendation is consistent with Dimension 4.

Dimension 4 currently has one indicator:

- **D4.1. Number of AI-related risk cases/incidents & the GDP ratio**
    - Data Source: The AI incidents data are from multiple sources including the AI Incident Database

(AIID)[4], the AI, Algorithmic, and Automation Incidents and Controversies Repository (AIAAIC)[5], the AI Governance Observatory from AI Governance Online (AIGO)[6], and the OECD AI Incidents Monitor (AIM)[7].

**D5. AI Governance Readiness**

AI Governance Readiness refers to the favourable conditions in a country for governing AI and utilizing AI to achieve the United Nations Sustainable Development Goals. According to UNESCO's *Recommendation on the Ethics of Artificial Intelligence*, Article 54, "Member States should ensure that AI governance mechanisms are inclusive, transparent, multidisciplinary, multilateral (this includes the possibility of mitigation and redress of harm across borders) and multi-stakeholder. In particular, governance should include aspects of anticipation, and effective protection, monitoring of impact, enforcement and redress", this recommendation aligns with Dimension 5.

Currently, there are two indicators under Dimension 5:

- **D5.1. Overall assessment of the level of governance in the country**
  - Data Source: World Bank (WGI, GTMI)
- **D5.2. The overall process of achieving sustainable development goals in the country**
  - Data Source: 2023 Sustainable Development Index.

## c) P3. AI Governance Instruments

**D6. AI Strategy & Planning**

AI Strategy & Planning refers to the overall plans formulated by governments of various countries for the development and application of artificial intelligence. In Article 56 of UNESCO's *Recommendation on the Ethics of Artificial Intelligence*, it is stated: "Member States are encouraged to develop national and regional AI strategies..."; and in Article 71: "Member States should work to develop data governance strategies...". This recommendation is aligned with the direction assessed in Dimension 7, which evaluates whether AI-related strategies have been established.

Dimension 6 currently covers four indicators:

- **D6.1. Whether an AI strategy has been released in the country**
  - Data Source: Survey with Local Expert Data Assistance.
- **D6.2. Whether the AI strategy has measurable goals**
  - Data Source: The Tortoise Media's Global AI Index indicators (Government has Measurable AI Targets).
- **D6.3. Whether the AI strategy mentions training or skills upgrading**
  - Data Source: The Tortoise Media's Global AI Index indicators (Dedicated Strategy mentions Training or upskilling).
- **D6.4. Budget scale & the GDP ratio for AI-specific expenditure**

---

[4] https://incidentdatabase.ai/

[5] https://www.aiaaic.org/aiaaic-repository

[6] https://www.ai-governance.online/ai-governance-observatory

[7] https://oecd.ai/en/incidents

- Data Source: The Tortoise Media's Global AI Index indicators (Dedicated Spending on Artificial Intelligence).

**D7. AI Governance Bodies**

AI Governance Bodies refer to specialized agencies established by governments of various countries to oversee AI governance affairs. According to UNESCO's *Recommendation on the Ethics of Artificial Intelligence*, Article 58, countries should "......consider adding the role of an independent AI Ethics Officer or some other mechanism to oversee ethical impact assessment, auditing and continuous monitoring efforts and ensure ethical guidance of AI systems". This recommendation aligns with the direction assessed in Dimension 7, which evaluates whether specialized agencies responsible for AI governance have been established.

Dimension 7 currently covers one indicator:

- **D7.1. Whether AI governance bodies have been established in the country**
  - Data Source: Survey with Local Expert Data Assistance.

**D8. AI Principles & Norms**

AI Principles & Norms refer to the principles and norms established by governments of various countries to guide the development, application, and governance of artificial intelligence. According to UNESCO's *Recommendation on the Ethics of Artificial Intelligence*, it is underlined that "......ensure that national AI strategies are guided by ethical principles", and in Article 48, "The main action is for Member States to put in place effective measures, including, for example, policy frameworks or mechanisms." This recommendation aligns with Dimension 8, which evaluates whether principles and norms for guiding AI have been established.

Dimension 8 currently covers one indicator:

- **D8.1. Whether governments have issued AI principles or norms**
  - Data Source: Survey with Local Expert Data Assistance.

**D9. AI Impact Assessment**

AI Impact Assessment refers to the evaluation of the potential impacts of artificial intelligence systems, including their effects on individuals, society, and the environment. According to UNESCO's *Recommendation on the Ethics of Artificial Intelligence*, Article 50, countries should "introduce frameworks for impact assessments, such as ethical impact assessment, to identify and assess benefits, concerns and risks of AI systems." This recommendation aligns with Dimension 9, which assesses whether tools/frameworks for assessing the impact of artificial intelligence have been developed.

Dimension 9 currently covers two indicators:

- **D9.1. Whether governments have introduced AI impact assessment mechanisms**
  - Data Source: Survey with Local Expert Data Assistance.
- **D9.2. Number of regulatory sandboxes for safety test of (financial) AI**
  - Data Source: World Bank (Global Experiences from Regulatory Sandboxes)

**D10. AI Standards & Certification**

AI Standards & Certification refer to mechanisms for assessing artificial intelligence systems to ensure compliance with relevant ethical and safety standards and to issue certification marks for compliance. According to UNESCO's *Recommendation on the Ethics of Artificial Intelligence*, Article 64, "Member States, international organizations and other relevant bodies should develop international standards that describe measurable, testable levels of safety and transparency, so that systems can be objectively assessed, and levels of compliance determined." This recommendation aligns with Dimension 10, which assesses whether mechanisms for assessing AI systems against standards have been developed.

Dimension 10 currently covers one indicator:

- **D10.1. Whether governments have developed standards and certification mechanisms for AI**
  - Data Source: Survey with Local Expert Data Assistance.

**D11. AI Legislation Status**

Different countries operate under varying legal systems, such as common law, civil law, and systems that are either inquisitorial or adversarial in nature. Furthermore, in some countries, an executive order or a Supreme Court judgment can carry the same legal weight as a law passed by the legislative assembly. Moreover, our categorization acknowledges the complex journey of legal document enactment, which often involves multiple stages and varied legal interpretations. This diversity in legal structures influences how AI is governed. Therefore, our categorization takes into account these diverse legal mechanisms, recognizing that any legal instrument can have significant implications for AI governance at a national level.

At present, AGILE Index focuses on four key legal areas: national-level artificial intelligence related laws, data protection laws containing AI provisions, consumer rights laws containing AI provisions, and advanced-stage AI legal instruments which are in the final stages of enactment. UNESCO's *Recommendation on the Ethics of Artificial Intelligence* "Recommends that Member States apply on a voluntary basis the provisions of this Recommendation by taking appropriate steps, including whatever legislative or other measures...," and its Article 133 states, "Data collection and processing should be conducted in accordance with international law, national legislation on data protection and data privacy, and the values and principles outlined in this Recommendation." This recommendation aligns with Dimension 11, which assesses the legal framework related to AI.

Dimension 11 currently covers four indicators:

- **D11.1. Whether countries have enacted a national-level law regarding AI**
  - Data Source: Survey with Local Expert Data Assistance.
- **D11.2. Whether countries have implemented data protection legislation specifically addressing AI**
  - Data Source: Survey with Local Expert Data Assistance.
- **D11.3. Whether countries have enacted consumer protection legislation specifically tailored to AI**
  - Data Source: Survey with Local Expert Data Assistance.
- **D11.4. Whether countries are working on AI legal instruments which are at a later stage of enactment**
  - Data Source: Survey with Local Expert Data Assistance.

**D12. Global AI Governance Engagement**

Global AI Governance Engagement refers to the participation of countries in international AI governance affairs through international mechanisms. According to UNESCO's *Recommendation on the Ethics of Artificial Intelligence*, Article 80, countries should "work through international organizations to provide platforms for international cooperation on AI for development." This recommendation aligns with Dimension 12, which assesses the degree of international participation of countries in the field of AI governance.

Dimension 12 currently covers two indicators:

- **D12.1. The participation level in international AI governance mechanisms**
  - Data Source: Survey with Local Expert Data Assistance.
- **D12.2. The participation level in ISO AI standardization**
  - Data Source: The Tortoise Media's Global AI Index indicators (ISO participation level).

## d) P4. AI Governance Effectiveness

**D13. Public Understanding of AI**

According to UNESCO's *Recommendation on the Ethics of Artificial Intelligence*, Article 101, member states should "work with international organizations, educational institutions and private and non-governmental entities to provide adequate AI literacy education to the public on all levels in all countries in order to empower people and reduce the digital divides and digital access inequalities resulting from the wide adoption of AI systems." This recommendation aligns with Dimension 13, which assesses whether efforts contribute to promoting public awareness of AI.

Dimension 13 currently covers two indicators:

- **D13.1. The AI-related skill proficiencies of the public**
  - Data Source: OECD PISA math scores; Coursera skill report (data science proficiency, number of Coursera learners).
- **D13.2. The level of the public's awareness of AI's impact**
  - Data Source: IPSOS OEAI report (number of life areas most expected to change because of AI); KPMG TAI report (use of AI technologies and awareness of their use; subjective knowledge of AI).

**D14. Public Trust in AI**

According to UNESCO's *Recommendation on the Ethics of Artificial Intelligence*, Article 39, "......allows for public scrutiny that can decrease corruption and discrimination, and can also help detect and prevent negative impacts on human rights. Transparency aims at providing appropriate information to the respective addressees to enable their understanding and foster trust." This value aligns with Dimension 13, which assesses whether there are surveys conducted regarding public attitudes towards AI.

Dimension 14 currently covers two indicators:

- **D14.1. The level of the public's positive attitude towards AI's development**
  - Data Source: Tortoise Media's Global AI Index indicators (Proportion of Population that Trusts AI, Proportion of Population who think AI is more helpful than harmful; IPSOS OEAI indicators (overall expectations for life improvement due to artificial intelligence, perception that the benefits of products and services using artificial intelligence outweigh the drawbacks).
- **D14.2. The level of enterprises' positive attitudes towards AI's adoption**

- Data Source: IBM AI Adoption Index (number of representative companies deploying AI, number of representative companies exploring AI).

**D15. AI Development Inclusivity**

According to UNESCO's *Recommendation on the Ethics of Artificial Intelligence*, Article 91 states, "Member States should encourage female entrepreneurship, participation and engagement in all stages of an AI system life cycle," and Article 105 states, "Member States should promote the participation and leadership of girls and women, diverse ethnicities and cultures, persons with disabilities, marginalized and vulnerable people or people in vulnerable situations, minorities and all persons not enjoying the full benefits of digital inclusion." These recommendations align with Dimension 15, which evaluates whether the development of AI is inclusive of different groups and genders.

Dimension 15 currently covers three indicators:

- **D15.1. Gender ratio of AI literature authors**
  - Data Source: Based on statistical analysis of the DBLP Computer Science Bibliography literature database.
- **D15.2. Gender ratio of graduates in AI-related majors**
  - Data Source: The Tortoise Media's Global AI Index indicators (science graduate gender diversity, IT graduate gender diversity).
- **D15.3. The level of AI accessibility by disadvantaged groups**
  - Data Source: The OECD GDT indicators (internet user aged 55-74 years, low-income internet user).

**D16. Data & Algorithm Openness**

According to UNESCO's *Recommendation on the Ethics of Artificial Intelligence*, Article 75 states, "Member States should promote open data," and Article 76 states, "Member States should promote and facilitate the use of quality and robust datasets for training, development and use of AI systems, and exercise vigilance in overseeing their collection and use." This recommendation aligns with Dimension 16, which evaluates whether data, algorithms, and models are open to the public.

Dimension 16 currently covers two indicators:

- **D16.1. Number of impactful open AI models and datasets released**
  - Data Source: Based on statistics from the Hugging Face community.
- **D16.2. The level of contributions in the AI developer community**
  - Data Source: Tortoise Media's Global AI Index indicators (Stack Overflow Answers to AI related Questions, total GitHub commits, total GitHub Commits on High-Popularity Open-Source AI Packages).

**D17. AI Governance Research Activity**

According to UNESCO's *Recommendation on the Ethics of Artificial Intelligence*, Article 131 states, "Member States should, according to their specific conditions, governing structures and constitutional provisions, credibly and transparently monitor and evaluate policies, programmes and mechanisms related to ethics of AI, using a combination of quantitative and qualitative approaches……(d) strengthening the research- and evidence-based analysis of and reporting on policies regarding AI ethics; (e) collecting and disseminating progress, innovations, research reports, scientific publications, data and statistics regarding policies for AI ethics……" This recommendation aligns with Dimension 17, which evaluates the quantitative analysis of relevant research on AI governance topics.

Dimension 17 currently covers one indicator:

- **D17.1. Total number & the proportion of literature on AI governance topics**
  - Data Source: Based on statistical analysis of the articles retrieved on Springer, IEEE Xplore, and ACM Digital Library; Based on statistical analysis of the DBLP Computer Science Bibliography literature database.

**D18. AI for SDGs (AI4SDGs) Activity**

According to UNESCO's *Recommendation on the Ethics of Artificial Intelligence*, Article 47 states, "Participation of different stakeholders throughout the AI system life cycle is necessary for inclusive approaches to AI governance, enabling the benefits to be shared by all, and to contribute to sustainable development." This recommendation aligns with Dimension 18, which evaluates the relevance of research outcomes related to AI to sustainable development goals.

Dimension 18 currently covers two indicators:

- **D18.1. Total number & the proportion of literature on AI and SDGs topics**
  - Data Source: Based on statistical analysis of the DBLP Computer Science Bibliography literature database.
- **D18.2. Number of reported cases of AI applications for SDGs & the GDP ratio**
  - Data Source: Based on data from the AI for SDGs Observatory of the AI for Sustainable Development Goals (AI4SDGs) Think Tank platform[8].

[8] https://ai-for-sdgs.academy/casebase

# 2. Scoring Methodology

## a) Scoring Methodology for AI Legislation

If a country does not have a dedicated national-level AI law, it scores 0 in the national-level AI law indicator. This indicates the lack of a formal AI legislative framework at the national level. Laws from different provinces or regions within a country and regional legislations are excluded from this evaluation. A country with a comprehensive and detailed national-level AI law scores 100. The law should cover all aspects of AI, including generative AI, algorithms, ethical considerations, national digital infrastructure, and social impacts of AI.

Countries without specific data protection provisions for AI score 0 in the data protection law indicator; those with partial provisions score 50; and those with comprehensive data protection legislation for AI score 100. The scoring method for consumer rights laws containing AI clauses is similar, based on whether the country's consumer rights law specifically targets AI and its extent of coverage.

If a country has enacted partial legislation involving AI, it scores 100 in the partially enacted AI legislation indicator. This might mean the law covers some aspects of AI but is not comprehensive in scope or depth. This also includes documents that are about to become law or are in the later stages of the legislative process.

*Table 6* ***Five national legislations pertaining AI***

| Country | Name | Year of enactment | Explanation |
|---|---|---|---|
| US | Executive Order on the Safe, Secure, and Trustworthy Development and Use of Artificial Intelligence | 2023 | The US lacks a unified federal AI law, with individual states taking the initiatives. However, in 2023, President Biden issued an Executive Order on Safe, Secure, and Trustworthy Artificial Intelligence. |
| China | Interim Measures for the Management of Generative Artificial Intelligence Services | 2023 | These measures showcase China's proactive approach to ensure responsible AI development, deployment, and use. These, alongside the existing Personal Information Protection Law, demonstrate China's unwavering commitment to ethical AI practices. |
| France | The Law for a Digital Republic | 2016 | The law touches upon areas such as open government data, digital data management, and transparency requirements for public algorithms that may influence AI-related activities. |
| India | Information Technology Act 2000 | 2000 | It provides a legal framework to regulate and manage online activities, data privacy, and electronic communications and also being applied to AI activities. |
| Canada | Canadian Artificial Intelligence and Data Act (AIDA) | Draft Legislation | Introduced as part of the Digital Charter Implementation Act, 2022, is expected to set the foundation for the responsible design, development and deployment of AI systems that impact the lives of Canadians |

## b) Literature Analysis

When analysing nationality and gender information in various databases, we used multiple methods. First, we judged the nationality of the authors. If the author provided an address in the paper, we used this information to determine nationality. Otherwise, we inferred it through the author's collaboration network. We used the *global_gender_predictor* package to determine the gender of authors, based on the *World Gender Name Dictionary Second Edition*. When necessary, titles, abstracts, authors, publication dates, author addresses, article categories, and links information are collected.

To determine if a scientific literature is related to AI, we combined information on publishers and keywords. First, we identified literature published in AI journals or conferences as AI-related. Names and abbreviations of AI-related journals or conferences were extracted from the AMiner literature database's AI journal rankings. Literature published by these publishers was identified as AI-related. Additionally, we compiled a list of keywords for various AI sub-fields (e.g., machine learning, neural networks, reinforcement learning, Bayesian, Markov learning, etc.). If these keywords appeared in the title, the literature was identified as AI-related. To determine if a scientific literature was related to AI governance, we look for keywords like 'governance', 'policy', or 'ethics' in scientific literatures. If these keywords existed in the abstract or title, the literature was identified as governance related. To further confirm whether the literature is related to artificial intelligence governance, we use large language models to generate vector representations of the literature and perform clustering to eliminate irrelevant literature.

## c) Score Calculation at Each Level

In processing raw scores, we incorporated data entries and statistics from multiple sources for triangulation, enhancing reliability. This is especially useful when small fluctuations in scarce data can significantly impact scores; multiple data sources can reduce bias. Strong correlations between different data elements allow for mutual supplementation in cases of missing data. Where appropriate, ratio scores were considered to ensure fair comparisons between countries with different baseline statistics (such as population and GDP). Finally, percentile-fit normalization (see below) was used to standardize and average various data. In identifying genders, we combined average level inference, allocating 22.9% of unidentified genders as female, and the identified proportion was then percentile-fit normalized and averaged.

Where appropriate, ratio scores were aggregated to ensure fair comparisons between countries at different baseline factor (such as population and GDP). To compute the indicator score, we will use the average normalization score of the total and the ratio. For example, if a country obtains a normalized score of 5 in total number and 3 in per capita number, then the country's score in this indicator will be 4.

After obtaining indicator scores, we averaged the scores within each dimension and then standardized them to obtain dimension scores. Simple standardization was used to readjust the mean to 50; due to the dispersion of scores based on survey indicators and tools, averages were used without further standardization. We then averaged the dimension scores to obtain pillar scores and averaged the pillar scores to obtain the index score. Here, D4. AI Risk Exposure is a negative factor in P2 aspect, so [100 - dimension score] was used for averaging.

## d) Score Normalization and Data Imputation

For simple normalization, we use:

$$25 * \frac{x_n - \mu}{\sigma} + 50$$

where $\mu$ is the statistical mean of all countries, $\sigma$ is the statistical variance, and $x_n$ is the raw data of the country. After standardization, scores exceeding 0 and 100 were truncated to ensure they remained within the 0-100 range. For percentile-fit normalization, after each simple standardization, we extracted and removed one percentile of scores, then repeated the standardization and extraction on the remaining data until four score quartiles were obtained. This was necessary due to significant clustering in the original data and large magnitude differences, requiring adjustment of the standard deviation for better comparison of data at lower scales.

In the case of missing data within an indicator, we imputed the average of other available data sources within the indicator, given their strong correlation. For missing data in indicator scores, we used rank-adjusted mean imputation. We temporarily imputed 50, calculated the belonging dimension score, estimated the ranking, and then calculated the average score of countries with a close estimated ranking for the final imputation. This imputed value was only used for calculating the dimension score. Experiments showed this method effectively prevents excessively lowering or raising the scores of countries being imputed.

| Country | Important ML systems | 1st SN | 2nd SN | 3rd SN | 4th SN | QN | LN | SN | Raw |
|---|---|---|---|---|---|---|---|---|---|
| **US** | 255 | **100** | | | | **100** | **100** | **100** | **255** |
| **China** | 44 | **54** | **65** | | | **65** | **17** | **54** | **44** |
| **UK** | 54 | **58** | **74** | | | **74** | **21** | **58** | **54** |
| **Germany** | 17 | **44** | **40** | **49** | | **49** | **7** | **44** | **17** |
| **France** | 16 | **44** | **38** | **47** | | **47** | **6** | **44** | **16** |
| **Japan** | 2 | **39** | **26** | **19** | **15** | **15** | **1** | **39** | **2** |
| **Canada** | 48 | **56** | **68** | | | **68** | **19** | **56** | **48** |
| **Italy** | 4 | **40** | **27** | **23** | **23** | **23** | **2** | **40** | **4** |
| **Russia** | 1 | **38** | **25** | **17** | **11** | **11** | **0** | **38** | **1** |
| **India** | 3 | **39** | **26** | **21** | **19** | **19** | **1** | **39** | **3** |
| **Brazil** | 0 | **38** | **23** | **15** | **7** | **0** | **0** | **0** | **0** |
| **S.Africa** | 0 | **38** | **23** | **15** | **7** | **0** | **0** | **0** | **0** |
| **Singapore** | 6 | **40** | **29** | **27** | | **27** | **2** | **40** | **6** |
| **UAE** | 0 | **38** | **23** | **15** | **7** | **0** | **0** | **0** | **0** |
| Mean | 32.1 | 47.6 | 37.5 | 24.7 | 12.6 | 45.3 | 16.0 | 50.2 | 40.9 |
| SDV | 67.0 | 16.7 | 18.8 | 12.6 | 6.4 | 28.9 | 28.9 | 18.1 | 73.7 |

Comparing the purple entries, the difference between the LN socre of US (255) and China (44) is much greater than the difference between the LN score of China (44) and Japan (2). However, 255 is around 6 times of 44, 44 is 22 times of 1, the LN score can not reflect the gap in propotion. In comparison, QN scores reflects propotional gap well.

Comparing the blue entries, SN can not distinguish small differences well, such as that of 0, 1 and 3. That is mainly because the scale of scoring is set by significantly larger values. In comparison, QN's scale reduces for small values, so the small differences can be reflected in the QN scores well.

*Figure 30 Comparison of Quantile-adapted normalization with other normalization Methods*

# 3. Other Related Indexes

In our survey of existing AI indexes, we compiled six evaluation systems or indices that related to the field of AI governance to some extent, which provide useful information on countries' AI capabilities and governance themes. These include UNESCO's Readiness Assessment Methodology for AI Governance (RAM), Tortoise Media's Global AI Index (GAII), Oxford Insights' Government Readiness Index (GRI), OECD's Going Digital Toolkit (GDT), AI Vibrancy Toolkit (AIVT) from Stanford Institute for Human-Centered Artificial Intelligence (HAI) , and the European council's Center for AI and Digital Policy Centre's AI and Democratic Values Index (CAIDP AIDVI). The following table, based on the AGILE Index framework, organizes, and compares the content of each assessment system.

*Table 7* ***Comparisons among AI-governance-related indexes***

| | Development | Governance Environment | | Governance Instruments | | | | |
|---|---|---|---|---|---|---|---|---|
| Index | AI Development | Risk Exposure | Government Quality | Strategy and Norms | Governance Institute | Laws and Standards | Impact Assessment | International Participation |
| AGILE Index | 2 | 2 | 2 | 2 | 1 | 2 | 2 | 2 |
| Unesco RAM | 2 | 1 | 2 | 3 | 1 | 3 | 2 | 1 |
| Tortoise Media GAII | 3 | 1 | 1 | 2 | 1 | 1 | 0 | 1 |
| Oxford Insights GRI | 2 | 1 | 3 | 2 | 0 | 1 | 0 | 2 |
| OECD GDT | 1 | 1 | 1 | 1 | 0 | 0 | 0 | 0 |
| Standford HAI AIVT | 2 | 0 | 0 | 0 | 0 | 0 | 0 | 0 |
| CAIDP AIDVI | 0 | 0 | 1 | 1 | 0 | 1 | 0 | 2 |

| | Governance Effectiveness | | | | | Evaluation Method | | |
|---|---|---|---|---|---|---|---|---|
| Index | Public Perception | Development inclusivity | Research Openness | AI Governance Research | AI4SDG Activity | AI Governance Topic Relevancy | Index Transparency | Unified Process and Management |
| AGILE Index | 2 | 2 | 2 | 2 | 2 | 3 | 3 | 2 |
| Unesco RAM | 2 | 2 | 2 | 2 | 2 | 3 | 1 | 1 |
| Tortoise Media GAII | 2 | 1 | 2 | 0 | 0 | 2 | 2 | 2 |
| Oxford Insights GRI | 2 | 1 | 2 | 0 | 0 | 2 | 2 | 2 |
| OECD GDT | 2 | 2 | 1 | 0 | 0 | 1 | 3 | 3 |
| Standford HAI AIVT | 1 | 0 | 1 | 0 | 0 | 1 | 3 | 2 |
| CAIDP AIDVI | 0 | 0 | 0 | 0 | 0 | 1 | 3 | 2 |

| Color : | 3 | Strong Coverage | 2 | Basic Coverage | 1 | Weak Coverage | 0 | No Coverage |
|---|---|---|---|---|---|---|---|---|

In the table, each row corresponds to a distinct index, while the columns are segmented into five categories based on the AGILE Index framework's four pillars plus the evaluation method used. The column titles, except for the last, align with AGILE Index's dimensions. Within the table:

- A score of 3 signifies a robust assessment in the dimension that the column represents, indicating

strong alignment with AGILE Index's criteria.

- A score of 2 reflects a basic or moderate level of assessment.
- A score of 1 indicates a minimal or inadequate assessment in the dimension.
- A score of 0 denotes the absence of assessment in that particular dimension.

The last section of the table examines four specific questions related to the assessment method, with each question assigned to one column. These scores, ranging from 0 to 3, gauge how comprehensively the index addresses the respective aspect of AI governance. For instance, a relevance score of 2 suggests that the index generally pertains to AI governance assessment but has gaps in several key governance areas.

### Index comparison 1: The Unique Position of RAM

As an official document of UNESCO, RAM is intended to be a key tool for assessing countries' AI governance readiness, making it the most relevant index to AI governance themes and thus occupying a unique position. However, RAM currently only provides a methodology, and the subsequent index, report, website, and rankings are still under development and have not been released. Currently, RAM focuses on assessing Small Island Developing States (SIDS), African countries, and Latin American countries; leading AI countries and regions like the EU, UK, USA, and China have not yet been included in the assessment. RAM is currently building local networks to support data collection. According to the methodology, not too much work will be conducted at the international institutional level. Most tasks, including creating national reports, will be carried out at the national level, with central agencies responsible for publishing them on the AI Observatory. Therefore, AGILE Index and RAM are complementary. RAM provides AGILE Index with assessment references and theoretical basis, while AGILE Index compensates for the complexity and limited country selection in RAM's publishing process.

### Index comparison 2: Assessment Dimensions of AGILE Index and RAM

Compared to RAM, AGILE Index provides a more comprehensive assessment of national governance environments by incorporating observed numbers of AI events in each country, WGI indices, etc. At the same time, AGILE Index uses a broader set of indicators to assess various forms of international participation, while RAM only assesses each country's participation in ISO standard development. Overall, RAM places strong emphasis on the existence of laws and strategies, whereas AGILE Index's set of indicators is broader and more balanced.

## Index comparison 3: Relevance of Other Indexes to AI Governance

Tortoise Media's index focuses on each country's AI capabilities, which is related to AI governance but primarily covers development pillar. Oxford Insights' GRI focuses on how governments use AI and related technologies to enhance capabilities, focusing on pillar on governance environment. However, effective governance also requires other instruments, such as legislation and ethical norms, which GRI does not consider. GDT's low relevance is because its main goal is to provide a tool for assessing the effectiveness of government digital transformation. Many of its indicators are designed for the internet and have no direct relation to AI governance. As shown in the table, Stanford's AIVT mainly focuses on development and effectiveness, while the AIDVI focuses on environment and instruments. The narrow focus limits their comprehensive reflection of each country's AI governance capabilities.

## Index comparison 4: The AGILE Index's Comprehensive Assessment Approach Relative to Other Indexes

The AGILE Index offers a thorough evaluation of governance instruments, setting a notable standard in this domain. Other indexes tend to have more focused scopes and may not cover certain aspects as extensively, such as national governance research and AI4SDGs activities. These elements are particularly emphasized in UNESCO's recommendations and represent an evolving area of interest that could enhance the comprehensiveness of AI governance assessments.

## Index comparison 5: Sharing of Meta-Information and Data Openness

All the indexes mentioned have published their methodologies. Additionally, all indexes except UNESCO RAM have published related scores and data. AGILE Index, Tortoise Media GAII, OECD GDT, and Stanford AIVT have also released interactive websites for querying. OECD, Stanford AIVT, EC CAIDP, and AGILE Index have published additional data and information by released reports.

# 4. Links to Illustrations

**AGILE Index**

Website: https://agile-index.ai/
Email: contact@long-term-ai.cn